\documentclass[onecolumn]{aastex62}

\usepackage{soul}
\usepackage{threeparttable}
\graphicspath{{./}{figures/}}

\shorttitle{Disk-Hosting Members of Young Stellar Associations}
\shortauthors{Higashio et al.}
\begin{document}

\title{Disks in Nearby Young Stellar Associations Found Via Virtual Reality}

\author[0000-0002-9470-7802]{Susan Higashio}
\affiliation{NASA Goddard Space Flight Center, Exoplanets and Stellar Astrophysics Laboratory, Code 667, Greenbelt, MD 20771}
\affiliation{Center for Research and Exploration in Space Science and Technology, NASA/GSFC, Greenbelt, MD 20771}

\author[0000-0002-2387-5489]{Marc J. Kuchner}
\affiliation{NASA Goddard Space Flight Center, Exoplanets and Stellar Astrophysics Laboratory, Code 667, Greenbelt, MD 20771}

\author[0000-0002-3741-4181]{Steven M. Silverberg}
\affiliation{MIT Kavli Institute}

\author{Matthew A. Brandt}
\affiliation{NASA Goddard Space Flight Center, Science Data Processing Branch, Code 587, Greenbelt, MD, 20771}

\author{Thomas G. Grubb}
\affiliation{NASA Goddard Space Flight Center, Science Data Processing Branch, Code 587, Greenbelt, MD, 20771}

\author[0000-0002-2592-9612]{Jonathan Gagn\'e}
\affiliation{Plan\'etarium Rio Tinto Alcan, Espace pour la Vie, 4801 av. Pierre-de Coubertin, Montr\'eal, Qu\'ebec, Canada}
\affiliation{Institute for Research on Exoplanets, Universit\'e de Montr\'eal, D\'epartement de Physique, C.P.~6128 Succ. Centre-ville, Montr\'eal, QC H3C~3J7, Canada}

%\author[0000-0002-2592-9612]{Jonathan Gagn{\'e}}
%\affiliation{Plan\'{e}tarium Rio Tinto Alcan, Espace pour la Vie, 4801 ave. Pierre-de Coubertin, Montr\'{e}al, QC H1V 3V4, Canada}

\author[0000-0002-1783-8817]{John H. Debes}
\affiliation{Space Telescope Science Institute, 3700 San Martin Dr., Baltimore, MD 21218, USA}

\author[0000-0001-5347-7062]{Joshua Schlieder}
\affiliation{NASA Goddard Space Flight Center, Exoplanets and Stellar Astrophysics Laboratory, Code 667, Greenbelt, MD 20771}

\author[0000-0001-9209-1808]{John P. Wisniewski}
\affiliation{Homer L. Dodge Department of Physics and Astronomy, University of Oklahoma, 440 W. Brooks Street, Norman, OK 73019, USA}

%\author{Kellen D. Lawson}
%\affiliation{Homer L. Dodge Department of Physics and Astronomy, University of Oklahoma, 440 W. Brooks Street, Norman, OK 73019, USA}

\author[0000-0003-1740-3407]{Stewart Slocum}
\affiliation{Department of Computer Science, Johns Hopkins University}

\author{Alissa S. Bans}
\affiliation{Department of Physics, Emory University, 201 Dowman Drive, Atlanta, GA 30322, USA}

\author[0000-0002-0862-9108]{Shambo Bhattacharjee}
\affiliation{French National Centre for Scientific Research}

\author[0000-0002-2405-6856]{Joseph R. Biggs}
\affiliation{Disk Detective Citizen Scientist}

\author[0000-0002-9766-2400]{Milton K.D. Bosch}
\affiliation{Disk Detective Citizen Scientist}

\author[0000-0002-9622-9605]{Tadeas Cernohous}
\affiliation{Disk Detective Citizen Scientist}

\author[0000-0002-2993-9869]{Katharina Doll}
\affiliation{Disk Detective Citizen Scientist}

\author[0000-0002-4143-2550]{Hugo A. Durantini Luca}
\affiliation{Disk Detective Citizen Scientist}
\affiliation{IATE-OAC, Universidad Nacional de C\'{o}rdoba-CONICET. Laprida 854, X5000 BGR, C\'{o}rdoba, Argentina}

\author{Alexandru Enachioaie}
\affiliation{Disk Detective Citizen Scientist}

\author{Phillip Griffith, Sr.}
\affiliation{Disk Detective Citizen Scientist}

\author{Joshua Hamilton}
\affiliation{Disk Detective Citizen Scientist}

\author[0000-0002-1104-4442]{Jonathan Holden}
\affiliation{Disk Detective Citizen Scientist}

\author[0000-0001-8343-0820]{Michiharu Hyogo}
\affiliation{Disk Detective Citizen Scientist}
\affiliation{Meisei University, 2-1-1 Hodokubo, Hino, Tokyo 191-0042, Japan}

\author{Dawoon Jung}
\affiliation{Disk Detective Citizen Scientist}

\author{Lily Lau}
\affiliation{Disk Detective Citizen Scientist}

\author{Fernanda Pi\~neiro}
\affiliation{Disk Detective Citizen Scientist}

\author{Art Piipuu}
\affiliation{Disk Detective Citizen Scientist}

\author[0000-0002-1825-7133]{Lisa Stiller}
\affiliation{Disk Detective Citizen Scientist}

\collaboration{The Disk Detective Collaboration}

\begin{abstract}

The Disk Detective citizen science project recently released a new catalog of disk candidates found by visual inspection of images from NASA's Wide-Field Infrared Survey Explorer (WISE) mission and other surveys.
%asked members of the public to help search for new stars with excess emission at 22 $\mu$m indicative of circumstellar material. We have combined the results of this project with a machine learning analysis, and 
%available as part of the Mikuski Archive for Space Telescopes (MAST).
We applied this new catalog of well-vetted disk candidates to search for new members of nearby young stellar associations (YSAs) using a novel technique based on Gaia data and virtual reality (VR). We examined AB Doradus, Argus, $\beta$ Pictoris, Carina, Columba, Octans-Near, Tucana-Horologium, and TW Hya by displaying them in VR together with other nearby stars, color-coded to show infrared excesses found via Disk Detective. Using this method allows us to find new association members in mass regimes where isochrones are degenerate.
%We examined the AB Doradus, Argus, $\beta$ Pictoris, Carina, Columba, Octans-Near, Tucana-Horologium, and TW Hya young stellar associations by displaying them in VR together with other nearby stars, color coded to show infrared excesses found via Disk Detective.
%Our technique uses VR to intuitively immerse Viewers immersed in the 3D data via VR can easily see spatial relationships, especially when animated over time.
%We propose twelve new members of these YSAs and re-assign several other stars, .
%This approach allows us to propose twelve new members of young stellar associations, and re-assign several other stars. 
%Many of these new members occupy a mass regime where stellar isochrones are degenerate, explaining why previous searches have missed them.
%We examined AB Doradus, Argus, $\beta$ Pictoris, Carina, Columba, Octans Near, Tucana Horologium, and TW Hydra and found XX new XX new candidate disk-hosting members of these young stellar associations (YSA)s.   
%Using BANYAN $\Sigma$, we find 7 new candidate YSA members with disks, 12 new moving group members with previously identified disks, and 33 new disks around previously identified moving group members.  WE NEED A TABLE OF THESE!  OR WE NEED TO POINT TO THEM IN TABLE 1.
We propose ten new YSA members with infrared excesses:
three of AB Doradus (HD 44775, HD 40540 and HD 44510), one of $\beta$ Pictoris (HD 198472), two of Octans-Near (HD 157165 and BD+35 2953), and four disk-hosting members of a combined population of Carina, Columba and Tucana-Horologium: CPD-57 937, HD 274311, HD 41992, and WISEA J092521.90-673224.8.
%HD44775, HD40540, HD44510, HD198472, HD157165, BD+35 2953, CPD-57 937, HD274311, HD41992, and WISEA J092521.90-673224.8.
%Ten of the new members we propose have infrared excesses. We propose three new disk-hosting members of AB Doradus (HD 44775, HD 40540 and HD 44510), one of $\beta$ Pictoris (HD 198472), and two of Octans-Near (HD 157165 and BD+35 2953). 
%We find Columba and Carina are overlapped, and contiguous with Tucana-Horologium, leading us back to earlier pictures where the three groups comprise different segments of one population. Considering this, we propose four new disk-hosting members of this combined population: CPD-57 937, HD274311, HD41992, and WISEA J092521.90-673224.8.
%We find that Columba and Carina are highly overlapped (we refer to them as Columba-Carina, or Col-Car), and contiguous with Tucana-Horologium, leading us back to earlier pictures of this region where the three groups comprise different segments of one population. With this interpretation in mind, we propose four new disk-hosting members of this combined population: CPD-57 937, HD 274311, HD 41992, and WISEA J092521.90-673224.8.
This last object (J0925) appears to be an extreme debris disk with a fractional infrared luminosity of $3.7 \times 10^{-2}$. 
%One of these, WISEA J092521.90-673224.8, appears to be an extreme debris disk, with a fractional infrared luminosity of $3.7 \times 10^{-2}$. 
We also propose two new members of AB Doradus that do not show infrared excesses: TYC 6518-1857-1 and CPD-25 1292.
We find HD 15115 appears to be a member of Tucana-Horologium rather than $\beta$ Pictoris. 
We advocate for membership in Columba-Carina of HD 30447, CPD-35 525, and HD 35841. Finally, we propose that three M dwarfs, previously considered members of Tuc-Hor are better considered a separate association, tentatively called ``Smethells 165''.

\end{abstract}

%% Keywords should appear after the \end{abstract} command. 
%% See the online documentation for the full list of available subject
%% keywords and the rules for their use.

%% From the front matter, we move on to the body of the paper.
%% Sections are demarcated by \section and \subsection, respectively.
%% Observe the use of the LaTeX \label
%% command after the \subsection to give a symbolic KEY to the
%% subsection for cross-referencing in a \ref command.
%% You can use LaTeX's \ref and \label commands to keep track of
%% cross-references to sections, equations, tables, and figures.
%% That way, if you change the order of any elements, LaTeX will
%% automatically renumber them.
%%
%% We recommend that authors also use the natbib \citep
%% and \citet commands to identify citations.  The citations are
%% tied to the reference list via symbolic KEYs. The KEY corresponds
%% to the KEY in the \bibitem in the reference list below. 

\section{Introduction}

Stars in young moving groups and associations often show excess emission at infrared (IR) wavelengths pointing to the presence of circumstellar disks \citep[e.g.][]{2011ApJ...732...61Z, 2016ApJ...826..123M, 2016ApJ...832...50B}: gas-rich protoplanetary and transitional disks, and gas-poor debris disks.  Since membership in these young moving groups and associations yields stellar age estimates, these disks in young moving groups have become central to our understanding of planet formation. They provide us with a timeline for disk evolution
between $\sim 3$ and 200 Myr, chronicling the transition from primordial and debris stages, and providing critical time constraints for gas and planetesimal accretion and orbit evolution \citep[e.g.][]{2009AIPC.1158....3M, 2017ApJ...836...34M}. Furthermore, many disks with resolved images, like $\beta$ Pictoris, HR 4796A, AU Microscopii, belong to young moving groups; knowing the ages of these disks allows us to examine and model their dynamics in detail \citep[e.g.][]{2015ApJ...798...83N, 2015ApJ...815...61N, 2020ApJ...889L..21G, 2020AJ....159..288M}.

%THIS PARAGRAPH IS TOO GENRAL
%The interest in young moving groups began upon the discovery of 4 stars that formed an association with TW Hya; this group of stars became known as the TW Hya Association \citep{1997Sci...277...67K}.  Soon after, the Horologium Association (HorA) was discovered by \citet{2000AJ....120.1410T}, and the Tucanae Association (TucA) was discovered at the same time by \citet{2000ApJ...535..959Z}.

%\citet{2001ASPC..244...43T} realized that there was a relationship between HorA and TucA.  This led to the Search for Associations Containing Young Stars (SACY) survey \citep{2001ASPC..244...37D,2001ASPC..244...43T,2001ASPC..244...49Q}.  Initially made up of a core of 44 stars containing HorA and part of TucA, it became known as the Great Austral Young Association (GAYA).   These stars were found around 50 pc, with an age of about 30 Myr.  As a result, the membership of GAYA has been further studied and has since evolved, and what is now known as the GAYA complex consists of the Tucana-Horologium (Tuc-Hor), Columba and Carina \citep{2008hsf2.book..757T}.

Thanks to recent releases of data from the European Space Agency's (ESA) Gaia telescope, Data Release 2 (DR2) and Early Data Release 3 (EDR3), there has been a flurry of activity to identify new young stellar association (YSA) candidates. Some of these studies used infrared excess as a clue to group membership, and others used Gaia-informed group membership as a way to find new disks. For example,  \citet{2018AJ....156...75E} announced 179 new disks in the Upper Scorpius Association based on Gaia data and photometry from NASA's Wide-field Infrared Survey Explorer (WISE) mission.
\citet{2020AJ....160...44L} refined this work using DR2 astrometry. \citet{2020AJ....159..200M} identified 27 candidate members of the Lupus star forming region by combining DR2 data with disk determinations based on photometry from NASA's {\it Spitzer Space Telescope}. 
\citet{2019ApJ...887...87Z} described the large percentage of M-type stars with warm excess infrared emission in the $\chi^1$ Fornacis cluster.
\citet{2019ApJ...870...27Z} argued for the validity of the nearby Argus association based in part on a membership list containing 80\% infrared-excess stars.  

Nonetheless, much more remains to be learned about disks in moving groups using Gaia data, especially in concert with infrared photometry from WISE. The presence of a disk can be an especially vital youth indicator for A and F type stars, a mass range where isochrones can be degenerate, and Lithium depletion methods for age assessment fail.

The Disk Detective citizen science project \citep{2016ApJ...830...84K}, launched in 2014, asked members of the public to examine images of WISE sources to search for reliable disk candidates: stars with infrared excess at 22$\mu$m. This effort has generated a catalog of roughly 30,000 strong disk candidates, many of which are known to be YSA members. A representative subset of these sources has been checked for background objects via higher resolution imaging \citep{2018ApJ...868...43S}.  Since possession of a disk is often a sign of youth \citep[but see also the discussion of Be stars in][]{2016ApJ...830...84K}, this new catalog of reliable disk candidates has great potential as a source of new YSA members.

Indeed, we have begun combing the Disk Detective catalog to identify new members of nearby YSAs (including young moving groups), using BANYAN $\Sigma$ \citep{2018ApJ...856...23G} (Silverberg et al., in prep.). Since Gaia DR2 became available in April 2018, several codes and techniques for identifying new moving group candidates have seen wide use. The multivariate Gaussian model  \citep{2018ApJ...856...23G} known as BANYAN $\Sigma$ uses a Bayesian algorithm to identify new YSA  members.  This tool calculates membership probabilities of stars for 27 YSAs within 150 pc of the sun using multivariate Gaussian distributions in six dimensions, that is, X, Y, Z, U, V, W space.  The LocAting Constituent mEmbers In Nearby Groups (LACEwING) algorithm \citep{2017AJ....153...95R} uses a frequentist approach combined with triaxial ellipsoid models and optionally an epicyclic traceback code.
 Additionally, \citet{2019MNRAS.486.3434L} developed a new 4-stage iterative method for evaluating the membership in nearby YSAs. The \citet{2019MNRAS.486.3434L} paper contains 3D ellipsoidal models in XYZ and in UVW for 9 moving groups that are younger than 200 Myr and within 100 pc of the sun.  

%\citet{2020AJ....159..166U} has applied LACEwING to Gaia DR2 data.

These tools have greatly streamlined the process of finding new candidate members of YSAs for most situations. 
However, all of the tools described above, including BANYAN $\Sigma$, are limited by the quality of their YSA models--e.g., the parameters of their tri-axial ellipsoids.  The quality of these models is in turn determined by the number and fidelity of the lists of  assumed bona-fide members of each YSA.  Since our process had the potential to add multiple new members to each moving group, and since Gaia represents such a dramatic shift in the quantity and quality of astrometric data for nearby stars, we sought an alternative, ab initio process for identifying new YSA members that would not rely on such pre-Gaia membership lists.

This paper describes our novel VR-based approach and the new candidate YSA members we have identified.
We describe the Disk Detective citizen science project, and the new database of disk candidates available in the Mikulski Archive for Space Telescopes (MAST) in Section \ref{DiskDetective}. We discuss the VR methodology in Section \ref{methodology}.  We discuss the newly identified YSA members with disks in Sections \ref{NewCandidateYSAs}, \ref{Discussion_Car-Col_and_THA} and \ref{discussion_SED}, with further discussion about WISEA J092521.90-673224.8 in Section \ref{discussion_PP}.

%We vizualized the positions and velocities of these candidates and compared them to known GAYA members using a new virtual reality tool based on the PointClouds VR application.  This approach allowed us to find new likely members of this association that have been previously overlooked because they were not previously understood to show signs of youth.

%We discuss the disk candidates whose motion is similar to a pair of known associations, Columba (COL) and Carina (CAR) in section \ref{candidates}.

%The method that we use is based on VR technology, which provides a different perspective that may have found moving group candidates that the other models may have missed.  These findings are discussed in section \ref{results}.

\section{The Disk Detective Citizen Science Project}
\label{DiskDetective}

The Disk Detective citizen science project, first launched in January 2014, asks members of the public to examine images of sources from the AllWISE catalog to search for reliable disk candidates. 
%The input catalog of 277,686 sources was pre-selected using AllWISE oolor cuts, flags and other computed criteria \citep[see][]{2016ApJ...830...84K} to have excess emission at 22 $\mu$m (the WISE mission's W4 band). 
Most objects that appear to have infrared excesses based on simple WISE color cuts are in fact false positives created by confusion or contamination \citep{2012MNRAS.426...91K, 2018ApJ...868...43S, 2020ApJ...891...97D}.  Disk Detective solves this problem through citizen science; volunteers at Disk Detective have worked to weed out most of the false positives by examining images of source from the AllWISE catalog, the 2MASS project, the Sloan Digital Sky Survey and the DSS \citep{2018ApJ...868...43S}.

The Disk Detective project has resulted in several noteworthy discoveries, including a new category of disks, the ``Peter Pan Disks" \citep{2016ApJ...830L..28S,2020ApJ...890..106S, 2020MNRAS.494...62L, 2020MNRAS.496L.111C}.  Peter Pan disks are so called because the stellar ages, based on moving group membership, indicate that these gas-rich disks have lasted far longer than the typical lifetimes assumed for such disks. Disk Detective citizen scientists discovered the first example of this phenomenon \citep{2016ApJ...830L..28S, 2018MNRAS.476.3290M} and several later examples \citep{2020ApJ...890..106S}. The Disk Detective project discovered the closest brown dwarf disk younger than 10 Myr \citep{2020AAS...23631906S} and twelve likely warm dust disks around candidate members of co-moving pairs based on Gaia astrometry \citep{2018ApJ...868...43S}, which supports the hypothesis that warm dust is associated with binary systems \citep{2015ApJ...798...86Z}.
Disk Detective citizen scientists also helped locate the oldest known disk around a white dwarf \citep{2019ApJ...872L..25D} and the first debris disk discovered around a star with a white dwarf companion \citep{2016ApJ...830...84K}.  Fifteen citizen scientists have become named co-authors on refereed publications as a result of their participation in Disk Detective (not including this paper).

To extend Disk Detective information to those objects that were not fully classified before the project went offline on April 30, 2019, we used the objects that had been fully classified to train a GoogLeNet \citep{2014arXiv1409.4842S} neural network. The full details of this process will be presented in Silverberg et al. (in prep). Briefly, we composited the nine images used on the Disk Detective site for each subject into one image, along with the classification information for that subject, and used this to train the neural network. We tested three methods to determine whether a given subject was a good disk candidate: if ``None of the above'' was a simple majority of classifications (\texttt{majGood}), a plurality of classifications (\texttt{pluralGood}), or made up at least 20\% of classifications (\texttt{twentyGood}). The neural network then assigned each new (unclassified) subject a score of 0 to 100 based on each of the above neural networks. We then asked four expert citizen scientists to blindly evaluate 476 subjects with a ``good'' score $>90$ out of 100 by the ``majority''-trained neural network. Of these 476 subjects that were identified as likely good disk candidates by the neural network, the citizen scientists found 118 to be false positives; i.e. 25\% of the subjects classified as good by the neural network would not have been classified as ``None of the Above/Good disk candidate'' by citizen scientists. While this false positive rate is non-negligible, it is low enough to be useful for some purposes---e.g. searching for candidate members of YSAs. We added the subjects classified by the neural net (those with the majGood parameter $> 90$) to the list of Disk Detective disk candidates we used in this paper. A catalog of Disk Detective data is available to the public in the MAST archive\footnote{\url{https://mast.stsci.edu}}. This catalog contains all of the disk candidates we used, including those selected by the neural network.

\section{Examining Young Stellar Associations Using Virtual Reality}
\label{methodology}

%all of these methods struggle when faced with a complex system of stars like the GAYA, which contains multiple components that overlap in position and velocity space.  

%For this alternative approach, we examined Gaia DR2 data using the PointClouds VR application combined with the HTC Vive system.  PointClouds VR, available on GitHub at \url{https://github.com/nasa/PointCloudsVR}, provides an immersive experience where the viewer uses two controllers to move around in and pan in and out of an interactive 360$^{\circ}$ scene.  
%We used a custom layer developed by the 
%Augmented Reality/Virtual Reality (AR/VR) Development Laboratory at NASA's Goddard Spaceflight Center (GSFC) to upload and visualize the Gaia data, and animate it so that the stars can be told to move according to the space velocities measured by Gaia.   VR has seen variety of applications in astrophysics during the last few years, including displaying Ly$\alpha$ maps \citep{2018ApJS..237...31L}, mapping the morphology of a molecular cloud \citep{2019MNRAS.484.2089R} and cataloging galaxy groups \citep{2020MNRAS.tmp.2068L}. The GSFC (AR/VR) Development Laboratory has developed VR software for a variety of space science applications, including heliophysics and planetary science applications.

 We have been examining Gaia DR2 data on positions and velocities of stars using the PointCloudsVR application combined with the HTC Vive system.  PointCloudsVR, available on GitHub at \url{https://github.com/nasa/PointCloudsVR}, leverages the advantages of virtual reality to intuitively immerse viewers in the 3D data and easily see spatial relationships.  It uses two controllers to move around in and zoom/pan in and out of an interactive 360$^{\circ}$ scene.   We  used  a  custom application layer  developed  by  the  Augmented  Reality/Virtual  Reality  (AR/VR)  Research \& Development Laboratory at NASA’s Goddard Spaceflight Center (GSFC) to load and visualize the Gaia data, and animate the stars according to the space velocities measured by Gaia.  This (AR/VR) Research \& Development Laboratory has developed VR software for a variety of space science and engineering applications, including heliophysics and planetary science applications.  VR in general has seen a variety of applications in astrophysics during the last few years, including displaying Ly$\alpha$ maps \citep{2018ApJS..237...31L}, mapping the morphology of a molecular cloud \citep{2019MNRAS.484.2089R} and cataloging galaxy groups \citep{2020MNRAS.tmp.2068L}.
 
%In VR, when the position and motion of these young stars are observed as they evolve over time, patterns of stars that move together start to appear.

%With the release of Gaia DR2 in April 2018, these stars were loaded into the PointClouds VR app. 

Of the 1.7 billion stars in DR2  \citep{2018AandA...616A...1G}, approximately 4 million stars have both Gaia-measured radial velocities and Gaia-measured parallaxes greater than 0.6. We loaded this sample into the PointCloudsVR app.  
We also loaded lists of the Gaia IDs of members of 40 YSAs from the literature (see below).

\begin{figure}[ht]
    \centering
    \includegraphics[width=0.8\linewidth]{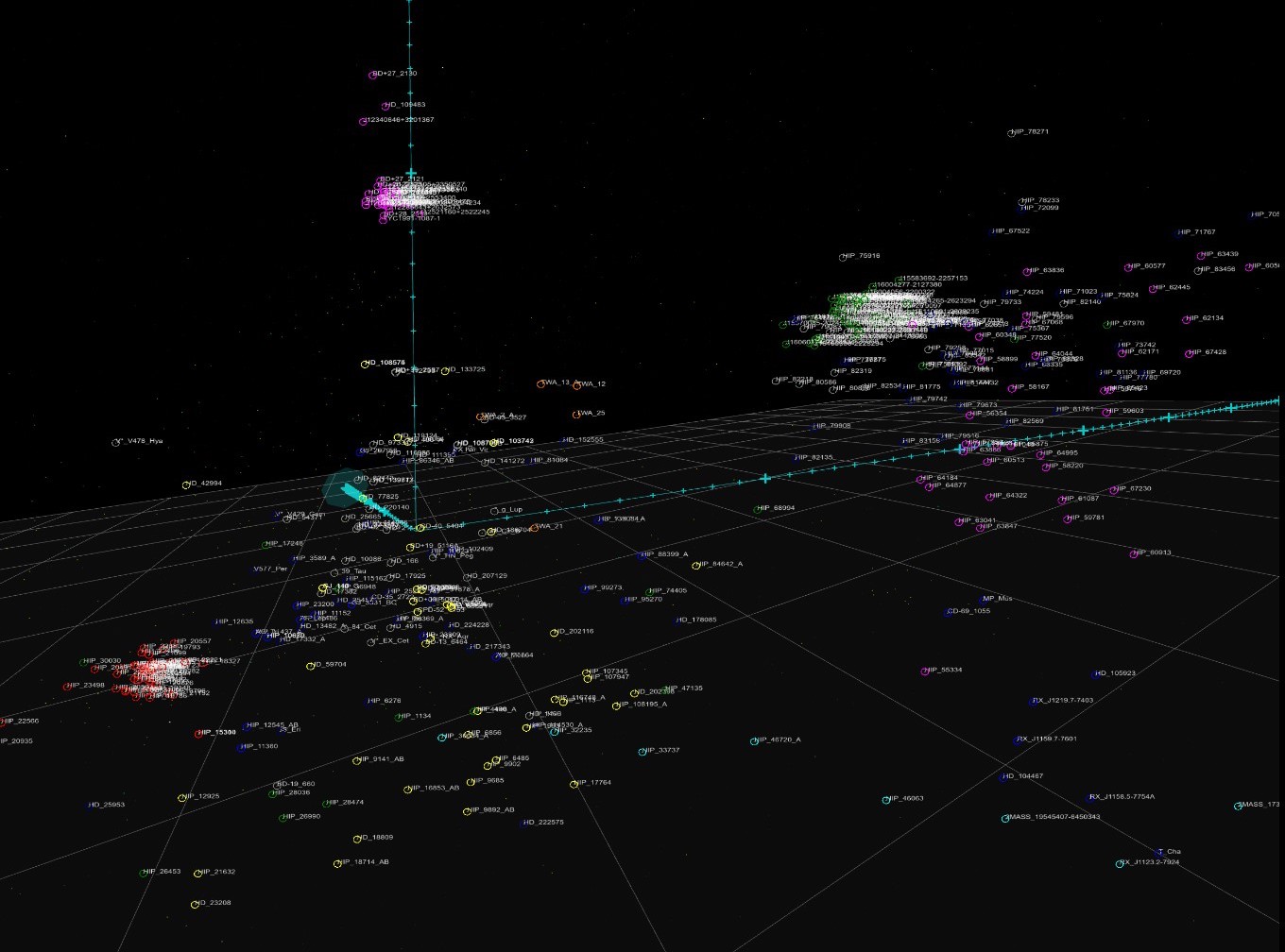}
    \caption{Screen shot of our  PointCloudsVR simulation. This view would ordinarily be experienced through a virtual reality headset. Points represent the positions of stars in YSAs plotted in Galactic (X, Y, Z) coordinates. The app allows us to advance the positions of all the points forward (and move them backward) in time using Gaia DR2 velocities together with newly identified disk hosting stars. Clearly visible are the Coma Ber (pink, top), Hyades (orange, lower left) Upper Scorpius (green, top right) associations, Octans (yellow). The X-axis points to the right, the Y-axis points to the left and the Z-axis points up. The big tick marks on the X, Y, Z axes are 100 pc apart; the small ticks are 10 pc apart.}
    \label{fig:pointclouds}
\end{figure}

We then loaded the Gaia IDs of our catalog of 48,965 disk candidates from Disk Detective, including 13,626 that were determined by the neural network, as described above.  We used these lists to highlight and color code the representations of the Gaia DR2 stars in our VR simulations. 
For example, in one run all the stars with disk candidates from Disk Detective were color-coded orange, and all the known members of the Tuc-Hor association green, and turned off all the rest of the Gaia DR2 catalog.  Figure \ref{fig:pointclouds} shows a representative screenshot of the PointCloudsVR simulation---collapsed to two dimensions of course. We used this color coding to highlight various lists of stars to search for new YSA members, one YSA at a time, using the following iterative process.

First, we highlighted the members of the YSA, observed them from various angles and distances to get a sense of its distribution in XYZ space, and evolved the stellar positions backwards and forwards in time to develop an intuition for the group's velocity distribution. Then we highlighted our disk candidates alongside the known YSA members and compared them to look for possible new kinematic YSA members.  Once we had recorded a preliminary list of the disk candidates associated with the YSA, we created a new catalog combining the new candidate members with the previously published members. We then highlighted only that new combined catalog, examined it, and removed any obvious interlopers.  Then we repeated this process, highlighting the whole Disk Detective catalog again, comparing it with the new combined catalog to find new candidate members, again combining any new candidates and checking for interlopers.  We repeated this inspection process for each YSA in our database until it stopped turning up new candidates.

%To conduct the search in VR, the COL catalog was illuminated and observed from various angles and distances to find the group's location in XYZ space, evolving the YSA backwards and forwards over time to try to understand the group's motion UVW. Then, the Disk Detective catalog was illuminated and objects were observed together.  The objects that were in both catalogs were recorded, and then all disk candidates were examined visually, starting with those closest in proximity, to find those that stayed together with COL as they evolved both backwards and forwards in time.

%Once the disk candidates that appeared to move together in COL  were recorded, a new catalog of these objects was created and reviewed in VR.  Any obvious interlopers were removed and the Disk Detective catalog was scanned again to find any objects that may have been missed.  This was repeated until we were satisfied that the visual inspection had captured all possible COL candidates.

%The same process was carried out with CAR. Two objects were recovered that appear to have a similar motion and direction as CAR, that had also been found in our search for COL. Since both COL and CAR are close in proximity and have similar motion, it was difficult to distinguish the difference between the two groups based on visual inspection alone, and the groups were often illuminated together to review the motion of disk candidates.

%Although numerical clustering algorithms can also be used to search for and scope out the size of these six-dimensional clumps, our VR approach has multiple advantages over clustering techniques.  First, YSAs vary greatly in their degrees of clustering, and some of them have multiple components, such as a tight core combined with a more diffuse halo.  The human eye has no difficulty recognizing these structures, which could easily fool a clustering algorithm unless it were repeatedly fussed with and re-tuned. Second, Gaia detection limits impose a luminosity-dependent edge to any large structure on the side away from the Sun.  A clustering algorithm would need to be taught about this effect, while a human recognizes it instinctively. 

Although numerical clustering algorithms can also be used to search for and scope out the size of these six-dimensional clumps, our VR approach has multiple advantages over clustering techniques.  First, YSAs vary greatly in their degrees of clustering, and some of them have multiple components, such as a tight core combined with a more diffuse halo.  The human eye has no difficulty recognizing these structures, which could easily fool a clustering algorithm unless it were repeatedly re-tuned. Second, Gaia detection limits impose a luminosity-dependent edge to any large structure on the side away from the Sun.  A clustering algorithm would need to be taught about this effect, while a human eye recognizes it instinctively.  

We worked our way through 8 different YSAs using VR, comparing the Disk Detective catalog bit by bit against the lists of known YSA members. Table \ref{table:DDinknownMGs} lists all of the proposed new candidate YSA members we found.  %Table \ref{table:DDReferences} lists the references for Table \ref{table:DDinknownMGs}.  
Figures \ref{fig:vectors_ABDor-LS} through \ref{fig:HRdiagCarColTucHorDD} show position/velocity and colour magnitude diagrams for each of the YSAs where we identified likely new members.  The assumed members are plotted in blue, and the sources of each membership list appear in the figure captions. The proposed new members are plotted in red. 

\section{New Candidate YSA Members And YSA Members with Disks}
\label{NewCandidateYSAs}

Here we discuss each of the YSAs we examined except for Carina, Columba and Tucana Horologium. We will examine those three together in a subsequent section. We also briefly discuss each star of interest.

\begin{longrotatetable}
%\begin{center}
\begin{longtable}{ llllllll }
\caption{Candidate YSA Members with Disks Assigned Using VR} \label{table:DDinknownMGs} \\

\hline 
\multicolumn{1}{l}{\textbf{Gaia DR2 ID}} & \multicolumn{1}{l}{\textbf{Common Name/}} & \multicolumn{1}{l}{\textbf{Proposed}} & \multicolumn{1}{l}{\textbf{Prev. Published}} & \multicolumn{1}{l}{\textbf{Literature YSA}} & \multicolumn{2}{c}{\textbf{BANYAN $\Sigma$}} & \multicolumn{1}{l}{\textbf{W1-W4}}   \\
\multicolumn{1}{c}{} & 
\multicolumn{1}{c}{WISE ID} & \multicolumn{1}{l}{\textbf{YSA}} & \multicolumn{1}{l}{\textbf{Infrared Excess}} &
\multicolumn{1}{c}{}&
\multicolumn{1}{l}{\textbf{YSA}} &
\multicolumn{1}{l}{\textbf{Prob}} & \multicolumn{1}{c}{}
\\

\hline 
\endfirsthead

\multicolumn{8}{c}%
{{\bfseries \tablename\ \thetable{} -- continued from previous page}} \\
\hline 
\multicolumn{1}{l}{\textbf{Gaia DR2 ID}} & \multicolumn{1}{l}{\textbf{Common Name/}} & \multicolumn{1}{l}{\textbf{Proposed}} & \multicolumn{1}{l}{\textbf{Prev. Published}} & \multicolumn{1}{l}{\textbf{Literature YSA}} & \multicolumn{2}{c}{\textbf{BANYAN $\Sigma$}}  & \multicolumn{1}{l}{\textbf{W1-W4}} 
\\ 
\multicolumn{1}{c}{} & 
\multicolumn{1}{c}{WISE ID} & \multicolumn{1}{l}{\textbf{YSA}} & \multicolumn{1}{l}{\textbf{Infrared Excess}} &
\multicolumn{1}{c}{}&
\multicolumn{1}{l}{\textbf{YSA}} &
\multicolumn{1}{l}{\textbf{Prob}}   & \multicolumn{1}{c}{} \\
\hline 
\endhead

\hline \multicolumn{8}{r}{{Continued on next page}} \\ \hline
\endfoot

\hline \hline
\endlastfoot
%\begin{table}[ht]
%\caption{Disk Candidates that appear to have similar velocity and position as Moving Groups as observed in VR
%\label{table:DDinknownMGs}}
%\tablewidth{\textwidth}
%\begin{center}
%\begin{tabular}{ lllllll }
    %\hline
    %Gaia DR2 ID & WISE ID & Possible MG & Known Disk? & Known MG? & BANYAN $\Sigma$ & Prob \\
    %\hline
    2898071312314111488	&	HD 44775 / 	&	AB Dor	&	1	&	None	&	ABDMG/FIELD	&	0.8/99.2    &   $0.517 \pm 0.07$ \\
        &   J062218.66-295134.7	&	&	&	 & 	&  \\
    2889357751382351872	&	HD 40540 / 	&	AB Dor	&	2, 3, 4, 5, 6, 7	&	None	&	ABDMG/FIELD	&	6.3/93.7    &   $1.143 \pm 0.07$ \\
        &   J055752.60-342834.1	&	&   9, 43	&	&	& \\
    5499867942228621056	&	HD 44510 / 	&	AB Dor	&	8	&	None	&	ABDMG/FIELD	&	1.9/98.1    &   $0.533 \pm 0.07$ \\
        &   J061903.93-535823.9	&	&	&	&   & \\
    6470519830886970880	&	HD 198472 / &	$\beta$ Pic	&	5, 9, 10   &	$\beta$ Pic (11$\dagger$)	&	BPMG/FIELD  &	70.6/29.4   &   $0.519 \pm 0.09$ \\
        &   J205241.67-531624.8 &   &   &   &   &   \\
%    353635070145291008	&	J022113.11+460006.7	&	Beta Pic	&   13  &   11   & BPMG/FIELD  &   0.1/99.9    &   $0.523 \pm 0.10$ \\
    1336772153854267392	&	HD 157165 / 	&	Oct Nr	&	14, 15	&   None	&   FIELD   &	99.9 &   $0.439 \pm 0.08$  \\
        &   J172007.53+354103.6 &   &   &   &   &   \\
    2517397846786452224	&	HD 15115 / 	& Tuc-Hor	&	3, 5, 9, 16, 17, 18	&	$\beta$ Pic (3, 19, 20, 29) &	THA/FIELD   &   98.6/1.4    &   $0.545 \pm 0.08$ \\
        &   J022616.32+061732.8 &   &   35, 36, 37, 38, 39,   &   AB Dor (21)  &   &   \\
        &   &   &   40, 43, 44, 45 &   Tuc-Hor (22, 40)    &   &   \\
        &   &   &   &   Col (23)    &   &   \\
    5247128354025136128	&	TYC 9196-2916-1 /   &	Tuc-Hor	&	No	&	None	&   LCC/FIELD   &	25.2/74.8 &   $3.735 \pm 0.04$  \\ % too far away?
        &   J092521.90-673224.8 &   &   & & & \\
    2896173349085130112	&	HD 41992 / 	&	Tuc-Hor	&	13	&	Col/Car (12)	&	THA/FIELD	&	1.4/98.6 &   $0.836 \pm 0.08$   \\ % too far away?
        &   J060652.79-313054.1 &   &   &   &   &   \\    
    4881312593414911616 &   HD 30447 /  &   Col-Car &   2, 3, 5, 9, 16, 17, & Col (12, 19, 22, 24, 25) &   COL/FIELD &    99.5/0.5 &   $0.872 \pm 0.08$ \\
        &   J044649.55-261808.8 &   &   34, 41, 42, 43, 44, 46  &   &   &   \\
    4867855155206070784 &   CPD-35 525 /  &   Col-Car &   5   &  Col (26)    &  COL/FIELD  &   99.9/0.1 &   $1.063 \pm 0.09$ \\
        &   J044115.76-351358.1	&	&	&   &   & \\
    5498909546046688256 &   CPD-57 937 /  &   Col-Car &   5  & None  &   COL/CAR & 15.8/0.2/ &   $1.507 \pm 0.07$ \\
        &   J060210.78-570142.1	&	&	&	&	FIELD	&   84.1    &    \\
    2958833623399371392 &   HD 35841 /    &   Col-Car &   3, 5, 27, 28  &   Col (29, 30)   &   COL/FIELD   &   52.0/48.0    &   $1.121 \pm 0.08$ \\
        &   J052636.59-222923.8 &	& 34, 41, 44, 46 	&	&   & \\
    4715896429133940352 &   HD 10472 /    &   Col-Car  &  2, 3, 4, 5, 9, 10,    &    Tuc-Hor     &    COL/FIELD   &   77.8/22.2   &   $0.579 \pm 0.08$ \\
        &   J014024.15-605956.7 &	&   16, 17, 43, 44, 46	&   (4, 20, 31, 32) & \\
    4807752860334734720 &   HD 37852 /    &   Col-Car &   9  &   Col (33$\dagger$) &   COL/FIELD   &   83.5/16.5  &   $0.677 \pm 0.07$ \\
        &   J053930.48-404102.4 &	&	&   &	& \\
    
    %\hline
%\end{tabular}

%\end{center}
%\end{table}

\end{longtable}
    \begin{tablenotes}
        \item \small Notes.  References:
        1) \citet{2015ApJ...804..146D}, 
        2) \citet{2007ApJ...660.1556R},
        3) \citet{2014ApJS..211...25C},
        4) \citet{2015ApJ...798...87M},
        5) \citet{2016ApJS..225...15C},
        6) \citet{2017MNRAS.470.3606H},
        7) \citet{2017MNRAS.469..521K},
        8) \citet{2014MNRAS.437..391C},
        9) \citet{2013ApJS..208...29W},
        10) \citet{2014ApJS..212...10P},
        11) \citet{2019MNRAS.489.3625C},
        12) \citet{2021ApJ...915L..29G},
        13) \citet{2016ApJ...830L..28S},
        14) \citet{2018ApJ...868...43S},    
        15) \citet{2016ApJ...830...84K},    
        16) \citet{2013ApJ...775...55B},    
        17) \citet{2017ApJ...845..120B},    
        18) \citet{Engler2019},    
        19) \citet{2013ApJ...762...88M},    
        20) \citet{2017AJ....154..245M},    
        21) \citet{2018ApJ...862..138G},    
        22) \citet{Vigan2017},    
        23) \citet{2015MNRAS.454..593B},    
        24) \citet{Elliott2016},    
        25) \citet{2018ApJ...856...23G},    
        26) \citet{2018ApJ...860...43G},
        27) \citet{RiviereMarichalar2016},
        28) \citet{2018AJ....156...47E},
        29) \citet{2008hsf2.book..757T}, 
        30) \citet{2009A&A...508..833D},
        31) \citet{2000AJ....120.1410T},      
        32) \citet{2001ApJ...559..388Z},  
        33) \citet{2018ApJ...863...91F},
        34) \citet{2011ApJS..193....4M},
        35) \citet{2008ApJ...684L..41D},
        36) \citet{2012ApJ...752...57R},
        37) \citet{2014A&A...569A..29M},
        38) \citet{2019A&A...622A.192E},
        39) \citet{2020AJ....160..163L},
        40) \citet{2019ApJ...877L..32M},
        41) \citet{2014ApJ...786L..23S},
        42) \citet{2018A&A...611A..43L},
        43) \citet{2021RAA....21...60L},
        44) \citet{2020AJ....160...24E},
        45) \citet{2021MNRAS.502.5390P},
        46) \citet{2006ApJ...644..525M}.
        Kinematic match only have been marked with a $\dagger$.
    \end{tablenotes} 
%\end{center}
\end{longrotatetable}

\begin{longrotatetable}
%\begin{center}
\begin{longtable}{ lllllll }
\caption{Comovers and Companions to Candidate YSA Members with Disks Assigned Using VR} \label{table:ComoverstoDDinknownMGs} \\

\hline 
\multicolumn{1}{l}{\textbf{Gaia DR2 ID}} & \multicolumn{1}{l}{\textbf{Common Name/}} & \multicolumn{1}{l}{\textbf{Proposed}} & \multicolumn{1}{l}{\textbf{Prev. Published}} & \multicolumn{1}{l}{\textbf{Literature YSA}} & \multicolumn{2}{c}{\textbf{BANYAN $\Sigma$}}   \\ \multicolumn{1}{c}{} & 
\multicolumn{1}{c}{WISE ID} & \multicolumn{1}{l}{\textbf{YSA}} & \multicolumn{1}{l}{\textbf{Infrared Excess}} &
\multicolumn{1}{c}{\textbf{}}&
\multicolumn{1}{l}{\textbf{YSA}} &
\multicolumn{1}{l}{\textbf{Prob}} 
\\

\hline 
\endfirsthead

\multicolumn{7}{c}%
{{\bfseries \tablename\ \thetable{} -- continued from previous page}} \\
\hline 
\multicolumn{1}{l}{\textbf{Gaia DR2 ID}} & \multicolumn{1}{l}{\textbf{Common Name/}} & \multicolumn{1}{l}{\textbf{Proposed}} & \multicolumn{1}{l}{\textbf{Prev. Published}} & \multicolumn{1}{l}{\textbf{Literature YSA}} & \multicolumn{2}{c}{\textbf{BANYAN $\Sigma$}}  
\\ 
\multicolumn{1}{c}{} & 
\multicolumn{1}{c}{WISE ID} & \multicolumn{1}{l}{\textbf{YSA}} & \multicolumn{1}{l}{\textbf{Infrared Excess}} &
\multicolumn{1}{c}{\textbf{}}&
\multicolumn{1}{l}{\textbf{YSA}} &
\multicolumn{1}{l}{\textbf{Prob}} \\
\hline 
\endhead

\hline \multicolumn{7}{r}{{Continued on next page}} \\ \hline
\endfoot

\hline \hline
\endlastfoot

    2898402643271131264 &	TYC 6518-1857-1 	&	AB Dor	&	No	&   1	&	ABDMG/FIELD	&   5.9/94.1    \\
%       &	&	&	&   &   &  \\
    2911909593862390912	&	CPD-25 1292 &	AB Dor	&	No	&	1	&	ABDMG/FIELD	&   5.1/94.9    \\
%        &	&	&	&	&	& \\
    1336772325652957952	&	BD+35 2953  &	Oct Nr	&	2	&   3	&	N/A	&	N/A \\    
    4806146576925723264 &   HD 274311 &   Col-Car & 4   &   1   &   COL/FIELD   &   52.8/47.2     \\
%       &	&	&	&	&	& \\

\end{longtable}
    \begin{tablenotes}
        \item \small Notes.  References:  
        1) \citet{2017AJ....153..257O}, 
        2) \citet{Guillout2009},
        3) \citet{2017MNRAS.472..675A},
        4) \citet{2009ApJS..184..138H}.
    \end{tablenotes}
%\end{center}
\end{longrotatetable}

\subsection{AB Doradus Moving Group}

\begin{figure}[ht]

\includegraphics[width=0.33\linewidth]{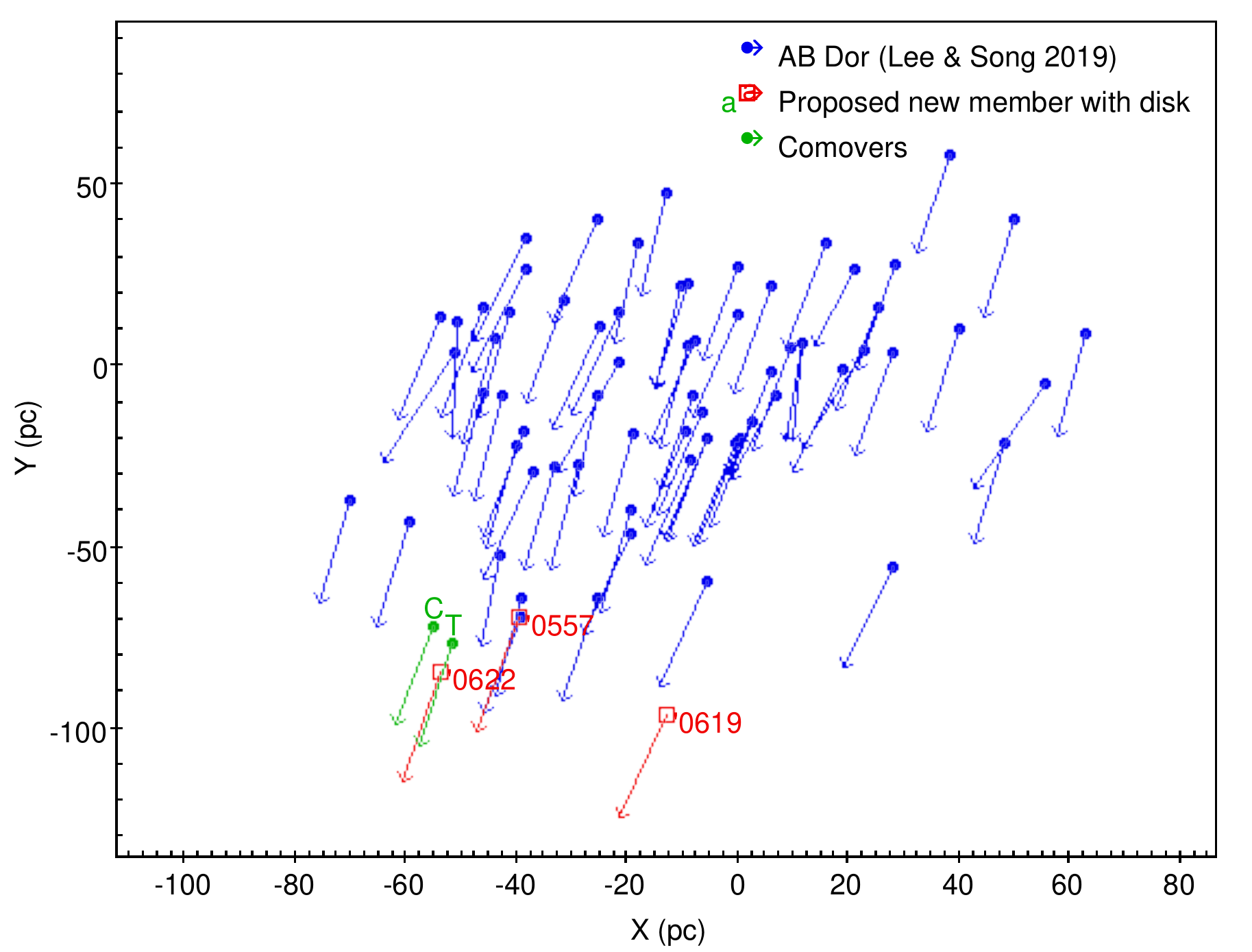}
\includegraphics[width=0.33\linewidth]{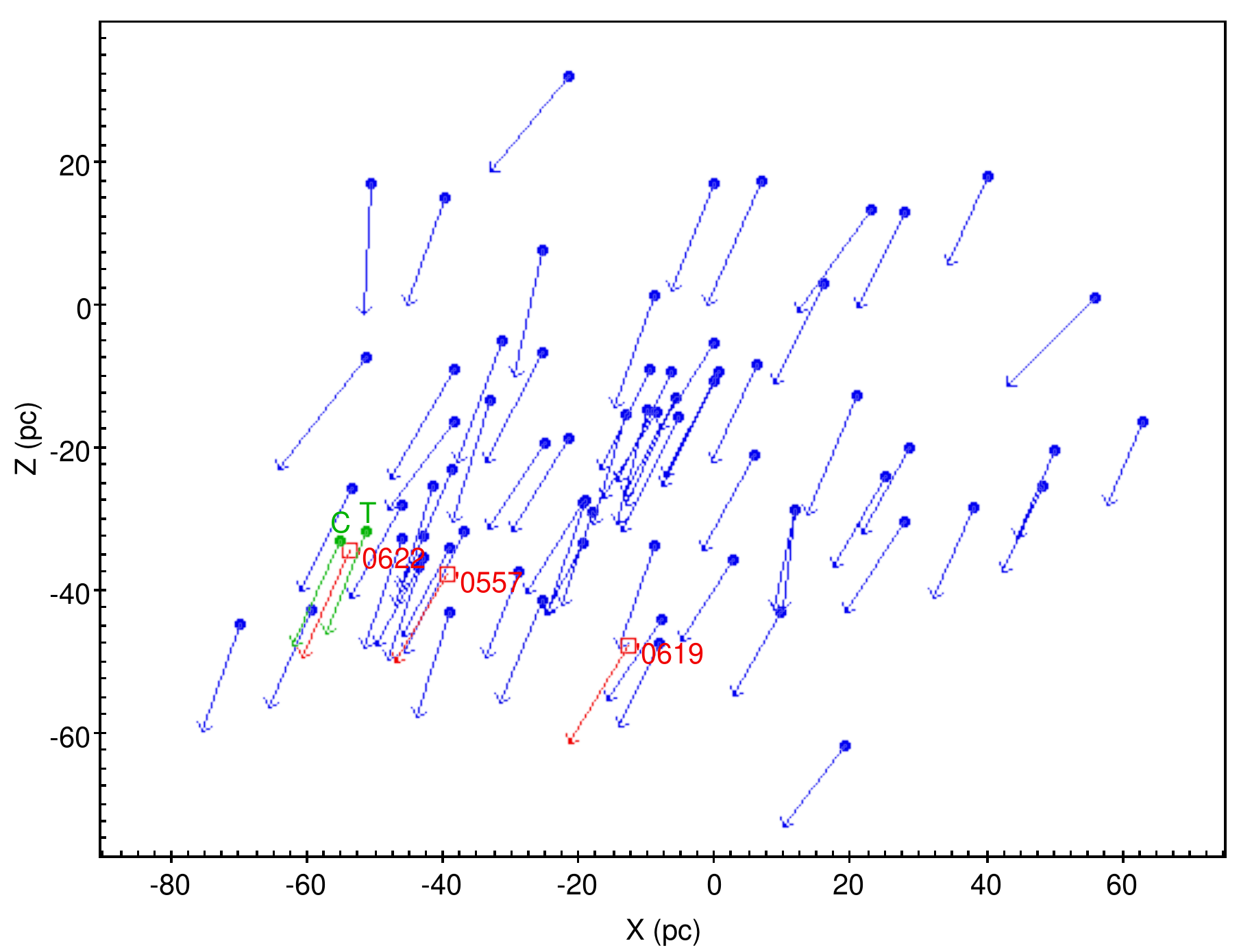}
\includegraphics[width=0.33\linewidth]{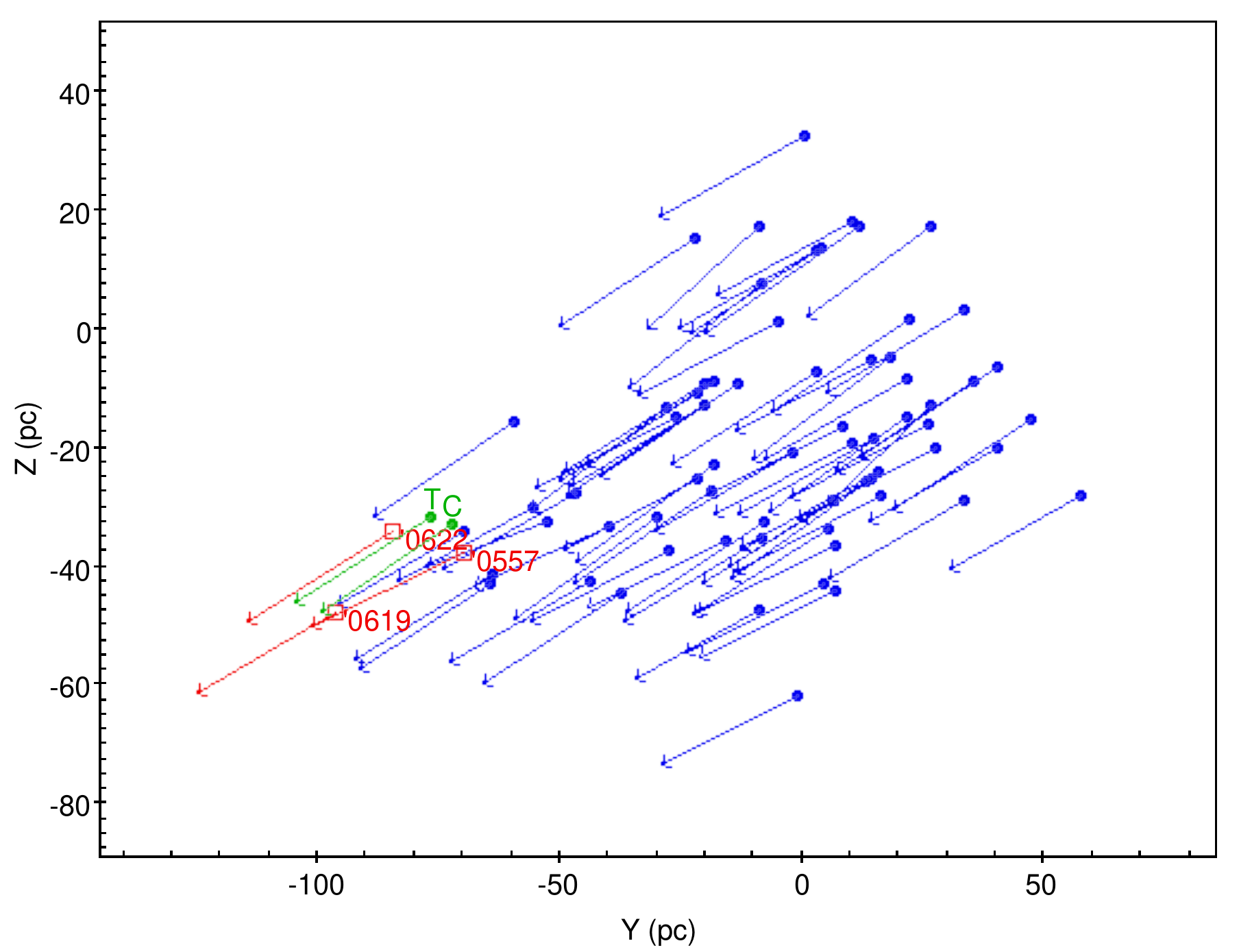}
\caption{Previously-identified members of AB Doradus from \citet{2019MNRAS.486.3434L} (blue), proposed new disk-hosting members of AB Doradus from Disk Detective (red) and comoving stars (green), in XYZ (parsecs from the sun) with vectors showing UVW (parsecs/1000 years).  Of these candidates, J0622 is found in Table 5 of \citet{2015ApJ...804..146D} as well as in comoving catalogs \citet{2018AJ....155..149B} and  \citet{2017AJ....153..257O} in a group of 3 comoving stars (labelled in green as C and T).  All three candidates were previously identified as infrared excess stars, but not as AB Dor members.}
\label{fig:vectors_ABDor-LS}

%\caption{Previously-identified members of AB Doradus from \citet{2019MNRAS.486.3434L} (blue), proposed new disk-hosting members of AB Doradus from Disk Detective (red) and comoving stars (green), in in XYZ (parsecs) with vectors showing UVW (parsecs/1000 years).  Of these candidates, J0622 is found in Table 5 of \citet{2015ApJ...804..146D} as well as in comoving catalogs \citet{2018AJ....155..149B} and  \citet{2017AJ....153..257O} in a group of 3 comoving stars (labelled in green as C and T).  All three candidates were previously identified as infrared excess stars, but not as AB Dor members.}
%\label{fig:vectors_ABDor-LS}

\end{figure}

The closest known moving group, AB Doradus (AB Dor) \citep{2004ApJ...613L..65Z}, contains a core or nucleus of $\sim 10$ stars at a distance of 20 pc, along with dozens of purported ``stream'' members distributed across the sky.
Figure~\ref{fig:vectors_ABDor-LS} shows a list of known members of AB Doradus from \citet{2019MNRAS.486.3434L}.
Several groups have recently proposed membership lists for this group, including \citet{2018ApJ...856...23G,2018ApJ...860...43G,2018ApJ...862..138G} and \citet{2019AJ....157..234S}. 
The new, more comprehensive, list by \citet{2019MNRAS.486.3434L} utilizes Gaia DR2 and machine learning. 

In this figure, the origin of each arrow is the star's current Galactic position from Gaia DR2. The arrows show how far each star moves in 1000 years based on its Gaia DR 2 velocity.  Based on the findings by \citep{2018ApJ...861L..13G} the age of AB Doradus has been refined at $133^{+15}_{-20}$ Myr. It has also been suggested that the ``stream" members do not all share a common age with the core members \citep{2013ApJ...766....6B}.  Note, however, that the \citet{2019MNRAS.486.3434L} membership list does not appear to show this core + stream structure. 

%The age of this group is not known precisely; a recent isochronal estimation by \citep{2015MNRAS.454..593B} yielded $149^{+51}_{-19}$ Myr. 

Our VR inspection revealed three new candidate disk-hosting members of this group to add to the list of possible AB Doradus members. Note that all of these stars have Gaia radial velocities (as do all the infrared-excess stars in this paper).
All three of these stars lie on one side of the association, in the -X, -Y direction (see the left panel of Figure~\ref{fig:vectors_ABDor-LS}).
They were assigned low ($<7$\%) probabilities of membership in AB Doradus by BANYAN $\Sigma$, probably because of the dearth of known members at that XYZ position in the BANYAN $\Sigma$ membership list.  But these new candidates have velocities resembling the average velocity of the known members (see Figure \ref{fig:vectors_ABDor-LS}) and the presence of infrared excesses is a sign of youth.  

%Therefore, we propose an extension of the group's "stream" in the in the -X, -Y direction (toward the Galactic center and away from the direction of Galactic rotation), pending further spectroscopic analysis of these three candidates.
%LIST THE VELOCITIES OF THESE PROPOSED MEMBERS, AND THE AVERAGE VELOCITY OF THE KNOWN MEMBERS

%DISCUSS THESE TWO COMOVING STARS AND HD 15115

\smallskip
\noindent
{\bf HD 44775  (WISEA J062218.66-295134.7)} is a proposed new AB Dor member, and an F3V at 105 pc.  After we identified this proposed member in VR, we read that \citet{2017AJ....153..257O} found this star to be part of a co-moving triple together with TYC 6518-1857-1 (98 pc) and CPD-25 1292 (96 pc). These two co-movers are shown as green arrows labelled T and C, respectively, in Figure~\ref{fig:vectors_ABDor-LS}.
%{\bf HD 44775  (WISEA J062218.66-295134.7)} is a proposed new AB Dor member, and an F3V at 105 pc.  \citet{2017AJ....153..257O} found this star to be part of a co-moving triple together with TYC 6518-1857-1 (98 pc) and CPD-25 1292 (96 pc). These two co-movers are shown as green arrows labelled T and C, respectively, in Figure~\ref{fig:vectors_ABDor-LS}.
%2020-07-25 Sue added reference for David in Table 1 bc it was missing--did Marc check it and remove it on purpose?  The co-moving papers are not listed in Table 1 on purpose bc they are not YSAs.

\smallskip
\noindent
{\bf HD 40540  (WISEA J055752.60-342834.1)} is a proposed new AB Dor member, and an A8IV at 88 pc, has a debris disk that was first recognized in IRAS data by  \citet{2007ApJ...660.1556R}.  Later {\it Spitzer} spectroscopy revealed that the dust has a negligible crystallinity fraction  \citep[$< 4$\%][]{2015ApJ...798...87M}.

\smallskip
\noindent
{\bf HD 44510  (WISEA J061903.93-535823.9)} is a proposed new AB Dor member, and an F3V at 108 pc. \citet{2014MNRAS.437..391C} previously identified this star as having an infrared excess.  %ADD THIS REF TO THE TABLE-Done!

%\begin{figure}[ht]
%\includegraphics[width=0.33\linewidth]{x-y_ABDor_DD_comover_2020-07.pdf}
%\includegraphics[width=0.33\linewidth]{x-z_ABDor_DD_comover_2020-07.pdf}
%\includegraphics[width=0.33\linewidth]{y-z_ABDor_DD_comover_2020-07.pdf}
%\caption{Disk Detective with AB Doradus where blue is previously identified members of AB Dor and red is new disk candidates.  The catalog of stars previously associated with AB Dor was derived from \citet{2018ApJ...856...23G,2018ApJ...860...43G,2018ApJ...862..138G,2019AJ....157..234S}.  The average XYZ for AB Dor in this figure is -10.7, -13.4, and -21.4 pc, respectively.  Of these candidates, J0622 is found in Table 5 of \citet{2015ApJ...804..146D} as well as in comoving catalogs \citet{2018AJ....155..149B} and  \citet{2017AJ....153..257O} in a group of 3 comoving stars.  All three candidates have been previously identified to have a disk.  We note that the motion of HD15115, a previously identified member, appears to be different than the rest, and has previously been identified as being a member of Columba in \citet{2013ApJ...762...88M}.}
%\label{fig:vectors_ABDor}
%\end{figure}

\begin{figure}[ht]
    \centering
    \includegraphics[width=0.5\linewidth]{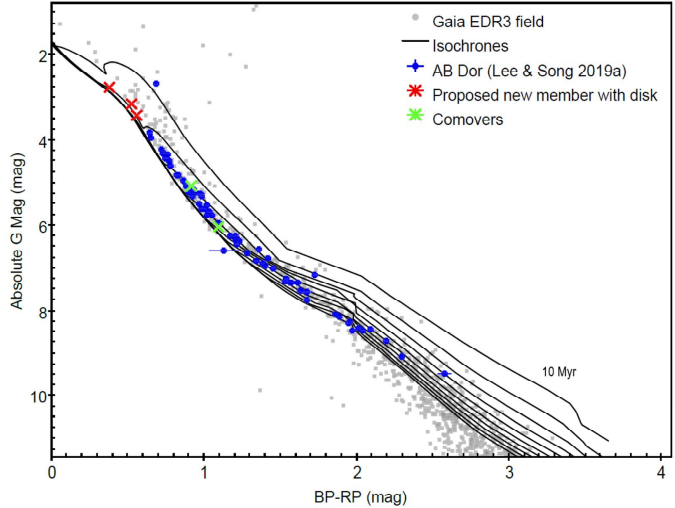}
    \caption{Colour magnitude diagram (CMD) of AB Doradus (blue) with proposed new members with disks (red) and comovers (green) with isochrones from 10 Myr (top) to 100 Myr (bottom) and Gaia EDR3 field stars. The proposed new members lie in a sequence with the other AB Dor members. Note how the isochrones tend to be degenerate toward the blue end of this diagram, where our proposed new disk-hosting members lie.}
    \label{fig:HRdiagABDorDD}
\end{figure}

%Figure \ref{fig:HRdiagABDorDD} shows a colour magnitude diagram (CMD) for AB Doradus, including the Gaia EDR3 field main sequence, shown in grey. The solid lines show isochrones at 10, 20, 30,...,100 Myr from \citet{2017ApJ...835...77M} available online\footnote{\url{http://stev.oapd.inaf.it/cmd}}. The blue points show the 
%previously-identified members of AB Doradus from \citet{2019MNRAS.486.3434L}. The red and green x's show our proposed new members: the red objects have infrared excess, the green are co-moving with them. 

Figure \ref{fig:HRdiagABDorDD} shows a colour magnitude diagram (CMD) for AB Doradus, including the Gaia EDR3 field main sequence, shown in grey. The solid lines show isochrones at 10, 20, 30,...,100 Myr from \citet{2017ApJ...835...77M} available online\footnote{\url{http://stev.oapd.inaf.it/cmd}}. The blue points show the previously-identified members of AB Doradus from \citet{2019MNRAS.486.3434L}. The red and green x's show our proposed new members: the red objects have infrared excess, the green are co-moving with them. 

The members from \citet{2019MNRAS.486.3434L} are barely distinguishable from the EDR3 field, consistent with the \citep{2015MNRAS.454..593B} age estimate of $149^{+51}_{-19}$ Myr, except perhaps at the reddest end of the CMD.  Toward the blue end, where the new proposed members are, the isochrones are degenerate, making it even harder to constrain the ages of the new proposed members. This CMD highlights an important trend; the new members we discover using VR are often in a mass regime where isochrones are degenerate, making them hard to identify as group members in an isochronal analysis.

%# File generated by CMD 3.0 (http://stev.oapd.inaf.it/cmd) on Fri Apr 20 17:59:28 CEST 2018
%# isochrones PARSEC release v1.2S + COLIBRI release PR16, as in Marigo et al. 2017
%# Basic reference: Marigo et al. (2017), ApJ, 835, 77
%# Photometric system: Gaia DR2 + Tycho2 + 2MASS (all Vegamags)
%# O-rich circumstellar dust ignored
%# C-rich circumstellar dust ignored
%# Kind of output: isochrone tables

%%%%%%%%%%%%%%%%%%%%%%%%%%%%%%%%%%%%%%%%%%%%%%%%%%

\subsection{Argus} 

%WE MUST MAKE SURE HD 192758 (J201815.83-425136.9) AND {\bf CD-47 2352} (J062542.92-470754.4) GO IN STEVEN"S PAPER.

The Argus association \citep{2008hsf2.book..757T} contains many stars within 100 pc of Earth, including the well-known debris disk hosts $\beta$ Leonis and 49 Ceti. \citet{2019ApJ...870...27Z} argued for the physicality of this association and provided the most recent membership list and age estimate for this association: 40-50 Myr.  We used this catalog of known Argus members, which includes nine stars with known infrared excesses. Our process recovered four previously identified members of Argus, including two disk-hosting stars previously identified by the Disk Detective project using the BANYAN $\Sigma$ tool (Silverberg et al. in prep).  However, our process did not recover any new candidate disk-hosting members of Argus.

\subsection{$\beta$ Pictoris} 

The well-studied $\beta$ Pictoris ($\beta$ Pic) moving group \citep{2001ApJ...562L..87Z} appears to be one of the closest YSAs to the Sun. \citet{2020A&A...642A.179M} provide a useful overview of age estimates for this group.   
\citet{2020A&A...642A.179M} themselves present a dynamical traceback age of $18.5^{+2.0}_{-2.4}$ Myr based on Gaia DR2. Recent estimates based on isochrone fitting  \citep[e.g.][]{2014MNRAS.445.2169M, 2016MNRAS.455.3345B} and 
estimates from lithium depletion boundary fitting \citep[e.g.][]{2014MNRAS.438L..11B, 2016A&A...596A..29M}, perhaps the most reliable technique, hover around 24 Myr.

We first examined membership lists for $\beta$ Pictoris from \citet{2018ApJ...856...23G, 2018ApJ...860...43G, 2018ApJ...862..138G} and \citet{2019AJ....157..234S}, but then decided to use the more comprehensive membership list from \citet{2019MNRAS.486.3434L}, derived using machine learning and Gaia DR2. 

%median BPMG age of 23 ± 3 Myr (overall 1σ uncertainty

\begin{figure}[ht]

\includegraphics[width=0.33\linewidth]{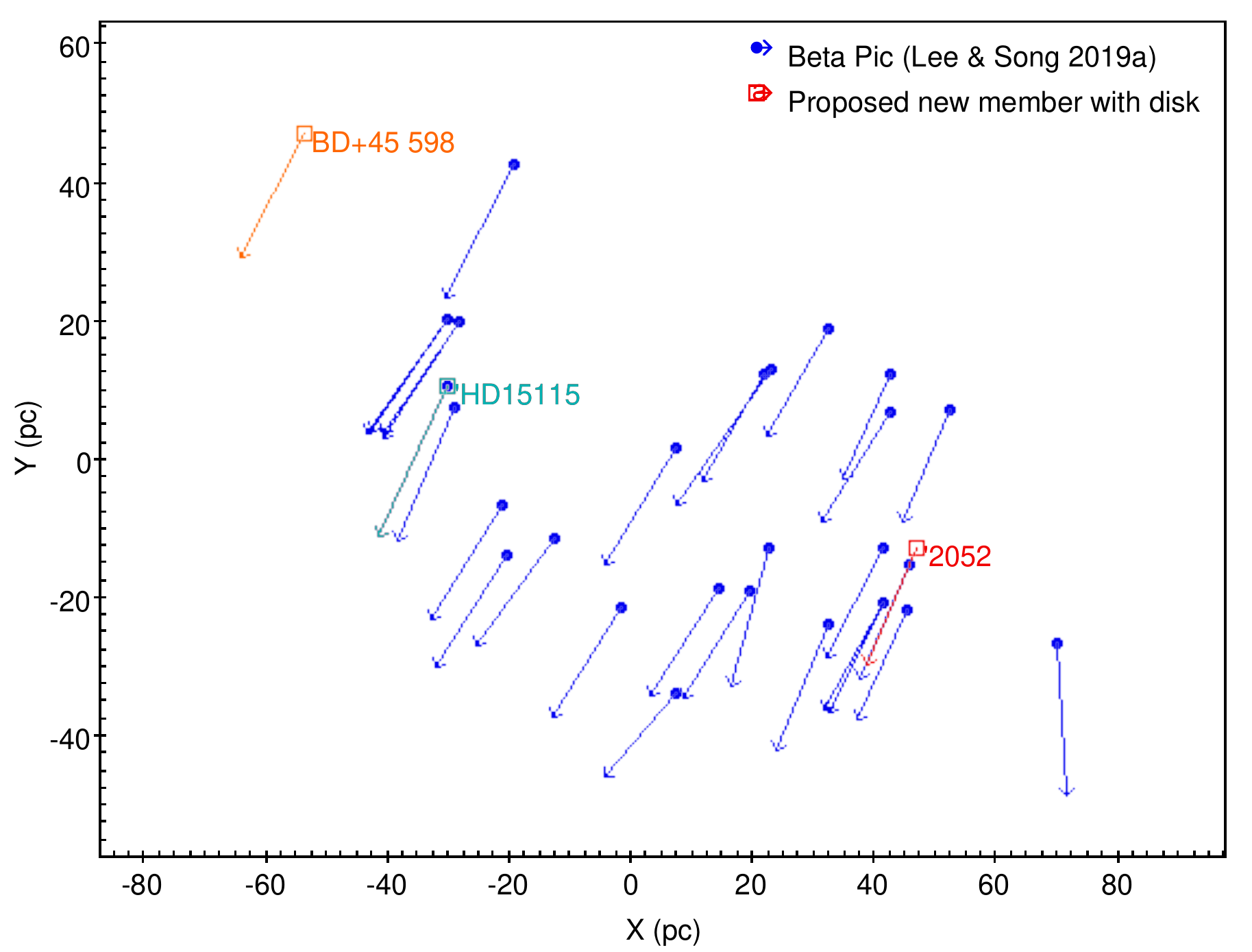}
\includegraphics[width=0.33\linewidth]{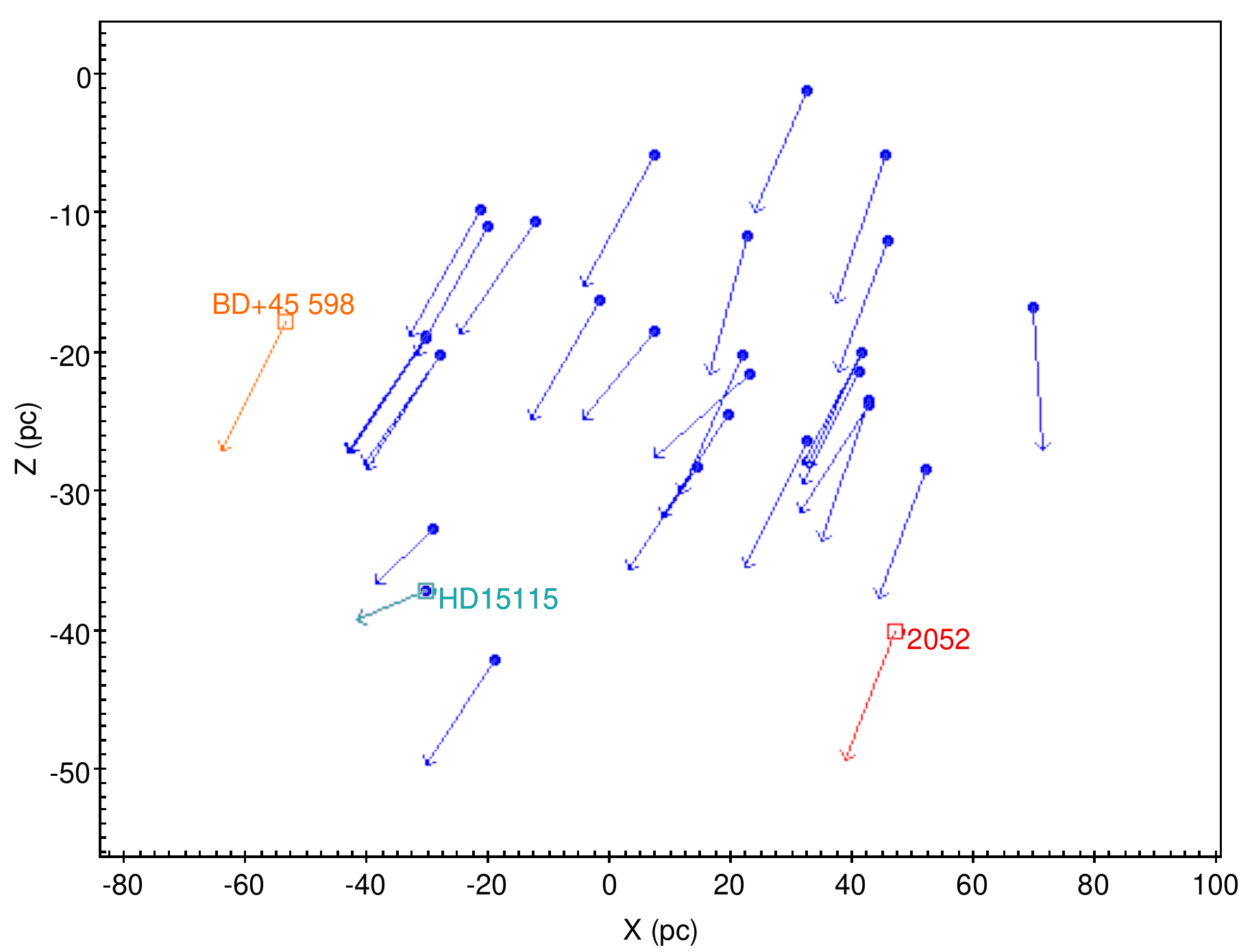}
\includegraphics[width=0.33\linewidth]{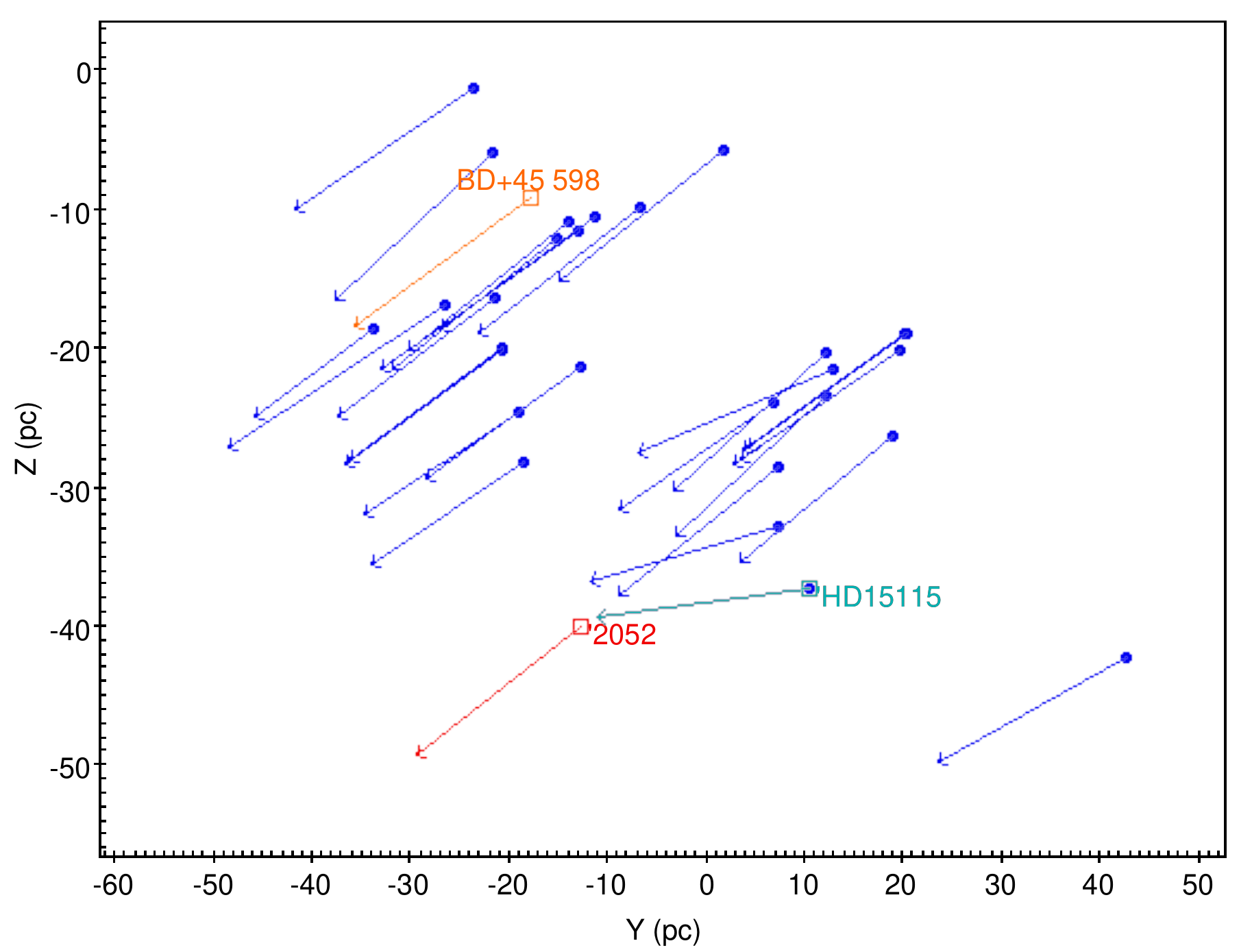}
\caption{Previously identified members of $\beta$ Pictoris (blue) from \citet{2019MNRAS.486.3434L} and one new proposed disk-hosting group member (J205241.67-532624.8, shown in red) in XYZ (parsecs from the sun) with vectors showing UVW (parsecs/1000 years). Our process also recovered BD+45 598 (J022113.11+460006) as a $\beta$ Pic member (orange).  The infrared excess star HD 15115 (J022616.32+061732.8) (teal) was identified as a member of $\beta$ Pic in the \citet{2019MNRAS.486.3434L} catalog. However, its velocity stands out from the velocities of the other stars here, as you can see (especially in the right panel).
%and has been a associated with AB Dor, Tuc-Hor and Columba.  
We prefer to identify it with Tuc-Hor; further discussion about this star is found in Figure \ref{fig:vectors_TucHor_LS}. }

%\caption{Previously identified members of $\beta$ Pictoris (blue) from \citet{2019MNRAS.486.3434L} and one new proposed disk-hosting group member (red) in in XYZ (parsecs) with vectors showing UVW (parsecs/1000 years). The infrared excess star HD 15115 (J022616.32+061732.8) (teal) was identified as a member of $\beta$ Pic in the \citet{2019MNRAS.486.3434L} catalog. However, its velocity stands out from the velocities of the other stars here, as you can see (especially in the right panel).
%and has been a associated with AB Dor, Tuc-Hor and Columba.  
%We prefer to identify it with Tuc-Hor; further discussion about this star is found in Figure \ref{fig:vectors_TucHor_LS}. }

\label{fig:vectors_BetaPic-LS}

\end{figure}

We propose to assign HD 198472 to this group; this star was previously known to have an infrared excess (prior to Disk Detective). Our examination also suggests a new group assignment for HD 15115, one of the disk-hosting members listed in \citet{2019MNRAS.486.3434L}. Figure~\ref{fig:vectors_BetaPic-LS} depicts the group members from \citet{2019MNRAS.486.3434L} in blue, and the two stars of interest (red and teal). 

%2020-07-20-Sue ARE WE SURE J2052 SHOULD BE EXCLUDED?  Notes in Table 1 say isochronal age 1.5Gyr but it is found in Cotten and Song and McDonald 2017, etc., and not previously identified in YSA, but BANYAN $\Sigma$ places it in 70$\%$ probability of being in Beta Pic

%The $\beta$ Pictoris moving group (Figure~\ref{fig:vectors_BetaPic})
%We derived our list of previously identified members from \citet{2018ApJ...856...23G,2018ApJ...860...43G,2018ApJ...862..138G,2019AJ....157..234S}. 
%We recovered three previously known disk-hosting members of this group. In addition, we found one candidate new disk-hosting member:

\smallskip
\noindent
{\bf HD 198472 (WISEA J205241.67-531624.8)} is a proposed $\beta$ Pictoris member, an F 5/6 V star at a distance of 63 pc.  This star was first robustly identified as having an infrared excess by \citet{2013ApJS..208...29W}. \citet{2015ApJ...804..146D} found an isochronal age for this star of $>1.5$ Gyr.  However, this star is kinematically identified with $\beta$ Pictoris in \citet{2019MNRAS.489.3625C}, but this paper did not mention the infrared excess or any other signs of youth.  Thus, the kinematics and the presence of an infrared excess suggest that an assignment of this star to $\beta$ Pictoris should be considered seriously.

%\smallskip
%\noindent
%{\bf BD+45 598 (J022113.11+460006)} is a rotationally variable K1 \citep{2009A&A...504..829G} at 73 pc. \citet{2011ApJS..193....4M} assigned this star to $\beta$ Pictoris, pointing out that the star's measured lithium equivalent width and fractional X-ray luminosity are consistent with the similar properties of known $\beta$ Pictoris members. But then \citet{2018ApJ...856...23G,2018ApJ...860...43G,2018ApJ...862..138G,2019AJ....157..234S} and \citet{2019MNRAS.486.3434L} excluded it from their membership lists. Figure \ref{fig:vectors_BetaPic-LS} shows it to be a bit of an outlier based on its position, but a good match to $\beta$ Pictoris based on its velocity. Overall, the quality of the match looks acceptable, so we support the conclusion of \citet{2011ApJS..193....4M}.  WISE photometry shows that the star has a modest 22$\mu$m infrared excess: $W1-W4 = 0.495 \pm 0.096$
%Gaia 353635070145291008

\smallskip
\noindent
{\bf HD 15115 (WISEA J022616.32+061732.8)} was identified as a member of $\beta$ Pictoris by \citet{2019MNRAS.486.3434L} but it has previously been associated with AB Dor, Tuc-Hor and Columba.  The well-known edge-on debris disk around this star has been resolved by HST NICMOS \citep{2008ApJ...684L..41D}, LBT \citep{2012ApJ...752...57R}, Gemini NICI  \citep{2014A&A...569A..29M} and VLT SPHERE \citep{2019A&A...622A.192E}, and has also been spatially resolved by Subaru's SCExAO/CHARIS \citep{2020AJ....160..163L}. ALMA imaging shows evidence for two rings in the disk separated by a cleared gap \citep{2019ApJ...877L..32M}. The center and right panels of Figure~\ref{fig:vectors_BetaPic-LS} show that its velocity is a poor match for $\beta$ Pictoris.  We prefer to identify this infrared excess star with Tuc-Hor; further discussion about this star can be found below and in Figure \ref{fig:vectors_TucHor_LS} under Section \ref{sect:Tuc-Hor}.

\smallskip

We also recovered BD+45 598 (WISEA J022113.11+460006) which \citet{2021arXiv210312824H} recently classified as a member of $\beta$ Pictoris, and which \citet{2018ApJ...856...23G,2018ApJ...860...43G,2018ApJ...862..138G,2019AJ....157..234S} and \citet{2019MNRAS.486.3434L} excluded from their membership lists.  We found it to be a bit of an outlier based on its position, but a good match to $\beta$ Pictoris based on its velocity. Overall, the quality of the match looks acceptable, so we support the conclusion of \citet{2011ApJS..193....4M} and \citet{2021arXiv210312824H}.

%\begin{figure}[ht]
%\includegraphics[width=0.33\linewidth]{figures/x-y_BetaPic_and_DD.pdf}
%\includegraphics[width=0.33\linewidth]{figures/x-z_BetaPic_and_DD.pdf}
%\includegraphics[width=0.33\linewidth]{figures/y-z_BetaPic_and_DD.pdf}
%\caption{New proposed disk-hosting members of $\beta$ Pictoris (red) and previously identified members (blue) from from %\citet{2018ApJ...856...23G,2018ApJ...860...43G,2018ApJ...862..138G,2019AJ....157..234S}.  The average XYZ for Beta Pic in this figure is 6.5, -7.0, and -16.9 pc, respectively.  All three Disk Detective candidates have been previously identifies as having a disk, and J022616.32+061732.8, seen in the middle of each figure, is a previously identified member of Beta Pic, and has been a associated with AB Dor, Tuc-Hor and Columba.  Further discussion about this candidate is found in Figure %\ref{fig:vectors_TucHor}. }
%\label{fig:vectors_BetaPic}
%\end{figure}

\begin{figure}[ht]
    \centering
    \includegraphics[width=0.5\linewidth]{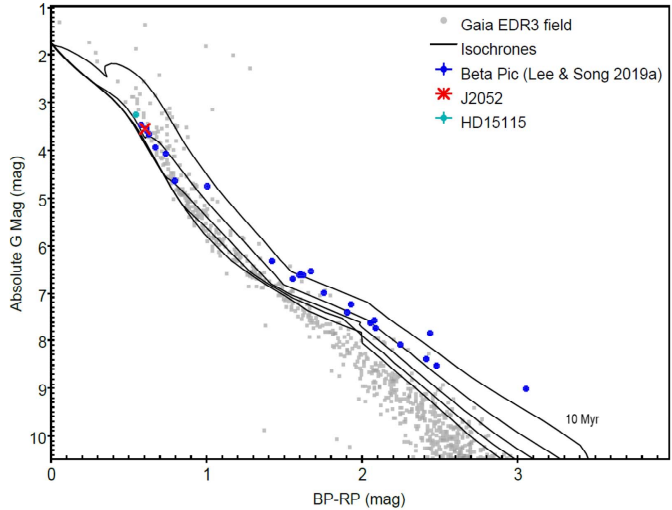}
    \caption{Colour magnitude diagram (CMD) of $\beta$ Pictoris (blue) with proposed new member with disk J0252 (red) and HD 15115 (teal) with isochrones from 10 Myr (top) to 50 Myr (bottom) and Gaia EDR3 field stars.}
    \label{fig:HRdiagBPMGDD}
\end{figure}

%Figure \ref{fig:HRdiagBPMGDD} shows a colour magnitude diagram (CMD) for $\beta$ Pictoris, including the Gaia EDR3 field main sequence, shown in grey. The solid lines show isochrones at 10, 20, 30, 40, 50 Myr from \citet{2017ApJ...835...77M} available online\footnote{\url{http://stev.oapd.inaf.it/cmd}}. The blue points show the previously-identified members of $\beta$ Pictoris from \citet{2019MNRAS.486.3434L}. The red x shows our proposed new member which has infrared excess while the teal dot shows HD 15115. 

Figure \ref{fig:HRdiagBPMGDD} shows a colour magnitude diagram (CMD) for $\beta$ Pictoris, including the Gaia EDR3 field main sequence, shown in grey. The solid lines show isochrones at 10, 20, 30, 40, 50 Myr from \citet{2017ApJ...835...77M} available online\footnote{\url{http://stev.oapd.inaf.it/cmd}}. The blue points show the previously-identified members of $\beta$ Pictoris from \citet{2019MNRAS.486.3434L}. The red x shows our proposed new member which has infrared excess while the teal dot shows HD 15115.

The members from \citet{2019MNRAS.486.3434L} are found above the main sequence EDR3 field, consistent with \citep{2014MNRAS.445.2169M, 2016MNRAS.455.3345B, 2014MNRAS.438L..11B, 2016A&A...596A..29M} at around 24 Myr and our proposed new member appears to be consistent with the previously-identified members of $\beta$ Pictoris.

%%%%%%%%%%%%%%%%%%%%%%%%%%%%%%%%%%%%%%%%%%%%%%%%%%

\subsection{Octans-Near}

\begin{figure}[ht]

\includegraphics[width=0.33\linewidth]{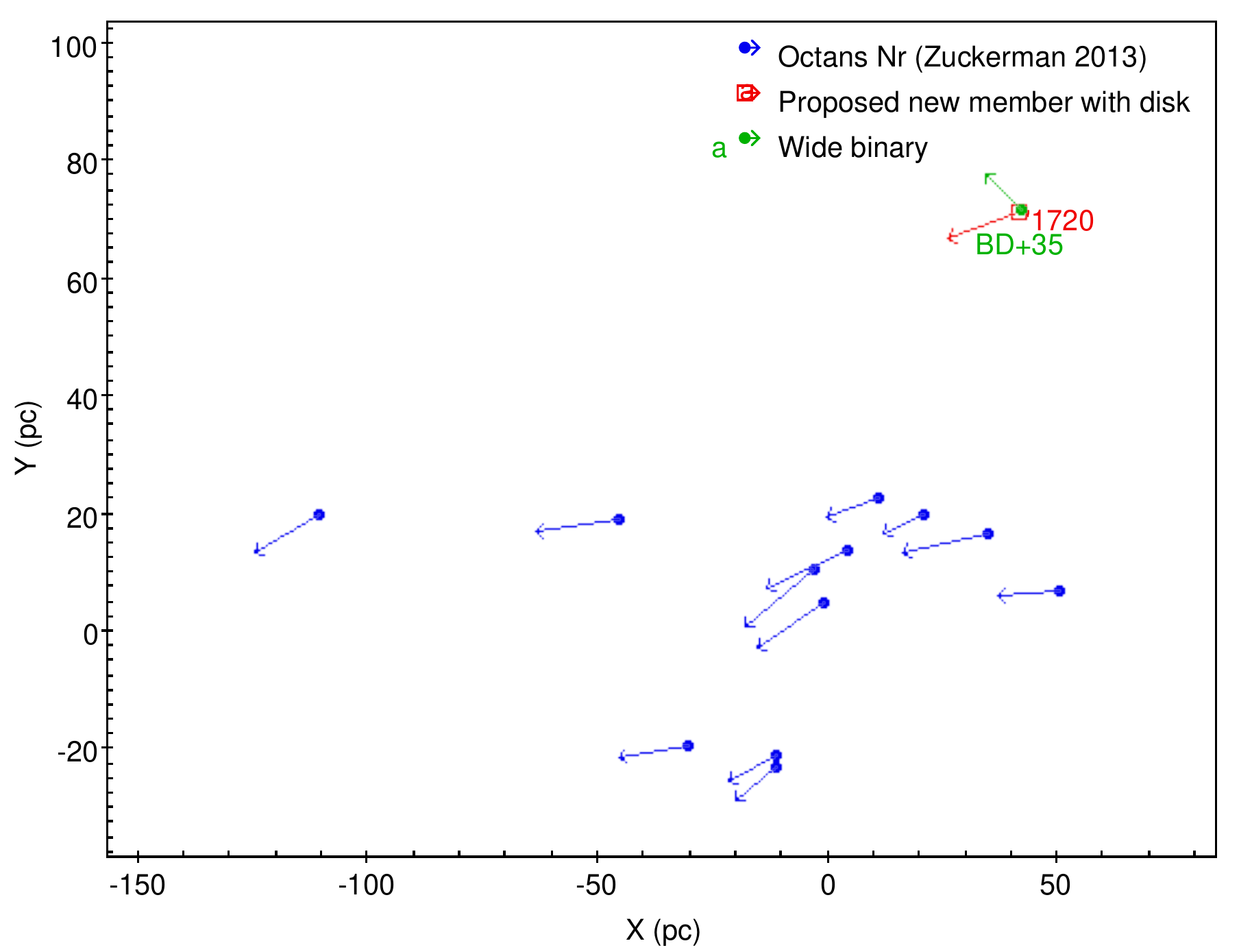}
\includegraphics[width=0.33\linewidth]{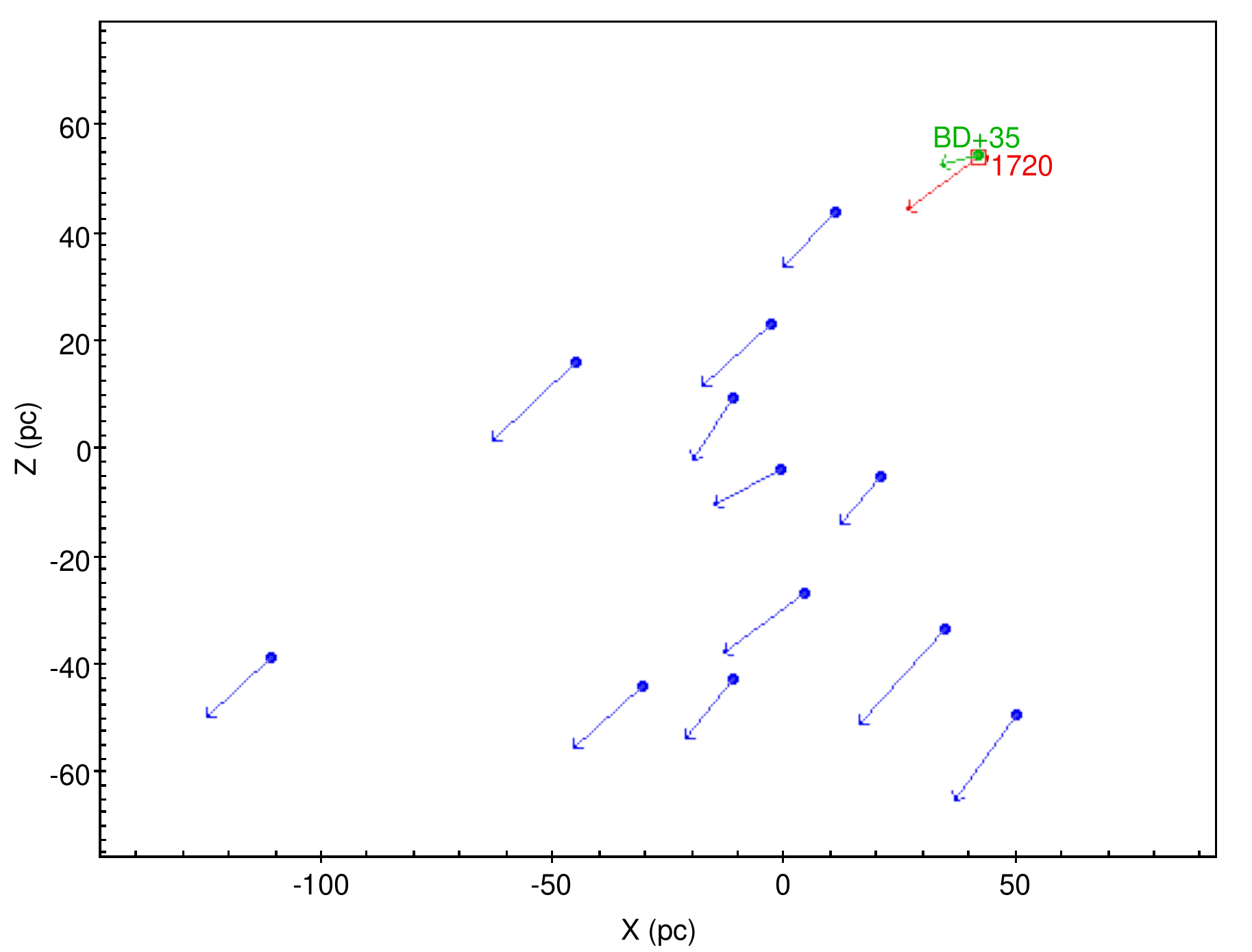}
\includegraphics[width=0.33\linewidth]{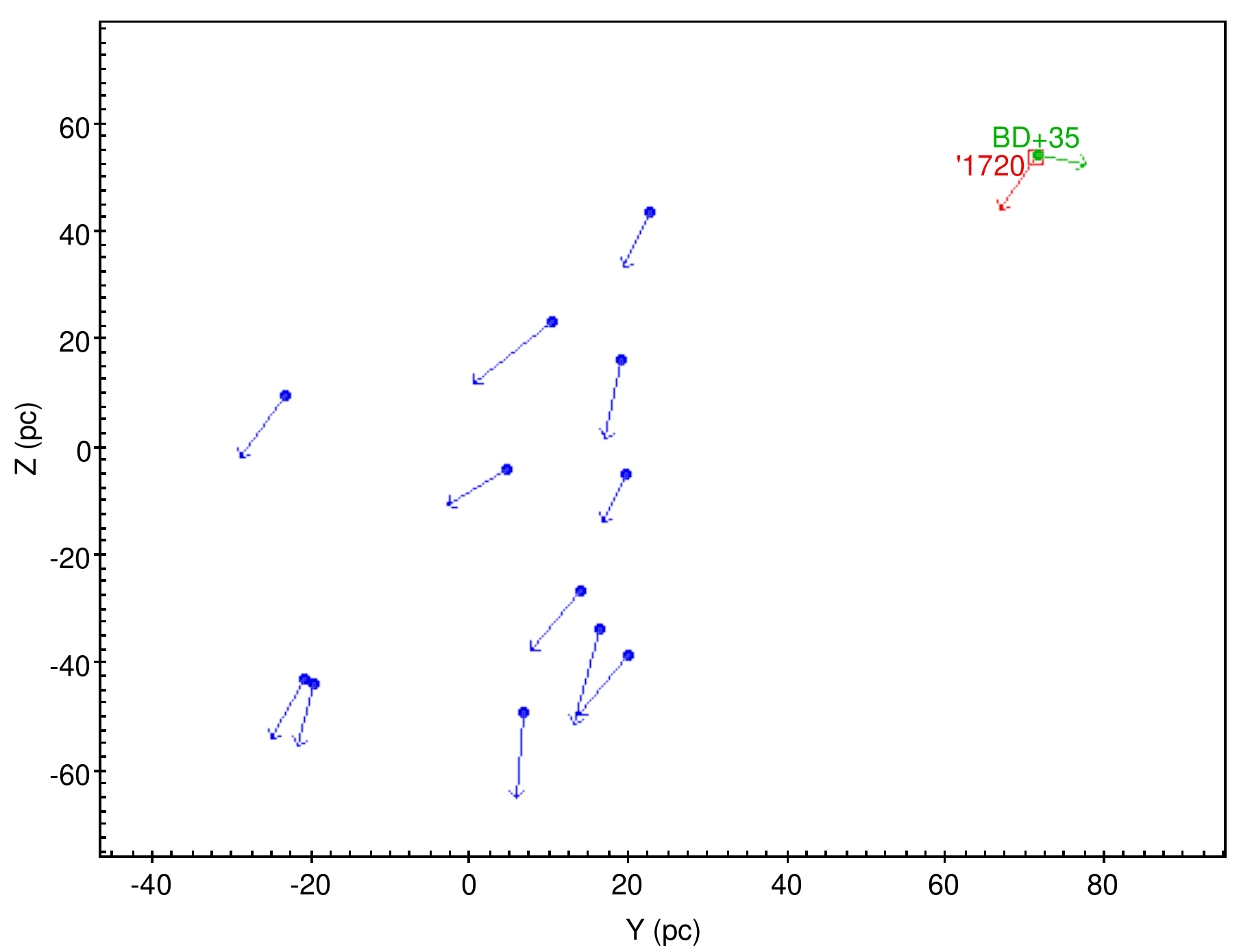}
\caption{Members of Octans-Near (blue) previously proposed by \citet{2013ApJ...778....5Z}, in XYZ (parsecs from the sun) with vectors showing UVW (parsecs/1000 years).   
%\caption{Members of Octans-Near (blue) previously proposed by \citet{2013ApJ...778....5Z}, in XYZ (parsecs) with vectors showing UVW (parsecs/1000 years).  %We recovered one known member of Oct Nr (J0342);  
%We propose two additional candidate members: 
We identified two other stars that could be related to this group:  the infrared excess star HD 157165 (J172007.53+354103.6), and a wide binary companion to this star (BD+35 2953) found by \citet{2017MNRAS.472..675A}. Note that BD+35 2953 (wide binary) has no measured radial velocity so the length and direction of the green arrow is only based on two velocity components. }
\label{fig:vectors_OctansNr}

\end{figure}

\citet{2013ApJ...778....5Z} presented a list of 14 stars with UVW velocities resembling Octans, but located close to the Sun, called ``Octans-Near" (Oct Nr) (Figure~\ref{fig:vectors_OctansNr}).  \citet{2016IAUS..314...21M} asserted that this group ``probably warrants stream status,'' meaning that evidence for the co-evolution of the Octans-Near stars was still thin.  
Indeed, the ages of the stars in this group range from 30 to 100 Myr according to  \citet{2013ApJ...778....5Z}. 
Both of these studies predate Gaia, however.

We applied our process to the \citet{2013ApJ...778....5Z} star list.
We recovered one previously-known member listed by \citet{2013ApJ...778....5Z}: J034239.80-203243.3, which was discovered by \citet{2014ApJS..212...10P} to have an infrared excess. Additionally we identify two other stars which may be related to Octans-Near, which reside somewhat farther from the Sun than the rest of Octans-Near, in the opposite direction from Octans.  These, nonetheless, resemble Octans-Near in velocity and show signs of youth (infrared excess and x-ray emission). 
These stars are roughly 25 pc away from the edge of Octans-Near (in the Y direction); for comparison, Octans-Near spans about 40 pc in this direction.

%The average XYZ for Octans Near in this figure is -7.5, 5.8, and -12.6 pc, respectively.

\smallskip
\noindent
{\bf HD 157165 (WISEA J172007.53+354103.6)}  This F8 star at 99 pc was one of the first 37 disk candidates published by the Disk Detective project \citep{2016ApJ...830...84K}.  Follow-up imaging by the project using Palomar/Robo-AO searched for background objects in the Sloan-i filter within the 12 arcsecond radius of the WISE 4 Point Spread Function and found none \citep{2018ApJ...868...43S}.  Modeling of the spectral energy distribution of this system found the disk temperature of $196 \pm 30$K and a fractional infrared excess of $L_{IR}/L_{\star} \approx 9.0 \pm 1.5 \times 10^{ -5}$  This star has a wide companion, BD+35 2953, 89.33 arcseconds away, inferred by \citep{2017MNRAS.472..675A} from the Tycho-Gaia astrometric solution.  See below.

\smallskip
\noindent
{\bf BD+35 2953}, another F8 star 99 pc distant from the Sun, is a wide companion to HD 157165. It is also the optical counterpart to a ROSAT All-Sky Survey X-ray source \citet{Guillout2009}.  This star does not have a measured radial velocity in the Gaia DR2 catalog, therefore, the green arrow representing it in the figure shows only the proper motion components of its motion.

%so we have excluded it from Figure~\ref{fig:vectors_OctansNr}.

%\citet{Guillout2009} measured the stellar spectrum to have a Lithium equivalent width of 7.56 picomeeters and a H$\alpha$ equivalent width of 21.86 picmoeters.

%ADD THIS COMPANION STAR TO THE TABLE AND MAYBE TO THE FIGURE 
%2020-07-26 Sue-this companion star unfortunately doesn't have an RV so it cannot be included on the figure.

\begin{figure}[ht]
    \centering
    \includegraphics[width=0.5\linewidth]{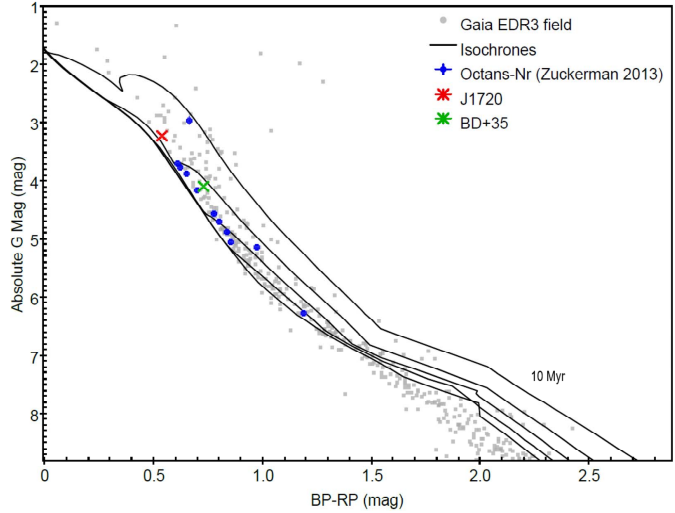}
    \caption{Colour magnitude diagram (CMD) of Octans-Near (blue) previously proposed by \citet{2013ApJ...778....5Z}, with one proposed new member with disk J1720 (red) and wide binary companion (BD+35 2953) to this star (green) found by \citet{2017MNRAS.472..675A} with isochrones from 10 Myr (top) to 50 Myr (bottom) and Gaia EDR3 field stars.}
    \label{fig:HRdiagOctansNrDD}
\end{figure}

%Figure \ref{fig:HRdiagOctansNrDD} shows a CMD for Octans-Near, including the Gaia EDR3 field main sequence, shown in grey. The solid lines show isochrones at 10, 20, 30, 40, 50 Myr from \citet{2017ApJ...835...77M} available online\footnote{\url{http://stev.oapd.inaf.it/cmd}}. The blue points show the 
%previously-identified members of Octans-Near from \citet{2013ApJ...778....5Z}. The red and green x's show our proposed new members: the red object has infrared excess, the green is a wide companion co-moving with it. 

Figure \ref{fig:HRdiagOctansNrDD} shows a CMD for Octans-Near, including the Gaia EDR3 field main sequence, shown in grey. The solid lines show isochrones at 10, 20, 30, 40, 50 Myr from \citet{2017ApJ...835...77M} available online\footnote{\url{http://stev.oapd.inaf.it/cmd}}. The blue points show the 
previously-identified members of Octans-Near from \citet{2013ApJ...778....5Z}. The red and green x's show our proposed new members: the red object has infrared excess, the green is a wide companion co-moving with it. 

The members from \citet{2013ApJ...778....5Z} are barely distinguishable from the EDR3 field, consistent with the age estimate of 30 to 100 Myr, with our proposed new member and comover consistent with the age as well.

\subsection{TW Hya Association}

Our process did not yield any new members of the TW Hya Association \citep{2016IAUS..314...16K}.
We searched for new members of TW Hya, starting with a list of known members from \citet{2018ApJ...856...23G}.  There were no nearby Disk Detective stars so our choice of this list of known members did not matter much.

%%%%%%%%%%%%%%%%%%%%%%%%%%%%%%%%%%%%%%%%%%%%%%%%%%

%The disk candidates that appear to be in a similar space and have similar motion to known young moving groups are found in figures \ref{fig:vectors_32Or} through \ref{fig:vectors_USCo}, with the data found in \ref{table:DDinknownMGs}.  The following is merely a listing of candidates that were found in VR to have similar motion and position when compared to a moving group catalog, and would need to be investigated further for additional properties for consideration.

%Since 2016, the Augmented Reality/Virtual Reality (AR/VR) Development Laboratory at NASA's Goddard Spaceflight Center (GSFC) has been developing software for various applications, including engineering, science (including heliophysics and astrophysics), simulations, modeling and outreach, etc.  Of these projects, 

% Thomas G. Grubb, AR/VR Product Development Lead

%The PointClouds VR app was loaded with Gaia DR2, and over 4 million stars that have a radial velocity and a parallax of over 0.6 can be found in VR, along with the Disk Detective catalog of disk candidates.  To conduct a search for young moving groups, catalogs of over 40 known moving groups were created and also loaded into the VR system.  The process described below was carried out for all the catalogs of known moving groups that were created, but this paper focuses mainly on the COL and CAR moving groups.

\section{Columba-Carina, Tucana-Horologium and 32 Orionis}
\label{Discussion_Car-Col_and_THA}

The Carina (Car), Columba (Col) and Tucana-Horologium (Tuc-Hor) groups overlap in position and velocity space. The 32 Orionis (32 Or) group lurks nearby these as well, at a similar velocity. Let us discuss these together.

\citet{2015MNRAS.454..593B} determined the ages for three of these YSAs homogeneously through isochrone fitting: $45^{+11}_{-7}$ Myr for the Carina association, $42^{+6}_{-4}$ Myr for the Columba association, and $45\pm{4}$ Myr for the Tucana-Horologium moving group. 
In other words, the ages for these three groups are all consistent with one common age around 45 Myr. In contrast, \citet{2017MNRAS.468.1198B} estimates the age of 32 Or using isochronal fitting at $24\pm{4}$ Myr.

%collectively called the "Great Austral Young Association" \citep{2001ASPC..244...43T, 2008hsf2.book..757T} 

%Carina and Columba \citep{2008hsf2.book..757T} in particular, seem nearly united into one single group when viewed in our VR simulation. 

%We assembled our initial catalog of previously published members of Columba from the list of bona fide members in \citet{2013ApJ...762...88M}.  Our initial Carina catalog was based on the membership list from \citet{2013ApJ...762...88M} plus additional members from \citet{2012ApJ...758...56S}.
%We assembled our initial catalog of previously identified Tuc-Hor members from \citet{2018ApJ...860...43G,2018ApJ...856...23G,2018ApJ...862..138G} and \citet{2019AJ....157..234S}. 

%Many authors have assembled catalogs of previously members of Columba: \citet{2013ApJ...762...88M}.  Our initial Carina catalog was based on the membership list from \citet{2013ApJ...762...88M} plus additional members from \citet{2012ApJ...758...56S}.
%Many authors have assembled catalogs of members of Tuc-Hor: \citet{2018ApJ...860...43G,2018ApJ...856...23G,2018ApJ...862..138G} and \citet{2019AJ....157..234S}. 

%DISCUSS LEE AND SONG 2019A VS LEE AND SONG 2019B HERE. \citet{2019MNRAS.489.2189L}

Recently, \citet{2019MNRAS.486.3434L}
proposed new membership lists for nine YSAs, including Carina, Columba and Tuc-Hor.  Then \citet{2019MNRAS.489.2189L} reexamined the group assignments using two unsupervised machine
learning algorithms (K-means and Agglomerative Clustering), and argued for two new groups by re-combining the memberships of 32 Or (ThOr), Columba, and Tuc-Hor.  The discrepancies between these catalogs assembled by the same authors using different numerical tools highlights the need for a new ab initio approach, and the need for the panoramic view afforded by our VR examination.

\begin{figure}[ht]

\includegraphics[width=0.33\linewidth]{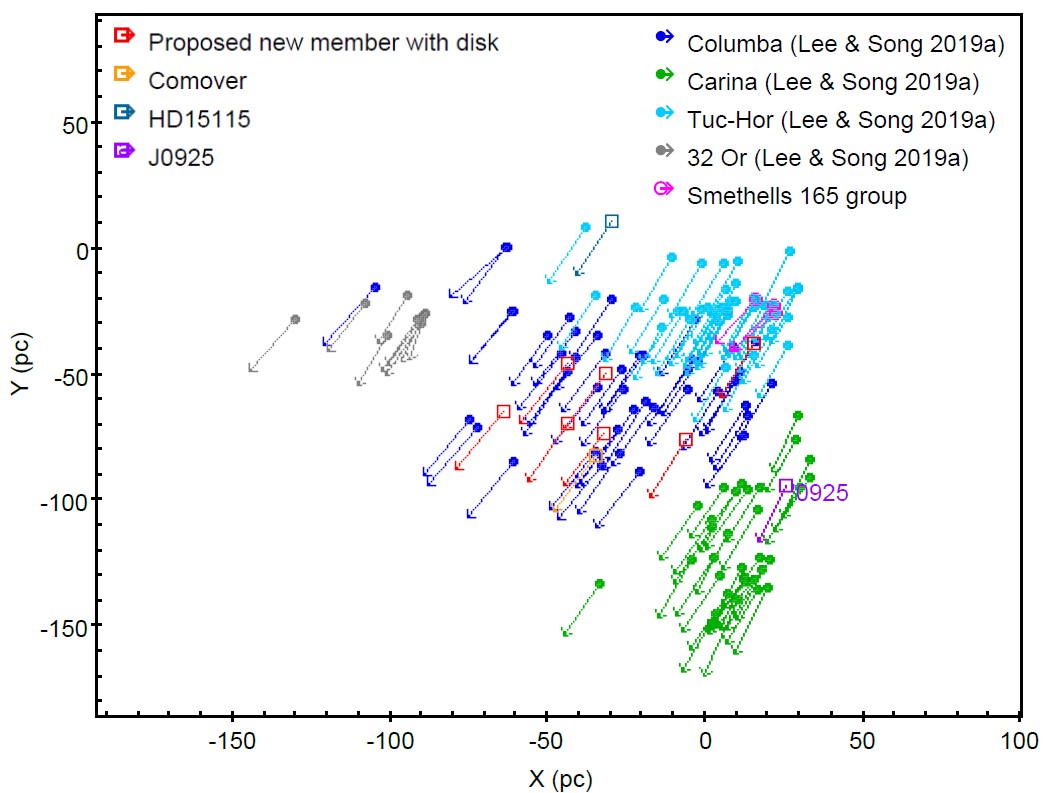}
\includegraphics[width=0.33\linewidth]{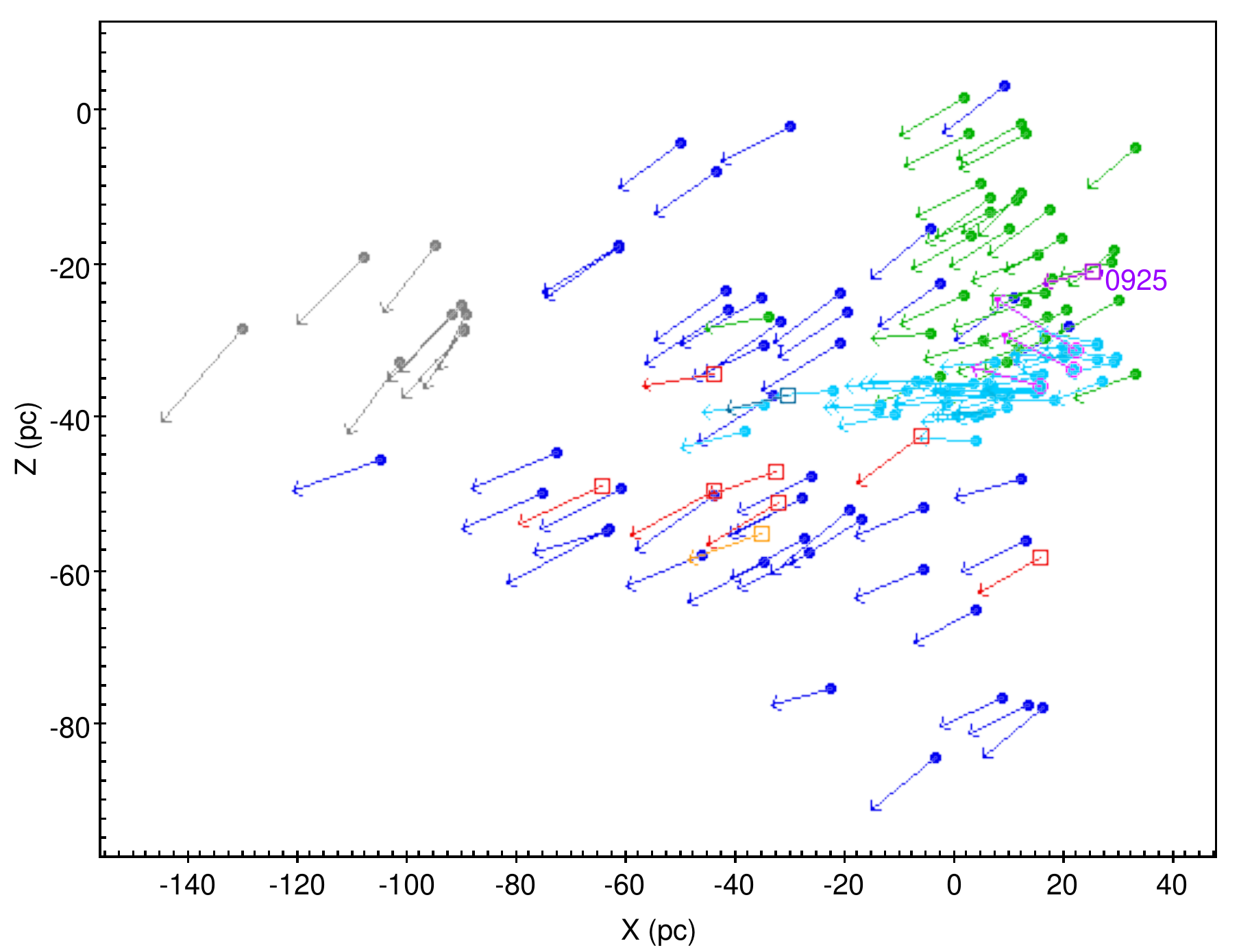}
\includegraphics[width=0.33\linewidth]{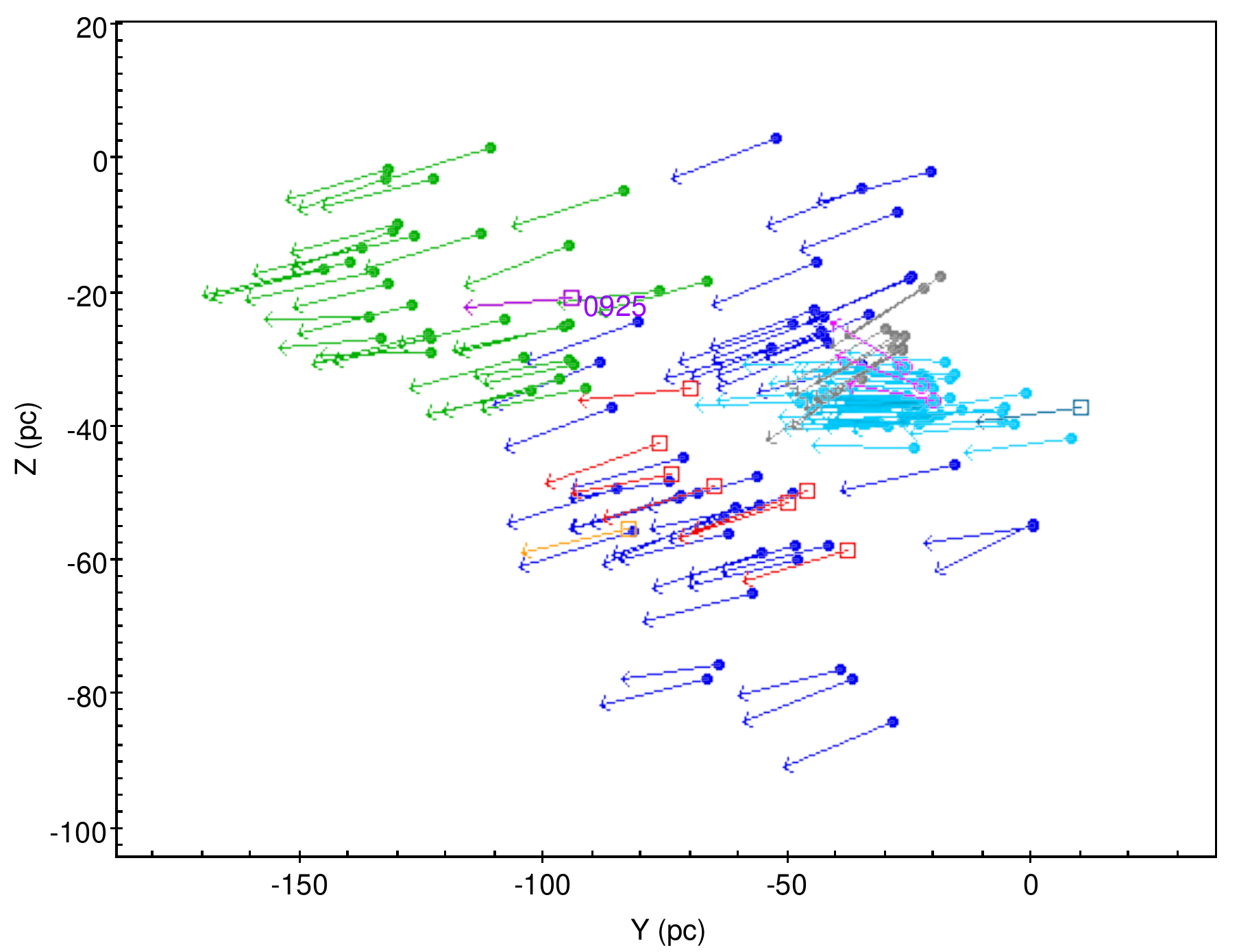}
\caption{Proposed new members of Carina, Columba and Tuc-Hor, together with known members according to \citet{2019MNRAS.486.3434L} in XYZ (parsecs from the sun). The proposed new members of Columba-Carina and Tuc-Hor are the red squares and the comover is the orange square. The extreme debris disk candidate, TYC 9196-2916-1 (J092521.90-673224.8), is the purple square.  HD15115 is the dark teal square.  Columba are the filled blue dots, Carina are the filled green dots, Tuc-Hor are the filled turquoise dots, and 32 Or are the filled grey dots.  The proposed Smethells 165 group members discussed in \ref{Smethells165} are the magenta rings.}
\label{fig:vectors_Car_Col_Tuc-Hor_32Or}
\end{figure}

Figure \ref{fig:vectors_Car_Col_Tuc-Hor_32Or} shows the positions and velocities for stars in these four groups. We suggest that while 32 Or is similar in position space and motion as Columba, Carina and Tuc-Hor, it is possible to distinguish it from these groups.  The mean velocity of Tuc-Hor with the Smethells 165 group removed is U = -9.944, V = -21.083, W = -0.524.  In comparison, the mean velocity of 32 Or is U = -10.3390794, V = -19.54700395, W = -8.901980704.  Its velocity is greater in the -W direction that that of the other groups and it does not overlap in position space in the middle panel. We also did not find any new candidate members of 32 Or, so we will not discuss it further.

\begin{figure}[ht]

\includegraphics[width=0.33\linewidth]{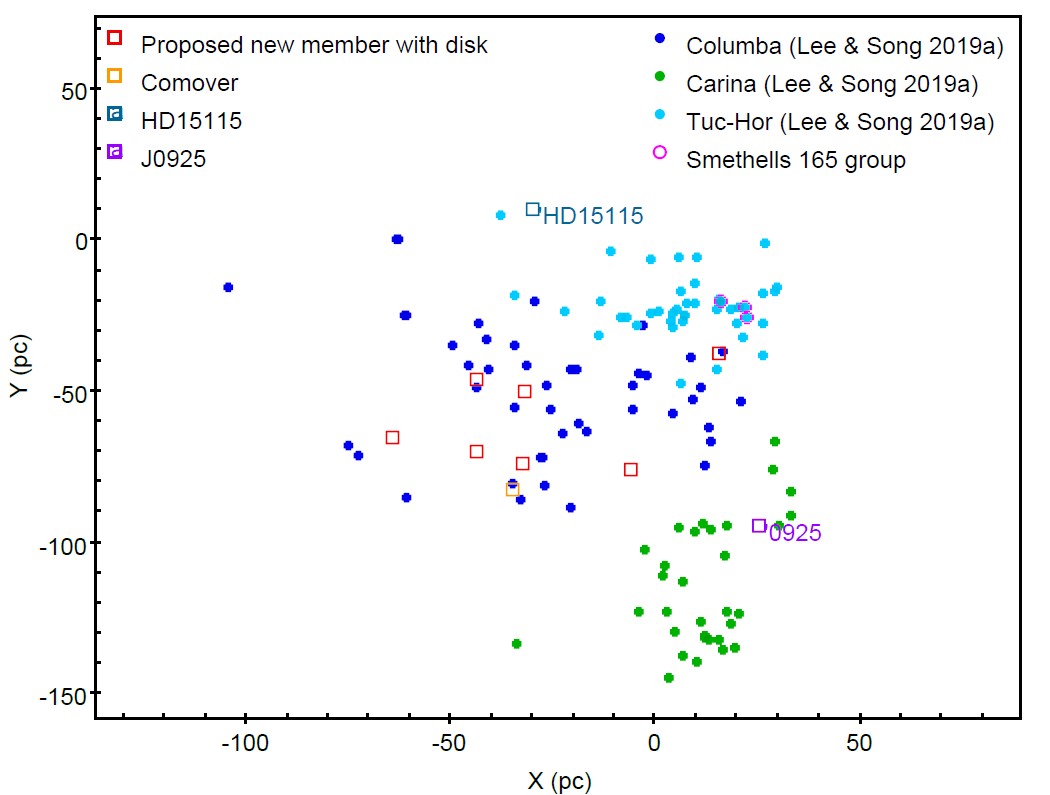}
\includegraphics[width=0.33\linewidth]{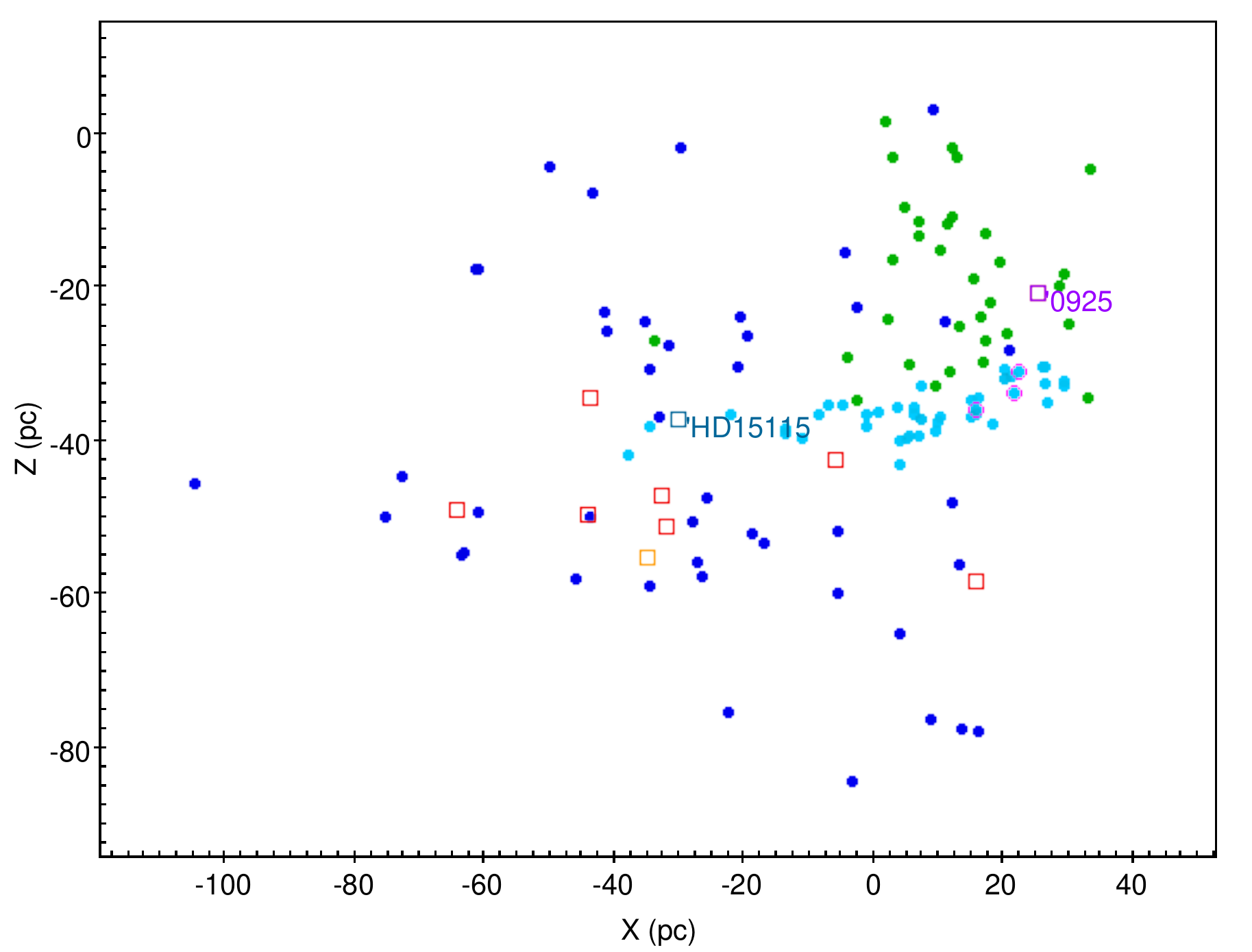}
\includegraphics[width=0.33\linewidth]{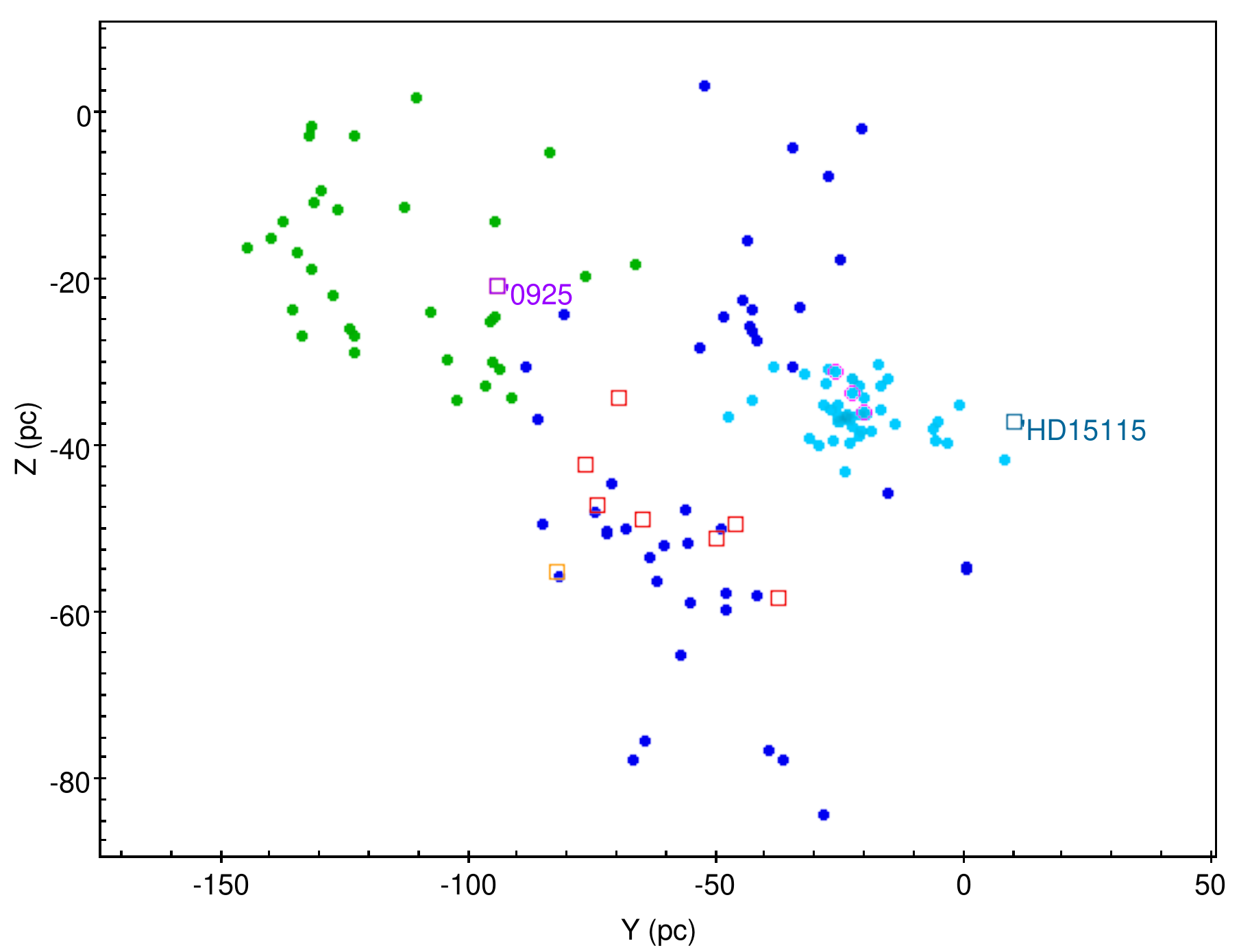}
\includegraphics[width=0.33\linewidth]{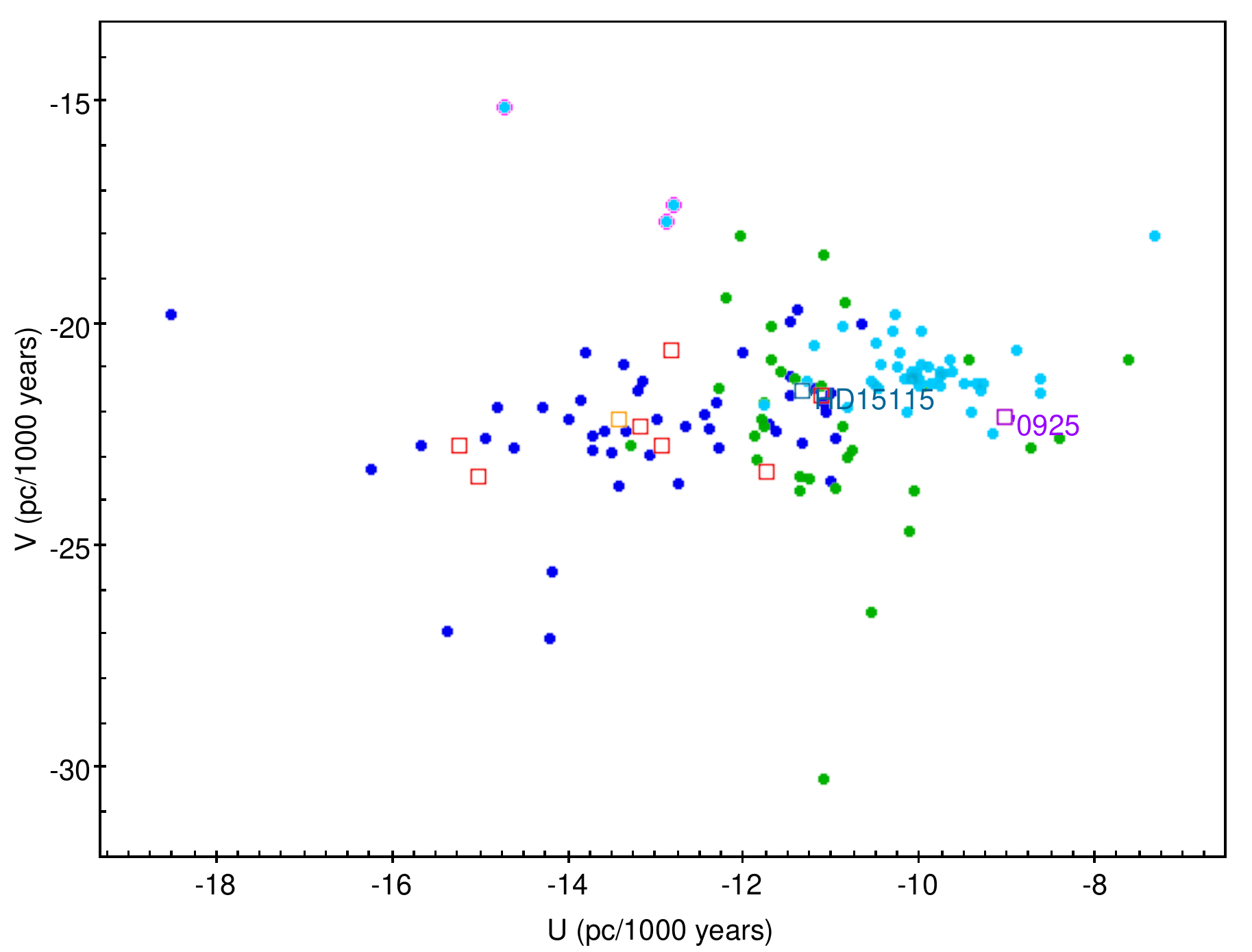}
\includegraphics[width=0.33\linewidth]{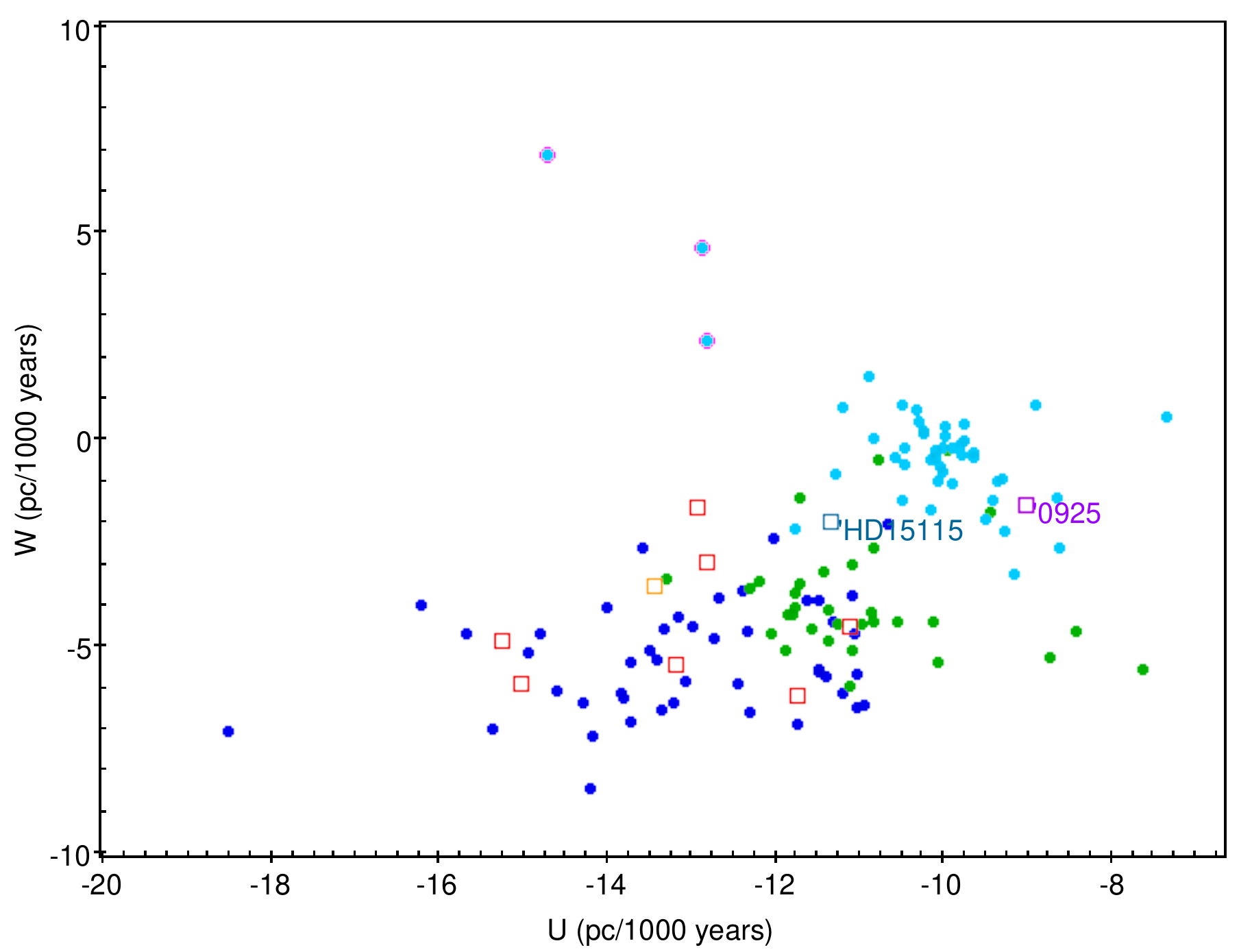}
\includegraphics[width=0.33\linewidth]{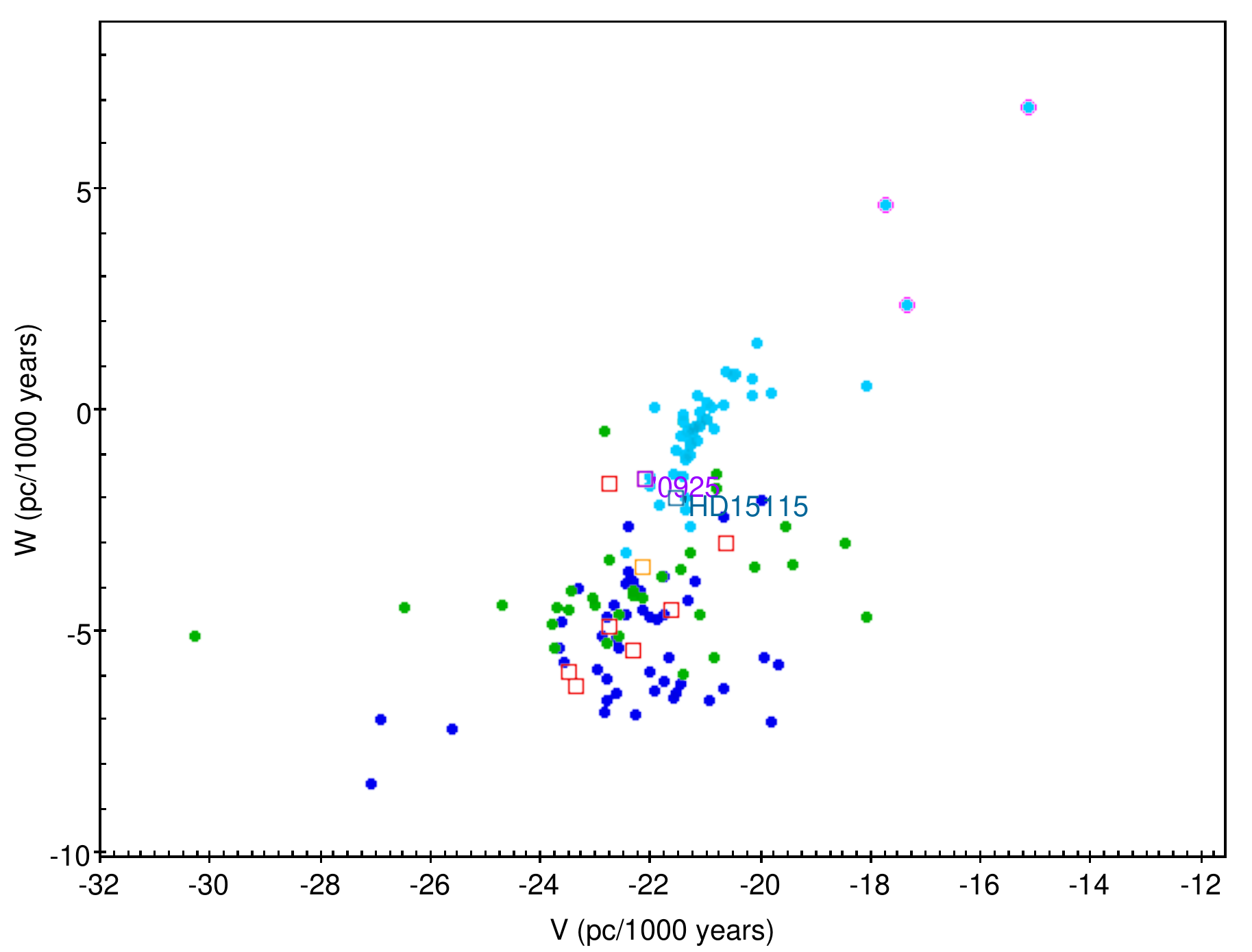}
\caption{Proposed new members of Carina, Columba and Tuc-Hor, together with known members according to \citet{2019MNRAS.486.3434L} in XYZ (parsecs from the sun) in the top 3 frames and UVW (parsecs/1000 years) in the bottom 3 frames. The proposed new members of Columba-Carina and Tuc-Hor are the red squares and the comover is the orange square. The extreme debris disk candidate, TYC 9196-2916-1 (J092521.90-673224.8), is the magenta square. HD15115 is the dark teal square.  Columba are the filled blue dots, Carina are the filled green dots, and Tuc-Hor are the filled turquoise dots.  The proposed Smethells 165 group members are the magenta rings.}
%\caption{Proposed new members of Carina, Columba and Tuc-Hor, together with known members according to \citet{2019MNRAS.486.3434L} in XYZ (parsecs \textbf{from the sun}) in the top 3 frames and UVW (parsecs/1000 years) in the bottom 3 frames. The proposed new members of Columba-Carina and Tuc-Hor are the red inverted triangles and the comover is the orange inverted triangle. The extreme debris disk candidate, TYC 9196-2916-1 (J092521.90-673224.8), is the magenta inverted triangle.  COL are the filled blue dots, CAR are the filled green dots, and Tuc-Hor are the filled turquoise dots.  The proposed loosely affiliated members of Tuc-Hor are the magenta rings.}
%\caption{Proposed new members of Carina, Columba and Tuc-Hor, together with known members according to \citet{2019MNRAS.486.3434L} in in XYZ (parsecs) in the top 3 frames and UVW (parsecs/1000 years) in the bottom 3 frames. The proposed new members of Columba-Carina and Tuc-Hor are the red inverted triangles and the comover is the orange inverted triangle. The extreme debris disk candidate, TYC 9196-2916-1 (J092521.90-673224.8), is the magenta inverted triangle.  COL are the filled blue dots, CAR are the filled green dots, and Tuc-Hor are the filled turquoise dots.  The proposed loosely affiliated members of Tuc-Hor are the magenta rings.}
\label{fig:DD_and_GAYA-L&S}

\end{figure}

Rather, based on Figures \ref{fig:vectors_Car_Col_Tuc-Hor_32Or} and \ref{fig:DD_and_GAYA-L&S}, along with our VR examination, we have decided to first discuss Tuc-Hor, and then discuss Carina and Columba as one combined association, Columba-Carina.

%Figure \ref{fig:DD_and_GAYA-L&S} 

Figure \ref{fig:DD_and_GAYA-L&S} shows the X, Y, Z, U, V, W positions and motion of the \citet{2019MNRAS.486.3434L} catalog.
Let us begin by ignoring the colors of the symbols, which may merely represent the history of how these YSAs were discovered.  Now, consider all the objects to be part of a single large group. With this perspective, we see a reasonably tight cluster in velocity space; all but a handful of outliers lie within about 6 pc/Myr of one another. In position space, the stars are spread over about 100 pc in Z, and 150 pc in X and Y.  The upper right panel and the lower middle panels show hints that the stars could be grouped into two separate clusters. 

When we allow ourselves once again to consider the colors of the symbols, we see that the hint of a separation between two groups places Tuc-Hor entirely into one group, but cleaves Columba in two in position space. In the upper right panel about half of the Columba (blue) stars lie in the upper right next to the Tuc-Hor (turquoise) stars, while the other half are towards the middle of the panel.
 
%Carina and Columba are not hard to distinguish from one another in position space (upper right and lower left), but they are strongly overlapped in velocity space. 

Of the three proposed groups, Tuc-Hor seems perhaps the most distinct, at least in velocity (it overlaps with Columba in position). The Tuc-Hor stars (turquoise) are more tightly clustered in both position and velocity than either of the other groups, especially once we ignore the four stars that we have labelled as ``Smethells 165 group'' (magenta).  Tuc-Hor stars clearly have the highest U, V, and W values (lower panels). The lower middle panel shows a possible slight gap between the Tuc-Hor stars and the others. They overlap somewhat with Columba stars in terms of position, but are relatively tightly constrained, occupying a range of only about 20 pc in Z. %Our observations in VR suggest that while 32 OR is similar in position space and motion as Columba-Carina and Tuc-Hor, it remains distinctly different from these groups. Based on this figure and our VR examination, we have decided to first discuss Tuc-Hor, and then discuss Carina and Columba as one combined association, Columba-Carina. %However, we will also hold onto the idea that perhaps these three groups form one single combined association, the GAYA.

%WHAT DO WE THINK ABOUT THE IDEA THAT 32 OR BELONGS IN THIS MIX? <--Based on our observations in VR, we find that while 32 OR is similar in position space and motion as Columba-Carina and Tuc-Hor, it remains distinctly different from these groups.

%Using the membership lists described above \citet{2013ApJ...762...88M,2013ApJ...762...88M,2012ApJ...758...56S, 2018ApJ...860...43G,2018ApJ...856...23G,2018ApJ...862..138G} and \citet{2019AJ....157..234S}, we found distinguishing the motion between COL and CAR to be difficult in VR.  When we switched to the new memberships contained in  \citet{2019MNRAS.486.3434L}, we found the motion of these two groups are more clearly defined. We continued our work using only the  \citet{2019MNRAS.486.3434L} membership lists. 

\subsection{Proposed Tucana-Horologium Association Members}
\label{sect:Tuc-Hor}

\begin{figure}[ht]

\includegraphics[width=0.33\linewidth]{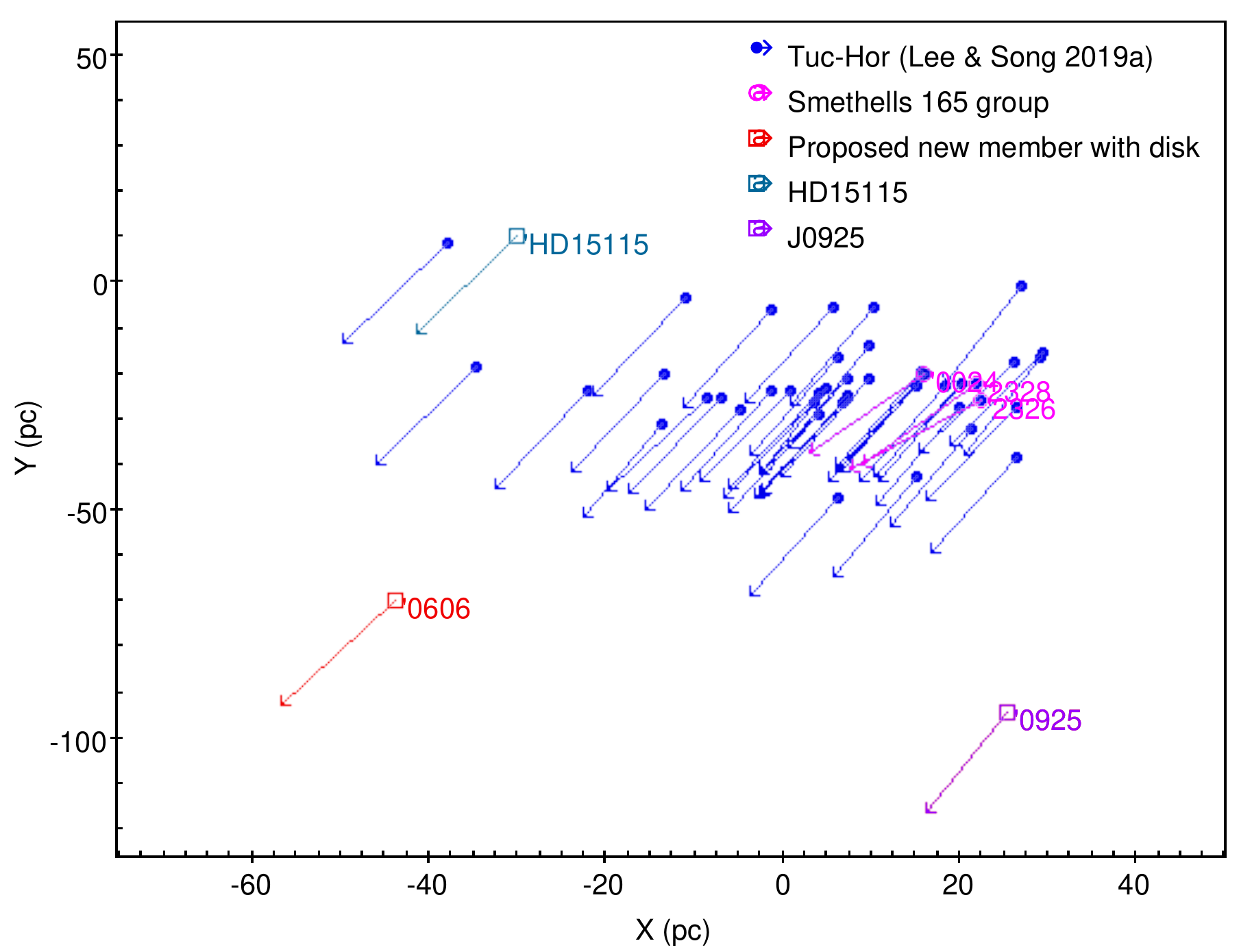}
\includegraphics[width=0.33\linewidth]{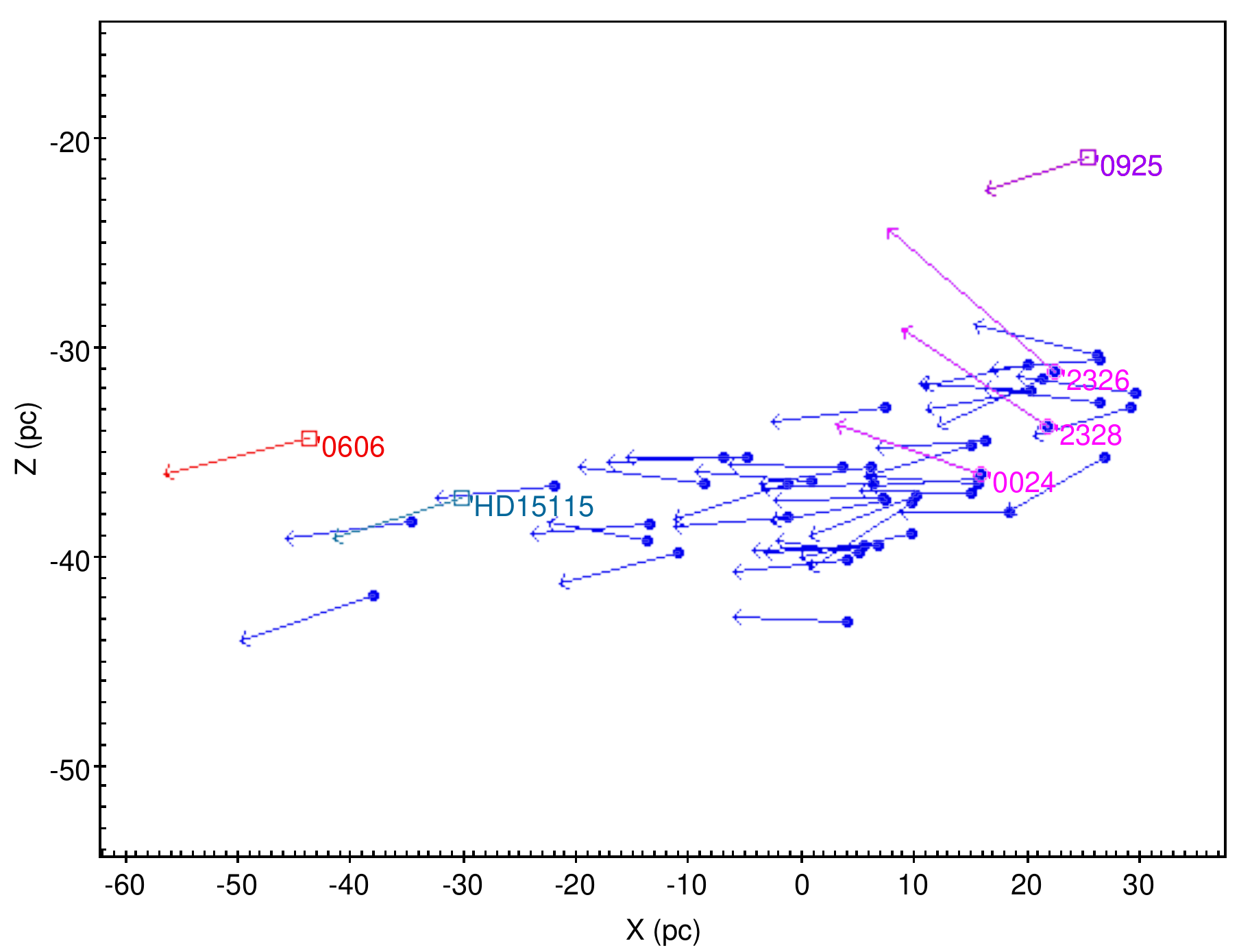}
\includegraphics[width=0.33\linewidth]{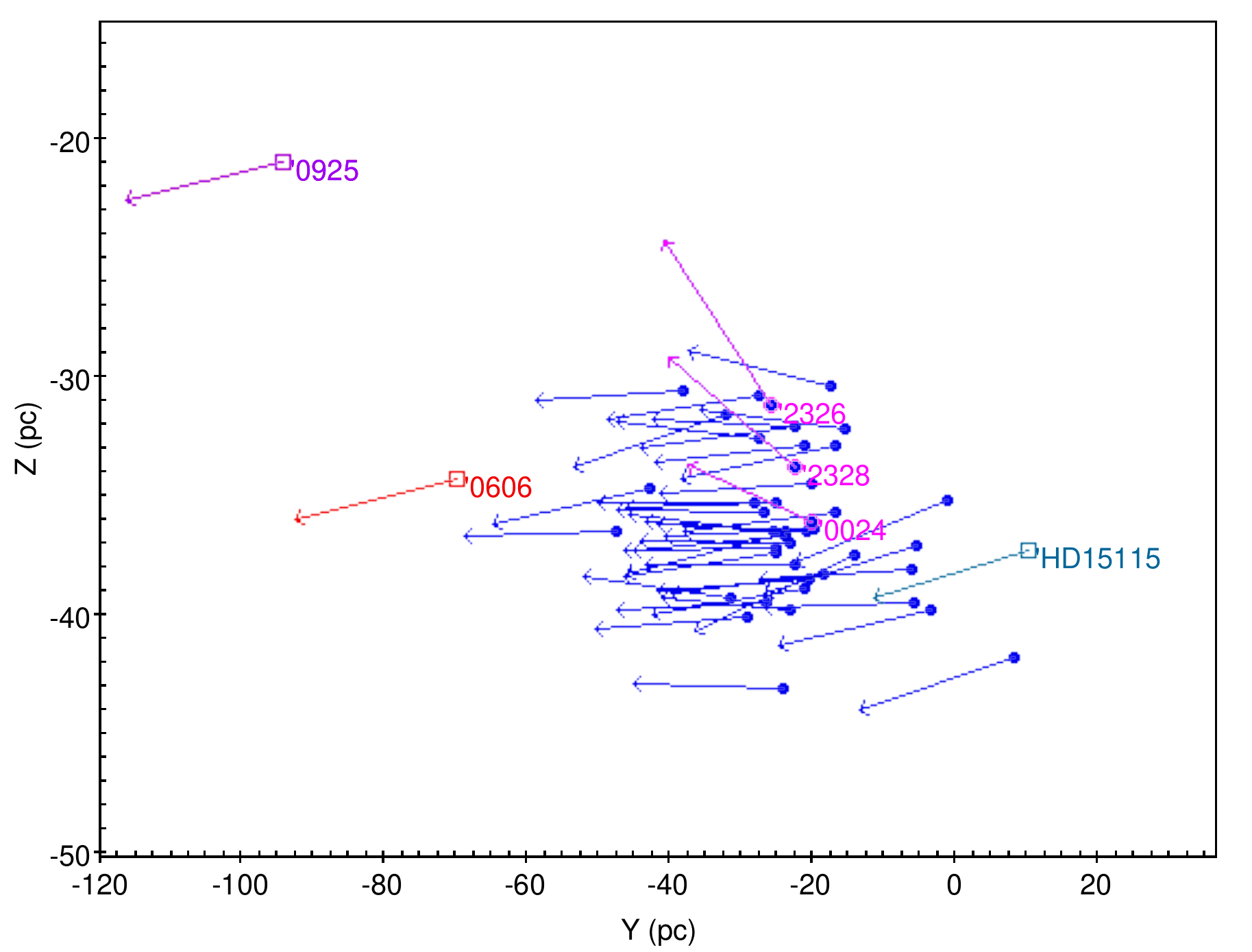}
\caption{Previously identified members of Tucana-Horologium (blue) from \citet{2019MNRAS.486.3434L} in XYZ (parsecs from the sun) with vectors showing UVW (parsecs/1000 years).  
A newly-identified loosely affiliated disk-hosting star (WISEA J060652.79-313054.1) is the red square, and another star, which is an extreme debris disk candidate (WISEA J092521.90-673224.8) is the purple square.  Three stars that had been called bona fide members are shown in magenta.  We propose that these three stars could be members of another group with an age similar to that of Tuc-Hor. We are calling this new group the ``Smethells 165 group'' after the brightest of the three stars.  %The blue star 2MASS J21551140-6153119 labelled 2155 appears to be an interloper, and does not appear to belong to {\bf Smethells 165 group} either.
HD 15115 (J022616.32+061732.8), which hosts a well-known edge-on debris disk, is the dark teal square.}
\label{fig:vectors_TucHor_LS}

%\caption{Previously identified members of Tucana-Horologium (blue) from \citet{2019MNRAS.486.3434L} in XYZ (parsecs \textbf{from the sun}) with vectors showing UVW (parsecs/1000 years).  
%Two newly-identified loosely affiliated disk-hosting stars are shown in red, and four loosely associated stars that had been called bona fide members are shown in magenta.  We wonder whether these four magenta stars could be members of another unknown group that is nearby and a similar age.
%HD 15115 (J022616.32+061732.8), which hosts a well-known edge-on debris disk, is shown in turquoise.}

\end{figure}

Many authors have assembled catalogs of members of Tuc-Hor: \citet{2018ApJ...860...43G,2018ApJ...856...23G,2018ApJ...862..138G} and \citet{2019AJ....157..234S}. 
Figure~\ref{fig:vectors_TucHor_LS} shows the new inventory of Tuc-Hor by \citet{2019MNRAS.486.3434L}. 
Thanks to this work, the core members of this group are relatively well defined, and we do not have any new stars to add to the list of core members.  
%CALCULATE MEAN VELOCITY OF INTERLOPERS AND MEAN VELOCITY OF THE GROUP.
%ANY CHANCE THE INTERLOPERS FIT IN CARINA COLUMBA?<--found on line 1607.  They do not fit in Car-Col.

%However, our VR examination did reveal some new loosely affiliated group members. First, we found that four previously identified Tuc-Hor members (2MASS J00240899-6211042, 2MASS J21551140-6153119, 2MASS J23261069-7323498 and 2MASS J23285763-6802338), while they match the group well in position space, are poor matches in velocity. We suggest that these stars, colored magenta in Figures \ref{fig:vectors_TucHor_LS} and \ref{fig:vectors_TucHor_and_BetaPic}, should likewise be considered "loosely affiliated".  The mean velocity of these 4 loosely affiliated stars in parsecs per 1000 years is U = -13.575, V = -18.078, W = 4.393. Meanwhile, the average velocity of the rest of Tuc-Hor is U = -9.944, V = -21.083, W = -0.524.  We looked to see if perhaps they might be better matched to another nearby YSA like $\beta$ Pictoris, but they do not appear to belong to any other YSA and we wonder whether could they be members of another unknown group that is nearby and a similar age.

%Second, 
Our VR examination revealed two infrared excess stars that are within about 60 pc of Tuc-Hor (HD 41992 and TYC 9196-2916-1), and have space velocities very similar to that of the group. We propose that they could be perhaps former group members or members of an extended group corona. The Tuc-Hor association is elongated in the X direction and compact in the Z direction; it is about 80 pc long in the X direction but only about 20 pc wide in Z. The two new loosely associated members we propose (HD 41992 and TYC 9196-2916-1) extend the group mostly in the Y direction.   

Additionally, we advocate for the kinematic membership of the well-known disk host HD 15115 in Tuc-Hor; this membership has been a matter of debate (see below). 

%Additionally, we can confirm the kinematic membership of the well-known disk host HD 15115 in Tuc-Hor; this membership has been a matter of debate (see below). 

\smallskip
\noindent
{\bf HD 41992  (WISEA J060652.79-313054.1)} is a F8V at 89 pc, first identified as an infrared excess star by Disk Detective, as mentioned in  \citet{2016ApJ...830L..28S}.   Our VR inspection suggests relationships between this star and both Columba-Carina and Tuc-Hor; the star's position places it closer to Columba-Carina, while its velocity resembles that of Tuc Hor (Figure \ref{fig:vectors_TucHor_LS}). 

\smallskip
\noindent
{\bf TYC 9196-2916-1 (WISEA J092521.90-673224.8)} is a star of previously unknown spectral type at 100 pc, with no previous mentions in the literature. Disk Detective identified an apparent infrared excess from this star, which appears loosely associated with Tuc-Hor in our inspections.  Our initial MCMC fit to the photometry yielded a best fit stellar temperature of 4400 K, implying a mid-K spectral type. It also revealed excess emission in all four WISE bands. Similar to HD 41992, our VR inspection suggests relationships between this star and both Columba-Carina and Tuc-Hor; the star's position places it closer to Columba-Carina, while its velocity resembles that of Tuc Hor (Figure \ref{fig:vectors_TucHor_LS}).
See Sections \ref{discussion_SED} and \ref{discussion_PP} for a more detailed discussion of this  candidate ``extreme debris disk'' and its infrared excess.

%which appears reasonably well fit by a single-temperature blackbody with a best fit temperature of 177K yielding a reasonably robust determination of  $L_{IR}/L_{\star}$, which is $1.9 \times 10^{-2}$. <--THIS NEEDS TO BE UPDATED BASED ON NEW DATA IN SECTION 6. While debris disks with $L_{IR}/L_{\star}$ of $2 ^{-2}$ <--I THINK THIS SHOULD BE $10^{-2}$--are known, they are rare \citep{2012ApJ...751L..17M,2019AJ....157..202S}.  In fact, \citep{2012ApJ...751L..17M} notes their 2 sources also exhibit 24 micron variability; J0925 also seems to exhibit variability based on WISE flags found in Section \ref{discussion_PP}. This level of infrared excess would be quite high for a debris disk, suggesting the object is perhaps a transitional disk. 

%It also reveals excess emission in all four WISE bands, which appears reasonably well fit by a single-temperature blackbody with a best fit temperature of 177K yielding a reasonably robust determination of  $L_{IR}/L_{\star}$, which is $1.9055 10^{-2}$. This level of infrared excess would be quite high for a debris disk, suggesting the object is perhaps a transitional disk. We discuss this candidate ``Peter Pan'' disk further below.

\begin{figure}[ht]

\includegraphics[width=0.33\linewidth]{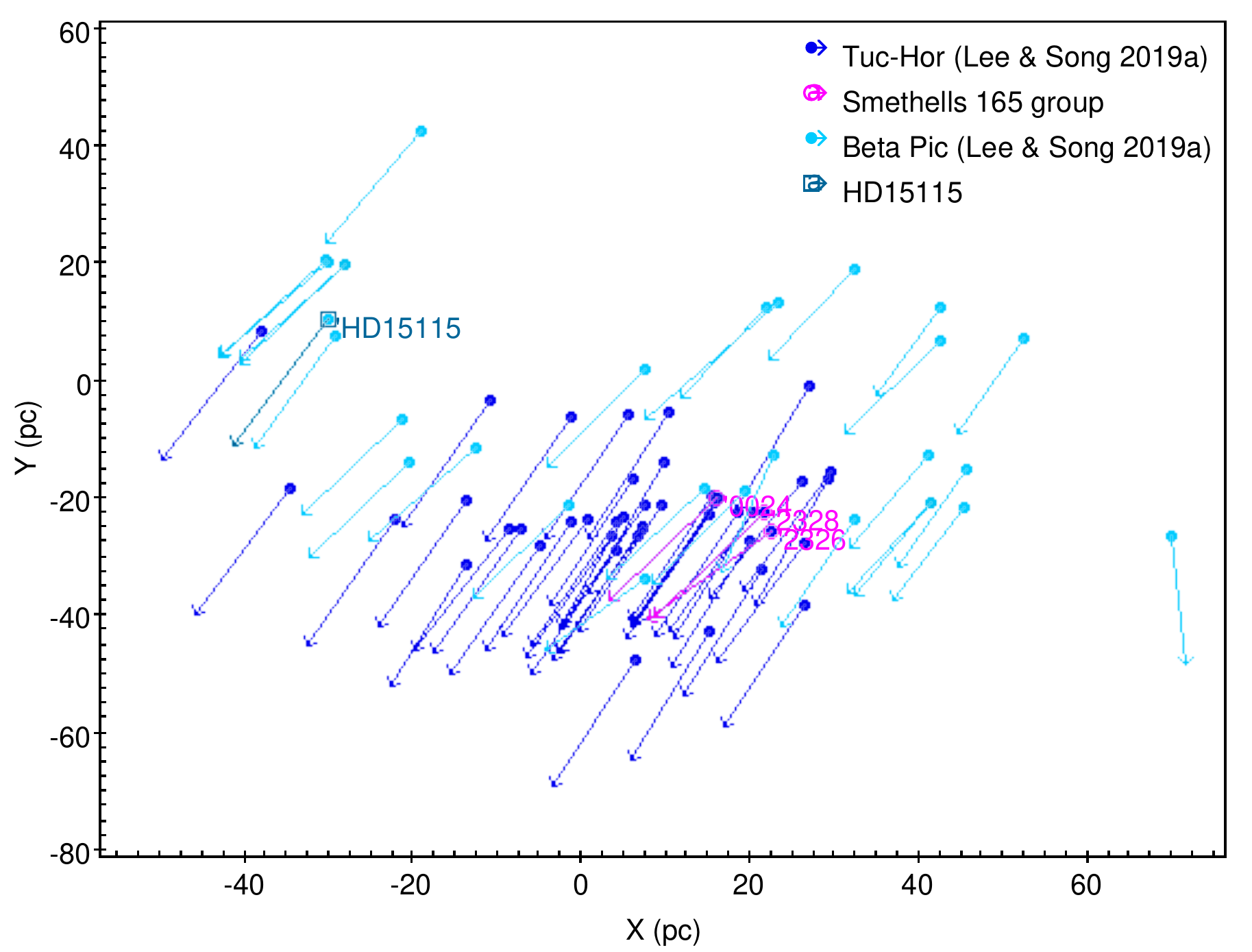}
\includegraphics[width=0.33\linewidth]{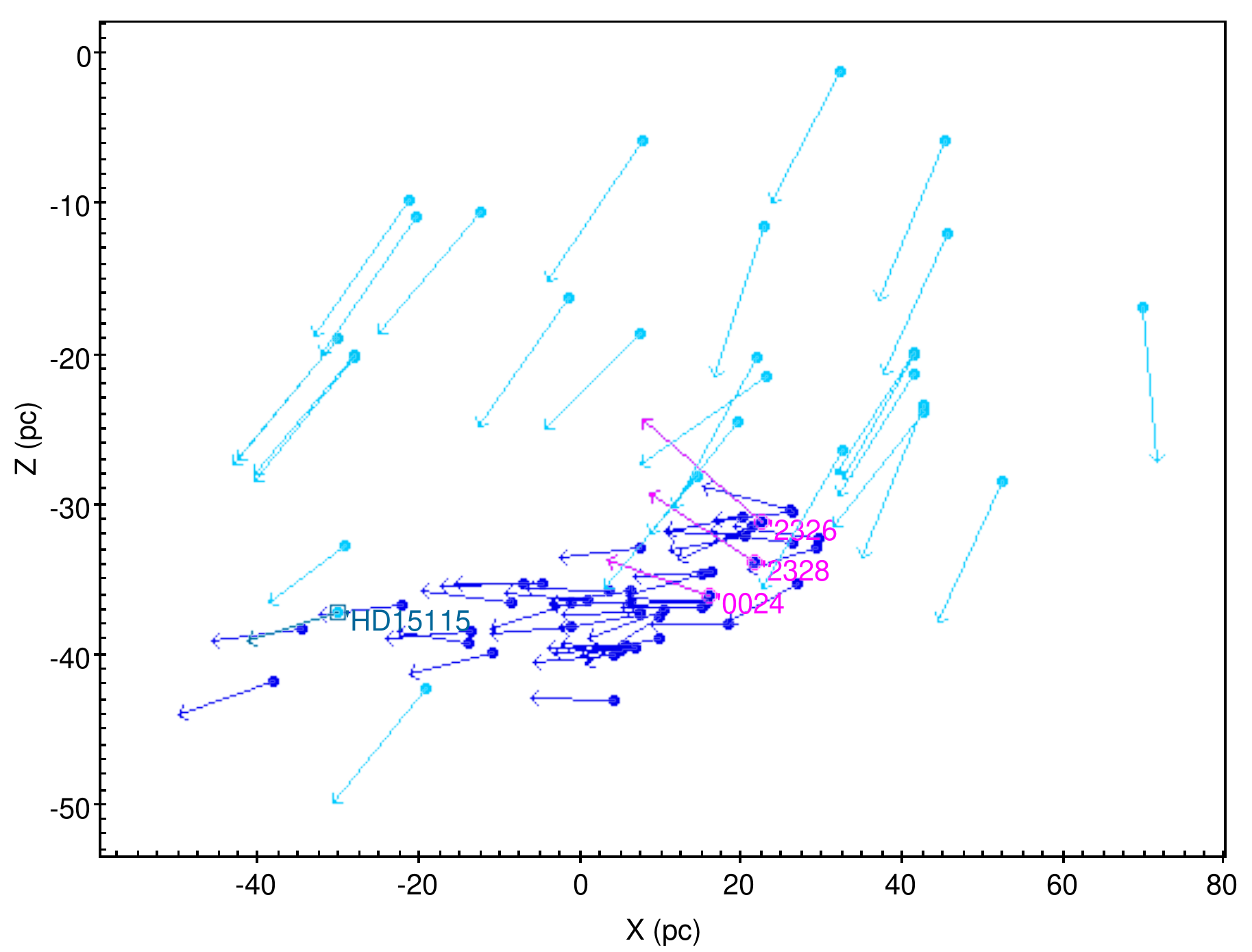}
\includegraphics[width=0.33\linewidth]{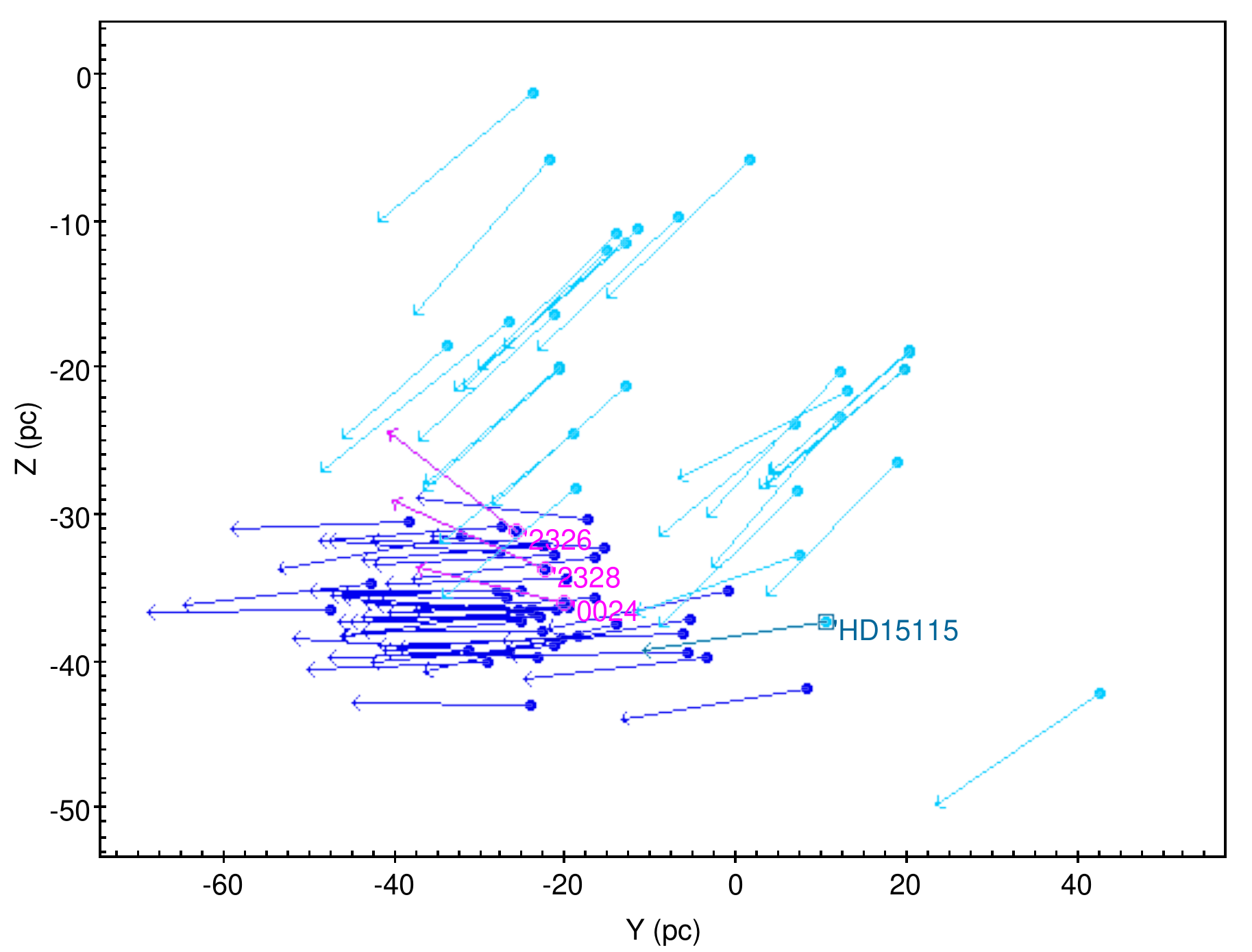}
\caption{The well-known disk-hosting star HD 15115 (J022616.32+061732.8), shown as the dark teal square, was previously identified as a member of $\beta$ Pictoris. However, its velocity and position are a better match for Tuc-Hor, as the middle and right panels of this figure reveal. Previously identified members of Tucana-Horologium (blue) and $\beta$ Pictoris (turquoise) are from \citet{2019MNRAS.486.3434L} in XYZ (parsecs from the sun) with vectors showing UVW (parsecs/1000 years). The proposed Smethells 165 group is shown in magenta.}
\label{fig:vectors_TucHor_and_BetaPic}

\end{figure}

\smallskip
\noindent
{\bf HD 15115 (WISEA J022616.32+061732.8)}  F4 IV  at 49 pc.  \citet{2008ApJ...684L..41D} estimated the star's age to be 100–-500 Myr, based on Ca ii H and K line indicators \citep{2000PhDT........17S}, isochrone fitting \citep{2004A&A...418..989N}, and other indicators. \citet{2008hsf2.book..757T} proposed the star as a high probability member of $\beta$ Pictoris, and \citet{2013ApJ...762...88M} declared it a member of the $\beta$ Pictoris moving group (age 24 Myr). Later,
\citet{2018ApJ...856...23G} examined it as a possible member of $\beta$ Pictoris, but then rejected it as a bona fide member of that group based on a visual inspection.  \citet{2019ApJ...877L..32M} assigned it to Tuc-Hor based on BANYAN combined with Gaia coordinates and radial velocity.  Our BANYAN-independent VR simulation supports this last assignment. Figure~\ref{fig:vectors_TucHor_and_BetaPic} shows the position and velocity of HD 15115 compared to the positions and velocities of $\beta$ Pictoris and Tuc-Hor members from \citet{2019MNRAS.486.3434L}. The middle and right and middle panels show how HD 15115 is a better fit for Tuc-Hor.

\smallskip

%\begin{figure}[ht]

%\includegraphics[width=0.33\linewidth]{figures/x-y_Tuc-Hor_and_DD_with_Beta_Pic_and_DD.pdf}
%\includegraphics[width=0.33\linewidth]{figures/x-z_Tuc-Hor_and_DD_with_Beta_Pic_and_DD.pdf}
%\includegraphics[width=0.33\linewidth]{figures/y-z_Tuc-Hor_and_DD_with_Beta_Pic_and_DD.pdf}
%\caption{New proposed disk-hosting kinematic members of Tucana-Horologium and Beta Pictoris (red) and previously identified members of Tucana-Horologium (blue) and Beta Pictoris (green).  The comparison of Disk Detective candidate J022616.32+061732.8, a previously identified member of Beta Pic, to Tuc-Hor and Beta Pic can be seen here. }
%\label{fig:vectors_TucHor_and_BetaPic}

%\end{figure}

%\begin{figure}[ht]

%\includegraphics[width=0.33\linewidth]{figures/x-y_Tuc-Hor_and_DD_with_Beta_Pic_and_DD_2020-12-21.pdf}
%\includegraphics[width=0.33\linewidth]{figures/x-z_Tuc-Hor_and_DD_with_Beta_Pic_and_DD_2020-12-21.pdf}
%\includegraphics[width=0.33\linewidth]{figures/y-z_Tuc-Hor_and_DD_with_Beta_Pic_and_DD_2020-12-21.pdf}
%\caption{New proposed disk-hosting kinematic members of Tucana-Horologium and Beta Pictoris (red) and previously identified members of Tucana-Horologium (blue) and Beta Pictoris (green) from \citet{2018ApJ...860...43G,2018ApJ...856...23G} and \citet{2018ApJ...862..138G,2019AJ....157..234S}.  The comparison of Disk Detective candidate J022616.32+061732.8, a previously identified member of Beta Pic, to Tuc-Hor and Beta Pic can be seen in turquoise. }
%\label{fig:vectors_TucHor_and_BetaPic}

%\end{figure}

\begin{figure}[ht]
    \centering
    \includegraphics[width=0.5\linewidth]{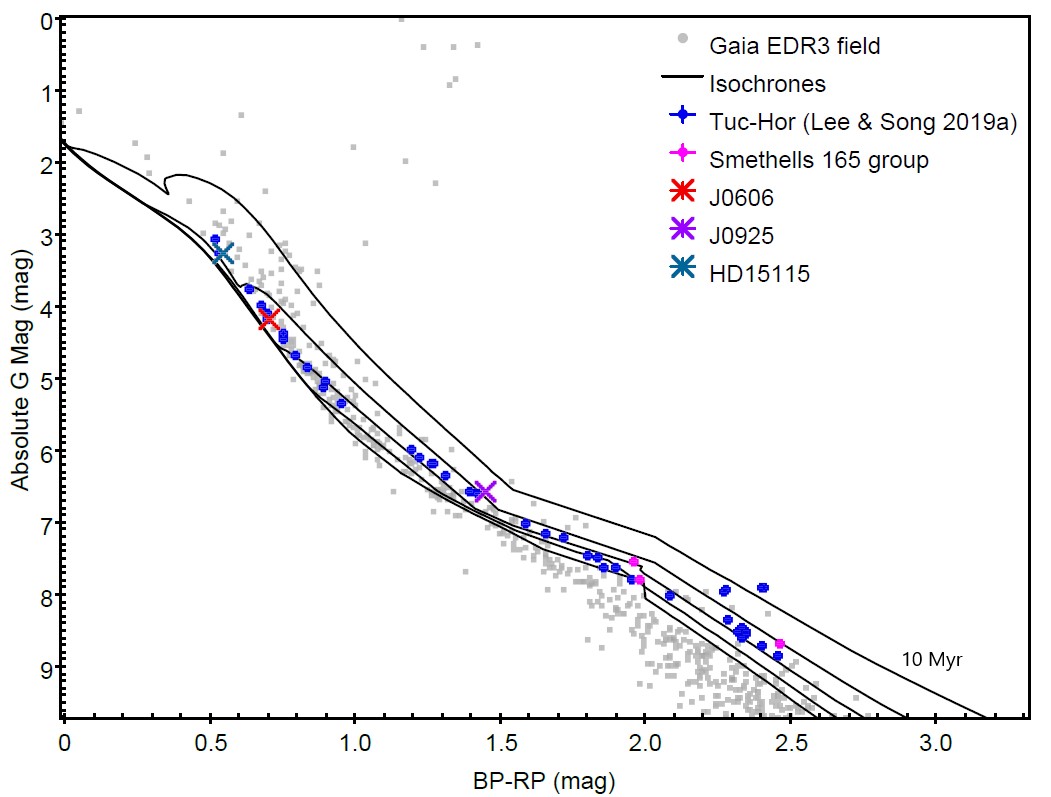}
    \caption{Colour magnitude diagram (CMD) of Tuc-Hor members, shown in blue. Proposed new member WISEA J060652.79-313054.1 is red, proposed extreme debris disk candidate J0925 is purple and HD 15115 (J022616.32+061732.8) is the dark teal. The proposed Smethells 165 group members are in magenta.  Also shown are isochrones from 10 Myr (top) to 50 Myr (bottom) and Gaia EDR3 field stars in grey.}
    \label{fig:HRdiagSmethells165}
\end{figure}

%WHAT ABOUT THE OTHER CATALOGS?

\subsection{A Group Containing Smethells 165}
\label{Smethells165}

Our VR examination revealed that four previously identified Tuc-Hor members (2MASS J00240899-6211042, 2MASS J21551140-6153119, 2MASS J23261069-7323498 and 2MASS J23285763-6802338) from the \citet{2019MNRAS.486.3434L} catalog match the group well in position space but do not match in velocity.
One of these stars, 2MASS J21551140-6153119, does not have parallax in EDR3 so we set it aside.
However, we suggest that the remaining three stars, listed in Table \ref{table:smethells165}, and colored magenta in Figures \textbf{\ref{fig:vectors_Car_Col_Tuc-Hor_32Or}, \ref{fig:DD_and_GAYA-L&S},} \ref{fig:vectors_TucHor_LS} and \ref{fig:vectors_TucHor_and_BetaPic}, may not belong to Tuc-Hor after all.  The mean velocity of these 3 stars in parsecs per 1000 years is U = -13.460, V = -16.742, W = 4.627.
%The mean velocity of these 4 stars in parsecs per 1000 years is U = -13.575, V = -18.078, W = 4.393.
%We suggest that these stars, colored magenta in Figures \textbf{\ref{fig:vectors_Car_Col_Tuc-Hor_32Or}, \ref{fig:DD_and_GAYA-L&S},} \ref{fig:vectors_TucHor_LS} and \ref{fig:vectors_TucHor_and_BetaPic}, should likewise be considered "loosely affiliated" \textbf{or possibly even that these stars may not belong to Tuc-Hor after all}.  The mean velocity of these 4 loosely affiliated stars in parsecs per 1000 years is U = -13.575, V = -18.078, W = 4.393. 
Meanwhile, the mean velocity of the rest of Tuc-Hor is U = -9.944, V = -21.083, W = -0.524.  

We looked to see if perhaps these three stars might be better matched to another nearby YSA like $\beta$ Pictoris, Columba-Carina or 32 Or. However, our visual inspection in VR suggests that they do not belong to any other YSA.  They seem more likely to represent another new unknown group, perhaps with an age similar to Tuc-Hor, based on the color magnitude diagram, Figure \ref{fig:HRdiagSmethells165}.  We are calling this group the ''Smethells 165 group''.  Notably, these 3 stars are all M dwarfs \citep{Torres2006,2017ApJ...840...87R}.

\begin{table}[ht]
\caption{Proposed Smethells 165 group \label{table:smethells165}}
\begin{center}
    \begin{tabular}{ lllcccc }
    \hline
    2MASS Name &   GAIA DR2 ID &   Common Name &   Spectral    &   U   &   V   &   W   \\
     &      &       &    Type   &   (pc/1000yrs)   &  (pc/1000yrs) & (pc/1000yrs)    \\
    \hline
    J00240899-6211042    &   4901926404913922816 &   Smethells 165   &   M0Ve$^{1}$  &   -12.805    &   -17.350    &  2.391   \\
    J23261069-7323498    &   6380514358792367232 &   PM J23261-7323  &   M0Ve$^{1}$  &   -14.706    &   -15.149    &  6.851 \\
    J23285763-6802338    &   6388014157668558080 &   UCAC4 110-129613    &   M2.5Ve$^{2}$  &   -12.868 &   -17.728    &  4.638 \\
    \hline
    \end{tabular}
    \begin{tablenotes}
        Notes.  References: $^{1}$ \citet{Torres2006}, $^{2}$ \citet{2017ApJ...840...87R}.
    \end{tablenotes}
\end{center}
\end{table}

%Our VR inspection revealed four stars that have previously been proposed as members whose velocities are similar to on another, but stand out from the rest of the association. We have labeled these in Figure~\ref{fig:vectors_TucHor_LS} as "Loosely affiliated".

%The list of these objects can be found in section \ref{candidates}. 

\subsection{Proposed Columba-Carina Members}
%\label{candidates}

\begin{figure}[ht]

\includegraphics[width=0.33\linewidth]{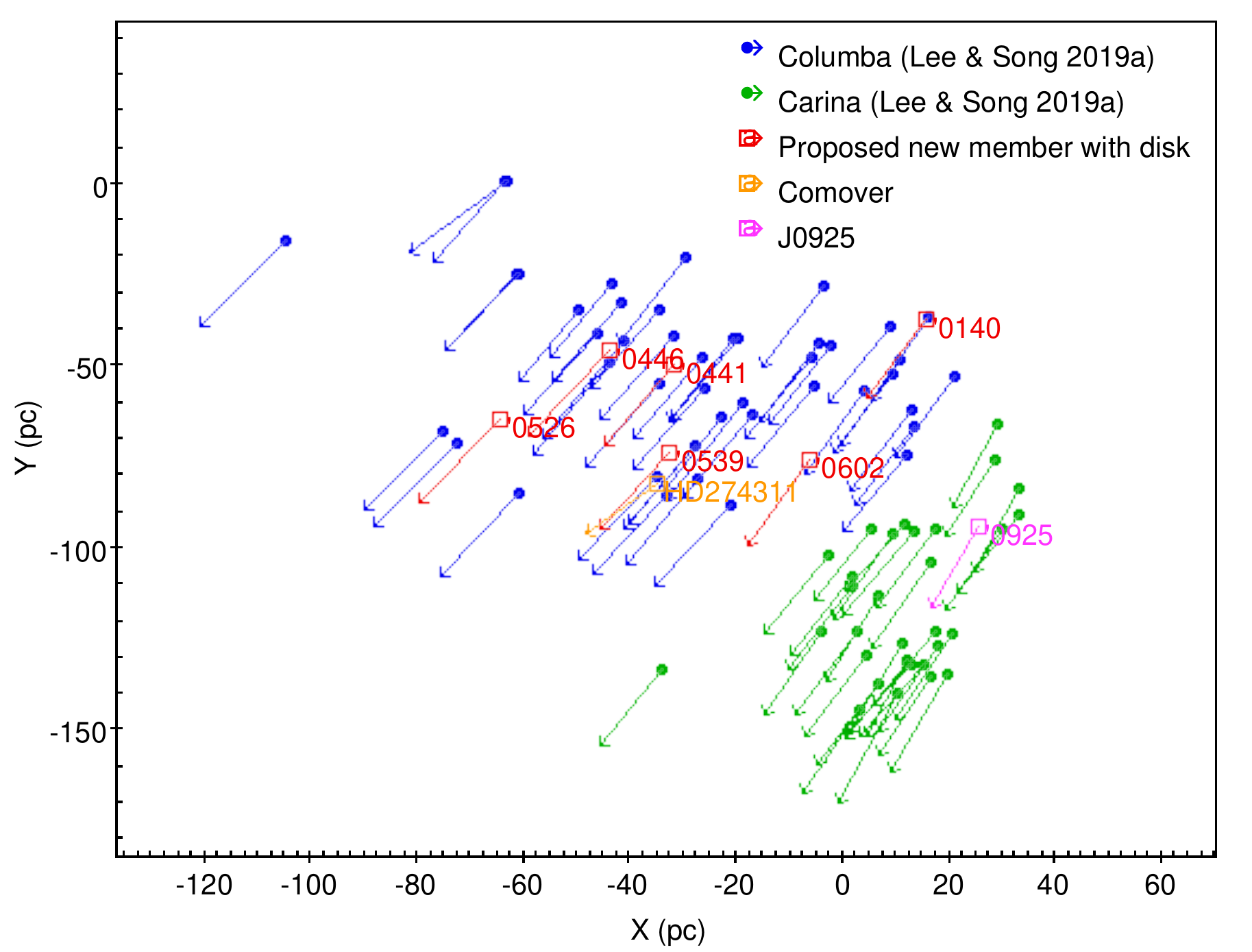}
\includegraphics[width=0.33\linewidth]{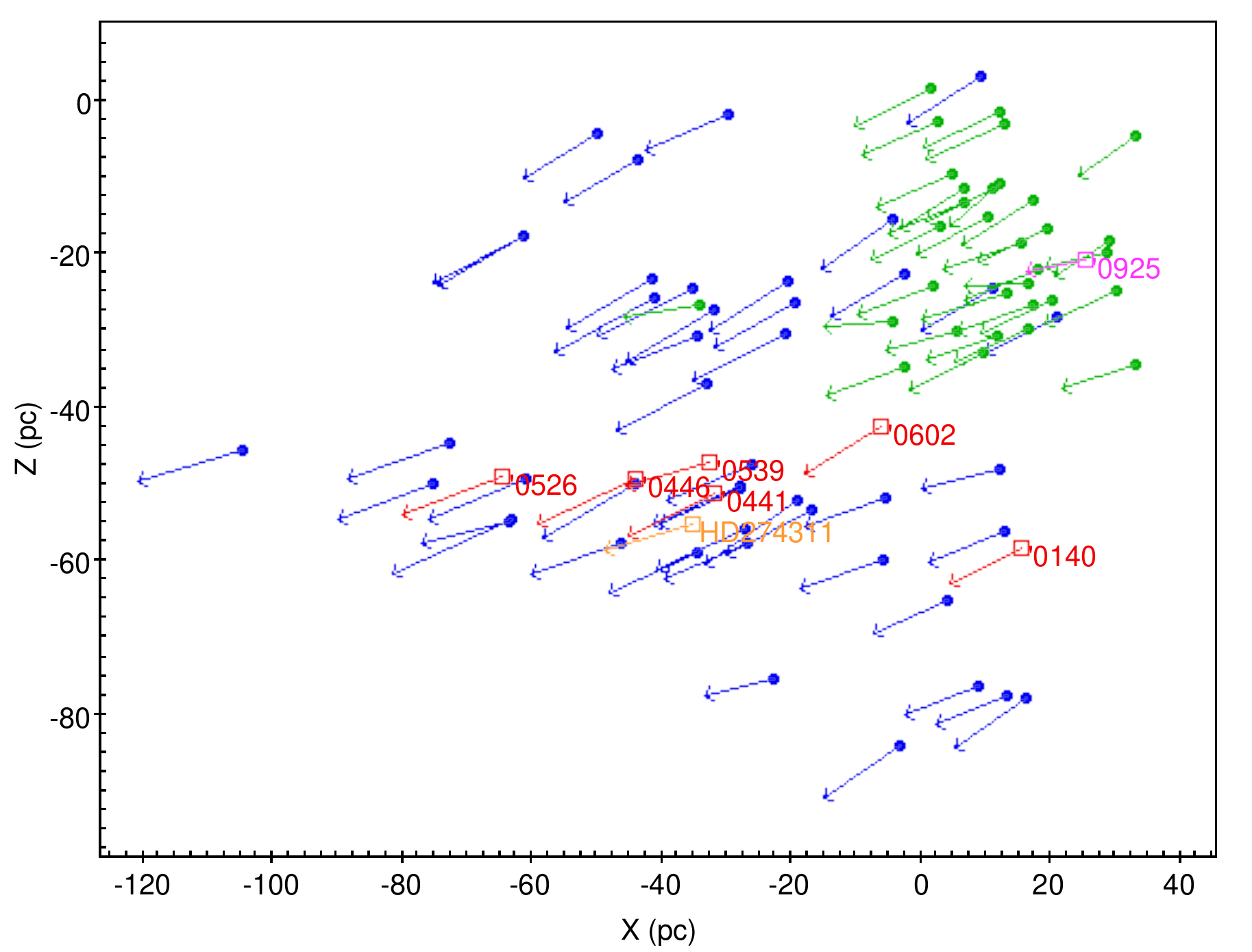}
\includegraphics[width=0.33\linewidth]{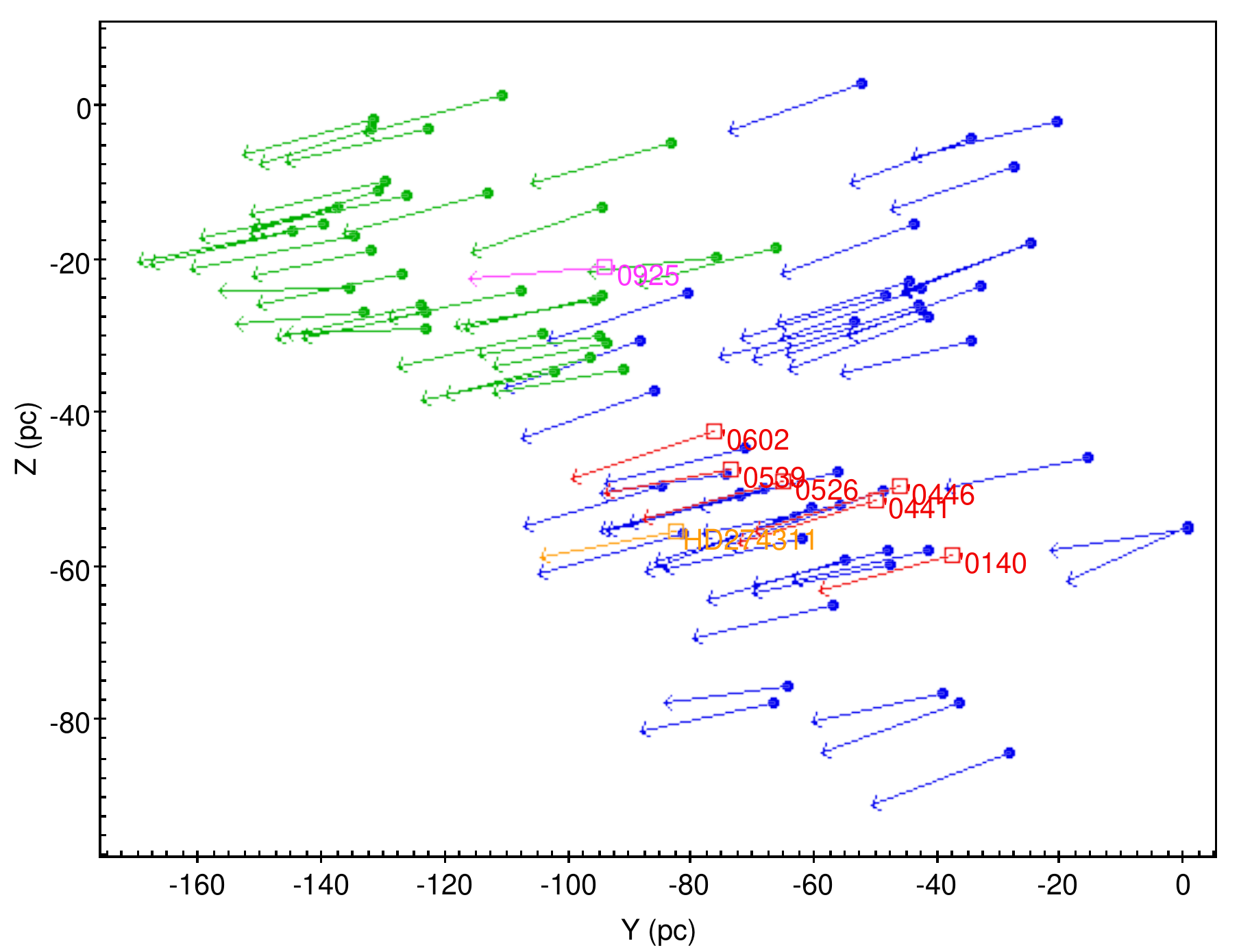}
\caption{Proposed new disk-hosting kinematic members of Columba-Carina (red), with comover (orange), extreme debris disk candidate J0925 (magenta), and previously identified members of Columba (blue) and Carina (green) from \citet{2019MNRAS.486.3434L} in XYZ (parsecs from the sun) with vectors showing UVW (parsecs/1000 years).}
\label{fig:vectors_Columba_Carina_DD_comover}

\end{figure}

Based on our VR examination, and analysis of Figures \ref{fig:vectors_Car_Col_Tuc-Hor_32Or} and \ref{fig:DD_and_GAYA-L&S}, we discuss Columba and Carina as one combined association, Columba-Carina.  Our VR inspections suggest that \citet{2019MNRAS.486.3434L} lists of Carina and Columba members need to be augmented with 6 additional members. These disk candidates are found in Table \ref{table:DDinknownMGs}, and the comover can be found in Table \ref{table:ComoverstoDDinknownMGs}.   Figure \ref{fig:vectors_Columba_Carina_DD_comover} shows the positions and motions of these disk candidates compared to the Columba and Carina members listed by \citet{2019MNRAS.486.3434L}.

One star with an infrared excess that we propose as a new member of Columba-Carina has never been previously identified with a YSA: CPD-57 937.
A second star with an infrared excess that we propose as a new member of Columba-Carina was previously proposed as a member of Tuc-Hor, but not of Columba or Carina: HD 10472.
We also found four other stars with infrared excesses that we assign to Columba-Carina that were previously proposed as Columba-Carina members prior to \citet{2019MNRAS.486.3434L}:  HD 30447, CPD-35 525, HD 35841, HD 10472 and HD 37852.
This last star, HD 37852, has an 83.5\% likelihood of membership in Columba according to BANYAN $\Sigma$ and a COL candidate in  \citet{2018ApJ...863...91F} and appears in Silverberg et al. in prep. It has four comovers according to \citet{2017AJ....153..257O}, which also likely belong to Columba-Carina.

\smallskip
\noindent
{\bf CPD-57 937 (WISEA J060210.78-570142.1)} This G5 star at 99 pc has not been previously proposed as a member of any YSA. Its infrared excess was previously noted by \citet{2016ApJS..225...15C} (on the ``reserved'' list, which consists of stars with IR excess from Tycho-2/ALLWISE without trigonometric parallaxes, and distance calculated using a SED fitting algorithm). Our inspection supports the assignment of this star to the combined Columba-Carina association.

\smallskip
\noindent
{\bf HD 10472 (WISEA J014024.15-605956.7)}
This star has never been assigned to Columba-Carina, but it has been previously associated with Tuc-Hor Association by \citet{2015ApJ...798...87M} and \citet{2017AJ....154..245M}.  We find Columba-Carina to be a better match.  Its infrared excess was previously noted by \citet{2016ApJS..225...15C} (on the ``prime'' list, which consist of stars within 120 pc of the sun with IR excess from Tycho-2/ALLWISE). (Note that a footnote in \citet{2020AJ....160...24E} suggests that others may have been considering re-assigning this object, but have not yet published their work.)

\smallskip
\noindent
{\bf HD 35841 (WISEA J052636.59-222923.8)} is an F3 V at 103.7 pc.  \citet{Elliott2016} listed this object as a ``candidate'' member of Columba.  Its infrared excess was previously noted by \citet{2016ApJS..225...15C} (on the ``prime'' list).

%The following two stars were not listed as members of Columba by \citet{2019MNRAS.486.3434L},

\smallskip
\noindent
{\bf HD 30447 (WISEA J044649.55-261808.8)}: an F3 V at 81 pc.  The debris disk around this star has been imaged by HST NICMOS \citep{2014ApJ...786L..23S} and VLT SPHERE \citep{2018A&A...611A..43L}. \citet{2018ApJ...856...23G, 2021ApJ...915L..29G} listed this star as a bona fide Columba member, as did \citet{2013ApJ...762...88M}.  BANYAN $\Sigma$ gives a probability of membership in Columba of 99.5\% However, \citet{2019MNRAS.486.3434L} did not list it as a member of this group.  Our inspection supports the assignment of this star to the combined Columba-Carina association.

\smallskip
\noindent
{\bf CPD-35 525 (WISEA J044115.76-351358.1)} a G7 at 76 pc. This star was proposed as a candidate Columba member in \citet{2018ApJ...860...43G} (and BANYAN $\Sigma$ gives a probability of membership in Columba of 99.9\%) but it is not included as a member by \citet{2019MNRAS.486.3434L}.  Its infrared excess was previously noted by \citet{2016ApJS..225...15C}(on their ``prime'' list). Our inspection supports the assignment of this star to the combined Columba-Carina association.

%\smallskip
%\noindent
%{\bf HD 37484 (J053739.64-283734.7)}.  ALREADY KNOWN TO BE MEMBER.  DELETE?

\smallskip
\noindent
%HD 37852 AND ITS 4 COMOVERS GO IN STEVEN'S PAPER

\smallskip
\noindent
{\bf HD 37852 (WISEA J053930.48-404102.4)} is a B8V at 93 pc. BANYAN $\Sigma$ provides a probability of membership in Columba of 83.5\% for this star, so we discuss it further in Silverberg et al. (in prep).  This star's WISE excess was first identified by \citet{2013ApJS..208...29W}. \citet{2017AJ....153..257O} found HD 37852 to be comoving with four other stars, three of which were included in \citet{2019MNRAS.486.3434L} as members of Columba.  The fourth of these comovers is HD 274311.

%THIS OBJECT IS IN STEVEN'S PAPER.  DON'T DELETE IT THOUGH BECASUE IT LOOKS LIKE WE HAVE SOMETHING NEW TO SAY ABOUT THE KINEMATICS.

%This last star is  comoving with HD 37852, . %ADD THESE TO THE PLOT. THEY MUST BE MEMBERS TOO! --added.

\smallskip

%\noindent
%{\bf HD 274561} K1V(e) at 102 pc ROSAT bright source from \citet{2009ApJS..184..138H}.  In SACY paper about YSAs. DELETE--FOUND IN \citet{2019MNRAS.486.3434L} as a BONA FIDE MEMBER.

\smallskip
\noindent
{\bf HD 274311} is a K5e at 105 pc.  This star is listed as a ROSAT bright source by \citet{2009ApJS..184..138H}. This is part of Group 55 in \citet{2017AJ....153..257O}, which has only five members.   

\smallskip

% I'm commenting out the HR diagrams for now so the referee doesn't ask us for isochrones for ALL the YSAs.

%\begin{figure}[ht]
%    \centering
%    \includegraphics[width=0.33\linewidth]{figures/HR_Diagram_Columba_Carina_and_DD.pdf}
%    \caption{HR diagram of disk candidates with Carina, Columba and Tuc-Hor where blue is COL, green is CAR, and red is Disk Detective.}
%    \label{fig:HRdiagDDwithColandCar}
%\end{figure}

\begin{figure}[ht]
    \centering
    \includegraphics[width=0.5\linewidth]{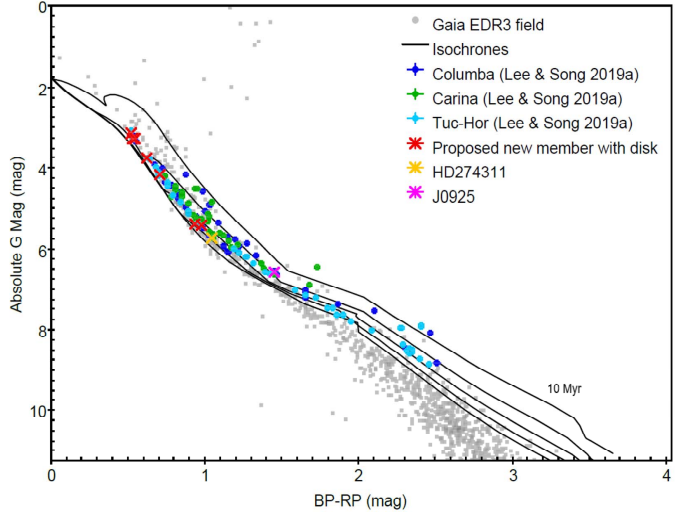}
    \caption{Colour magnitude diagram (CMD) of Carina (green), Columba (blue) and Tuc-Hor (turquoise) with proposed new members with disks (red), comover HD 274311 (orange) and proposed extreme debris disk candidate J0925 (magenta) with isochrones from 10 Myr (top) to 50 Myr (bottom) and Gaia EDR3 field stars. New association members we discover using VR are often in a mass regime where isochrones are degenerate.}
    \label{fig:HRdiagCarColTucHorDD}
\end{figure}

%Figure \ref{fig:HRdiagCarColTucHorDD} shows a CMD for Carina, Columba and Tuc-Hor, including the Gaia EDR3 field main sequence, shown in grey. The solid lines show isochrones at 10, 20, 30, 40, 50 Myr from \citet{2017ApJ...835...77M} available online\footnote{\url{http://stev.oapd.inaf.it/cmd}}. The blue dots show the 
%previously-identified members of Columba, the green dots show Carina, and the turquoise dots show Tuc-Hor, all from \citet{2019MNRAS.486.3434L}. The red x's show our proposed new members: the red objects have infrared excess, the orange x is  co-moving with them, and the magenta x shows the proposed extreme debris disk.

Figure \ref{fig:HRdiagCarColTucHorDD} shows a CMD for Carina, Columba and Tuc-Hor, including the Gaia EDR3 field main sequence, shown in grey. The solid lines show isochrones at 10, 20, 30, 40, 50 Myr from \citet{2017ApJ...835...77M} available online\footnote{\url{http://stev.oapd.inaf.it/cmd}}. The blue dots show the previously-identified members of Columba, the green dots show Carina, and the turquoise dots show Tuc-Hor, all from \citet{2019MNRAS.486.3434L}. The red x's show our proposed new members: the red objects have infrared excess, the orange x is  co-moving with them, and the magenta x shows the proposed extreme debris disk.

Toward the reddest end, the members from \citet{2019MNRAS.486.3434L} are found above the EDR3 field, consistent with the \citet{2015MNRAS.454..593B} common age estimate of around 45 Myr.  In general, the proposed new members are consistent with the previously identified members.  However, toward the blue end, where some new proposed members are, the isochrones are degenerate, making it even harder to constrain the ages of the new proposed members. Once again, this CMD highlights the trend that new members we discover using VR are often in a mass regime where isochrones are degenerate, making them hard to identify as group members in an isochronal analysis.

\section{Discussion: Spectral Energy Distributions}
\label{discussion_SED}

%Most of the infrared excess stars we identified as new YSA members only have significant excess emission in the W4 band as far as we can tell. The W1-W4 colors of all the stars are listed in Table 1.  However two of the newly identified YSA members show evidence for excess emission in multiple bands: J044115.76-351358.1 and J092521.90-673224.8. We modeled the spectral energy distributions (SEDs) for these two systems to help us better understand the distribution of circumstellar material in them (Figure~\ref{fig:SEDs}). 

\begin{figure}[ht]
    \centering
    \includegraphics[width=0.4\linewidth]{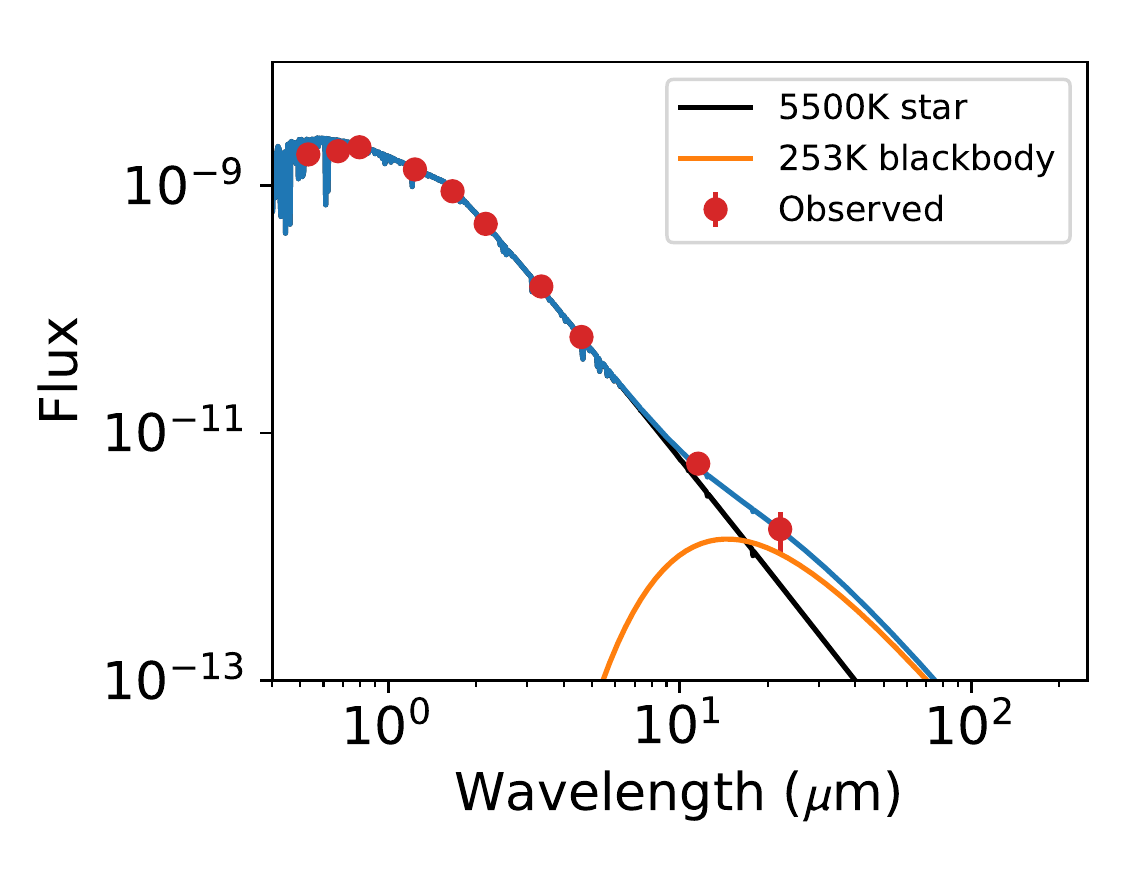}
     \includegraphics[width=0.4\linewidth]{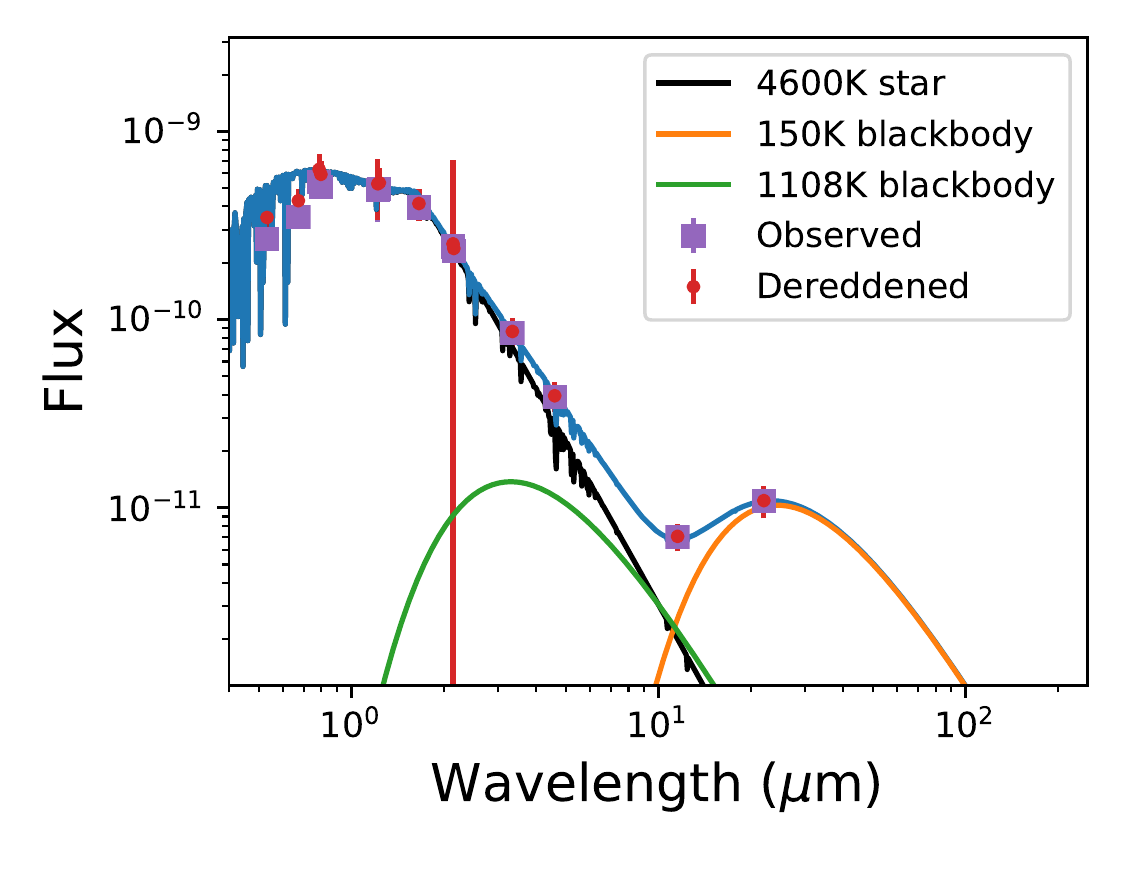}
    \caption{SEDs for J0441 (left) and J0925 (right). The photometry comes from \textit{Gaia DR2}, 2MASS, WISE, and (for J0925) DENIS. Each star is fit by a BT-Settl/CIFIST stellar model \citep{2015A&A...577A..42B}, and the infrared excess is modeled by one or two blackbody functions. Data for J0925 are dereddened following the \citet{1999PASP..111...63F} dereddening law as modified by \citet{2005ApJ...619..931I}. Note that the hot component of the excess emission from J0925 is variable, according to WISE photometry.}
    \label{fig:SEDs}
\end{figure}

Most of the infrared excess stars we identified as new YSA members only have significant excess emission in the W4 band. The W1-W4 colors of all the stars are listed in Table 1.  However two of the newly identified YSA members show evidence for excess emission in multiple bands: J044115.76-351358.1 and J092521.90-673224.8. We modeled the spectral energy distributions (SEDs) for these two systems to help us better understand the distribution of circumstellar material in them (Figure~\ref{fig:SEDs}). 

As an initial step in the modeling process, we  iteratively fit a BTSettl-CIFIST model \citep{2015A&A...577A..42B} to the observed photometry using maximum likelihood estimation (MLE). We initially fit only the three \textit{Gaia DR2} photometry points and the 2MASS $J$ point. Then we checked to see if the photometry in the next band (the shortest wavelength not included in the fit) was in excess (more than 5$\sigma$ above the flux predicted by the stellar model). If that next longer band did not show an excess, we redid the fit now including the photometry from this next longer wavelength band, We continued this process, adding new bands to the fit, until we either identified a photometric excess or all bands except W3 and W4 were included in the fit. This iterative method ensures that we did not accidentally incorporate points in our fit for the photosphere that contain excess emission. This model for the stellar photosphere yields $T_{\mathrm{eff}}$, log($g$), and $r_{\star}/d$, the ratio of stellar radius to distance.

Using these parameters derived from this model for the stellar photosphere, we then modeled the excess emission using one or two blackbody functions, corresponding to one or two single-temperature dust belts. We fit these to the excess emission using MLE as well, finding for each dust belt the temperature $T_{\mathrm{disk}}$ and $x_{\mathrm{disk}}$, a parameter that includes the number density of dust grains, the dust grain radius, and the optical depth of the dust. Once these dust parameters were derived, we then re-fit the entire system to find the optimum values of the continuous variables ($r_{\mathrm{star}}/d$, $T_{\mathrm{disk}}$, $x_{\mathrm{disk}}$, and in the case of a two-disk fit $T_{\mathrm{disk,2}}$, $x_{\mathrm{disk,2}}$) using maximum likelihood estimation, keeping our best-fit $T_{\mathrm{eff}}$ and log($g$) constant. We adopted the parameters derived from this new fit as our final values for the system. To estimate the uncertainties on these variables, we then ran a 5000-step Markov Chain Monte Carlo routine using \texttt{emcee} \citep{2013ascl.soft03002F}. This process yields the $5^{\mathrm{th}}$ and $95^{\mathrm{th}}$ percentile values for each of the variables.

%STEVEN: ADD DESCRIPTION OF HOW YOU COMPUTED THE SEDS.

For the first system, {\em J044115.76-351358.1}, the star is best fit as a $5500$ K G-type star, and the infrared excess is well-fit with a single-temperature blackbody. The excess has a best fit temperature of $253_{-23}^{+39}$ K, and best fit $L_{IR}/L{\star}$ of $6.27_{-0.74}^{+0.69} \times 10^{-4}$. These values are typical for warm debris disks. 

The second star, {\em TYC 9196-2916-1 (J092521.90-673224.8)}, merits a longer discussion. As Figure~\ref{fig:SEDs} shows, the infrared excess has multiple components. Additionally, the best-fit temperature and $\log(g)$ from our initial process ($T_{\mathrm{eff}} = 4400$ K, $\log(g) = 3.5$) significantly disagree with the spectroscopically-derived values for the star ($T_{\mathrm{eff}} = 4607 \pm 85$K, $\log(g) = 4.49 \pm 0.18$) from the third data release from the Galactic Archaeology with Hermes project \citep[GALAH DR3;][]{2021MNRAS.506..150B}, suggesting that interstellar extinction plays a significant role for this source. To account for this extinction, we used the Virtual Observatory SED Analyzer \citep[VOSA;][]{2008A&A...492..277B}, which incorporates extinction, to fit all available photometry at a wavelength shorter than 1 $\mu$m,  avoiding photometry that could be significantly impacted by the observed excess. We again used BT-Settl/CIFIST models, using only those models within 100 K of the $T_{\mathrm{eff}}$ and 1 dex of the $\log(g)$ for the source found in GALAH DR3, to obtain a best-fit stellar model and extinction. We then used this best-fit model and extinction, and fit the excess with blackbody functions as described above.

We found that the best-fit model for this source was an SED with a two-component blackbody excess, with temperatures $150_{-13}^{+37}$ K and $1108_{-393}^{+278}$ K and a total $L_{IR,warm}/L_{\star} \approx 1.6 \times 10^{-2}$ and $L_{IR,hot}/L_{\star} \approx 2.1 \times 10^{-2}$.
Using equation 3 from \citet{1993prpl.conf.1253B} and our best-fit temperatures and stellar luminosity ($L_{\star} \approx 0.267 L_{\odot}$), the blackbody radius of the outer disk for J0925 is $\sim 1.77$ AU, while the blackbody radius for the inner disk is $\sim 0.03$ AU. We also derive a surface area for the warm disk of $\sim 0.63$ AU$^2$, and for the hot disk of $\sim 3\times10^{-4}$ AU$^2$. Assuming 10-micron silicate grains, the dust mass of the warm disk is $\sim 8\times10^{-3}$ lunar masses, while the dust mass of the hot disk is $\sim 3\times10^{-6}$ lunar masses. This ignores larger bodies (e.g. comets) that could contain the bulk of the mass.

\section{J0925: AN EXTREME DEBRIS DISK}
\label{discussion_PP}

The Disk Detective citizen science project originally classified TYC 9196-2916-1 (J092521.90-673224.8) as a ``multiple,'' probably because of the DSS Blue, Red and IR images, which show two background sources, roughly 8 arcsec away. Disk Detective was constructed to search for stars with reliable excess at W4, and objects at this separation would be potential sources of confusion in W4 band.  However, this object has a strong excess at W3 and W4, and those background sources are not visible at all in either 2MASS or WISE images, so they could not be contributing at all to the excess at W3, and probably do not substantially contribute to the W4 excess either.

We triple checked the images of this star (also known as subject AWI00063mu in the Disk Detective v1 catalog) for contamination using IRSA finderchart, and found none. This star is too far south to have images in the Pan-STARRS archive. The nearest star in Gaia eDR3 is 11 arcsec; for comparison, WISE channels 1, 2, and 3 all have point spread functions with a Full Width at Half Maximum (FWHM) of about 6 arcsec, while WISE channel 4 has a FWHM of about 12 arcsec. The star was classified by citizen scientists Jonathan Holden and Phillip Griffith Sr.

%LETS SHOW IMAGES OF THIS OBJECT IN A FIGURE: DSS Blue, DSS WISE 3 and WISE 4,  Image size 24 arcsec.

\begin{figure}[ht]

\includegraphics[width=0.24\linewidth]{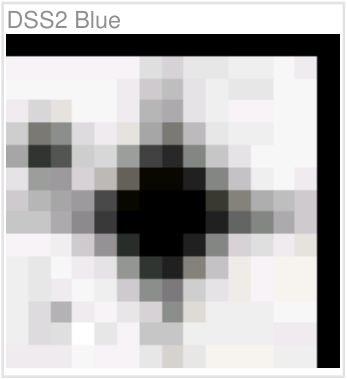}
\includegraphics[width=0.24\linewidth]{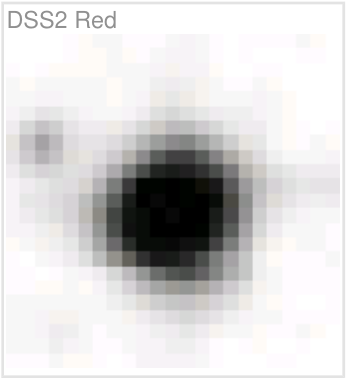}
\includegraphics[width=0.24\linewidth]{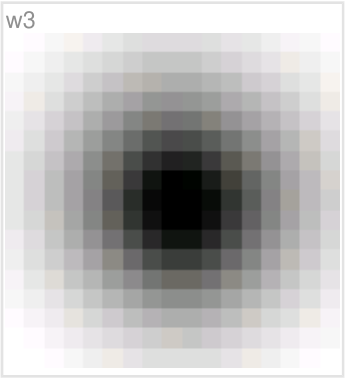}
\includegraphics[width=0.24\linewidth]{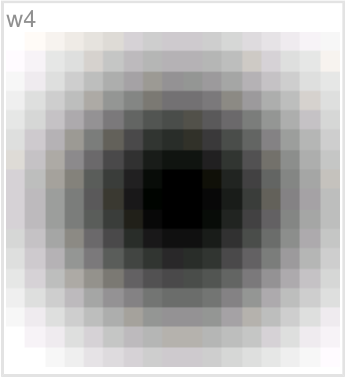}
\caption{Images of J092521.90-673224.8 in DSS blue, DSS red, WISE 3 and WISE 4. Each postage stamp is $24 \times 24$ arcseconds in angular extent. The background objects visible on the left of the DSS images do not seem to appear in the WISE images, implying that they do not explain the large excess emission measured in W3 and W4 bands.}
\label{fig:J0925_DSS_WISE}

\end{figure}

%PETER PAN DISK?
%LETS LOOK AT THE IMAGES AND MAKE SURE ITS NOT A VISUAL DOUBLE. DOES IT HAVE A SPECTRUM ANYWHERE?  E.G. LOOK FOR STARS IN GAIA THAT WOULD BE WITHIN THE 2MASS PSF.
%IF THIS IS A PETER PAN DISK IT GOES IN THE TITLE OF THE PAPER AND WE CONSIDER WRITING A PRESS RELEASE.

%It's listed as an emission line star in Cotar et al. 2021
%https://ui.adsabs.harvard.edu/abs/2021MNRAS.500.4849C/abstract

\begin{figure}[ht]
    \centering
    \includegraphics[width=0.5\linewidth]{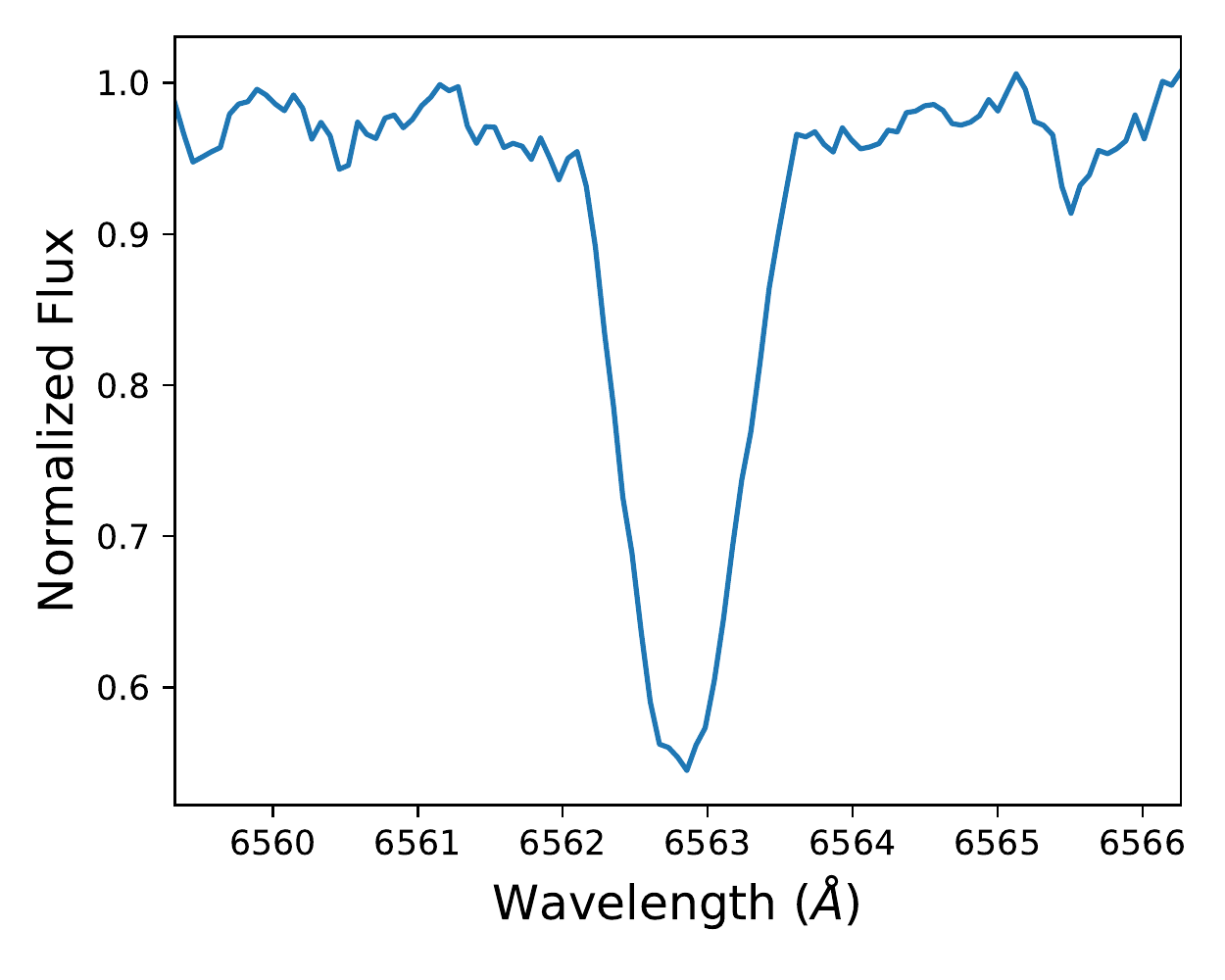}
    \caption{The spectrum of the H$\alpha$ line for J0925 from GALAH exhibits clear absorption, indicating no ongoing accretion.}
    \label{fig:J0925_Halpha}
\end{figure}

Intriguingly, this star is reported as an emission line star by \citet{2021MNRAS.500.4849C} based on a spectrum from the GALactic Archaeology with Hermes (GALAH) survey. This paper initially led us to speculate that J0925 could be hosting a Peter Pan disk, even though it is a K type star. Other known Peter Pan disk hosts are primarily M stars, which may be a consequences of disk dispersal physics  \citep{2021MNRAS.tmp.2286W}. However, an inspection of the GALAH spectrum (Figure \ref{fig:J0925_Halpha}) shows H$\alpha$ to clearly be in absorption rather than emission, indicating that there is likely no ongoing accretion.

%Assuming that this system is in fact a Peter Pan Disk, it does not stand out from other Peter Pan systems in terms of age. There are multiple other Peter Pan Disks in Carina, Columba, and Tuc-Hor. Moreover, there is another Peter Pan disk known to have a two-component excess: J0808.

Curiously, the ALLWISE variability flags for J092521.90-673224.8 indicate that this source is variable in $W1$, $W2$, and $W3$. This variability is similar to that seen for the hot-temperature component of the Peter Pan disk W0808 \citep{2018MNRAS.476.3290M}, in that it appears correlated in the $W1$ and $W2$ bands (with limited time-series data available at $W3$ compared with the data available at $W1$ and $W2$ from NEOWISE-R), with variations $\sim 0.392$ mags in both W1 and W2.  %DISCUSS VARIABILITY
However, such variability is also thought to be an inherent characteristic of extreme debris disks \citep{2012ApJ...751L..17M, 2019AJ....157..202S, 2021ApJ...910...27M}. Indeed, looking at the WISE data in terms of ``excess fraction'' (i.e. the stellar-photosphere-subtracted flux as a fraction of the flux from the photosphere) shows variations of $\sim 40\%$ at W1 and $\sim50\%$ of the underlying flux at W2. By comparison, \citet{2019AJ....157..202S} found variations of $\sim20\%$ at 3.6 $\mu$m and $\sim60\%$ at 4.5 $\mu$m for the ID8 system, which correspond to the lower temperature of the ``hot'' excess in that system compared to J0925. This indicates that J0925 can quite reasonably be explained as an extreme debris disk.

%In fact, \citep{2012ApJ...751L..17M} notes their 2 sources also exhibit 24 micron variability

\citet{2009ApJ...698.1989B}, \citet{2015ApJ...805...77M}, and \citet{2021ApJ...910...27M} describe extreme debris disks (EDDs) as warm disk-hosting stars between 10 and 200 Myr in age, with unusually high fractional luminosities and strong mid-IR excess emission.  
The dust temperature of these EDDs are found to be $>300$K, and the fractional luminosity $>0.01$ is about three orders of magnitude higher than normally found in debris disks (\cite{2009ApJ...698.1989B} in \citet{2021ApJ...910...27M}).  
The high dust luminosity could be due to recent collision events and their aftermath \citet{2013ApJ...777...78S}.
Relatively few EDDs are known;  \citep{2021ApJ...910...27M} lists a sample of ``all known EDDs'' containing only 17 stars.

We are not aware of any other K type extreme debris disks in Columba-Carina. However, the F5V Columba member HD 35841 hosts a debris disk with a fractional infrared excess of $L_{IR}/L_{\star} \approx 10^{-3}$.  Much of this excess arises from a narrow circular ring at r = 19-20 au, based on images from HST and Gemini \citep{2018AJ....156...47E}. For comparison, J0925 appears to have more dust with an $L_{IR}/L_{\star}  \approx 16$ times that of HD 35841.

Other young stellar associations do have K-type stars with extreme debris disks.
\citet{2015ApJ...798...86Z} discovered V488 Per, a K-type star with an age of 90 Myr.
With a characteristic temperature of $\sim 800$ K, this dust orbits only $\sim 0.06$ au from the star (assuming blackbody grains). This star appears to have more dust with a $L_{IR}/L_{\star}$ of $\sim 0.16$ which is $\approx$ 10 times that of J0925.

Another interesting K star with an extreme debris disk is TYC 8241-2652-1, a K2V in LCC  \citep{2018ApJ...868...32G}.
Curiously, the mid-IR luminosity of this star was observed to decay significantly over less than two years. It began with a warm ($\sim 450$K), unusually dust-rich disk with a $L_{IR}/L_{\star} \approx 0.11$, and evolved into a colder ($<$ 200K), more tenuous disk with a $L_{IR}/L_{\star} \sim 10^{-3}$ \citep{2012Natur.487...74M}. 
\citet{2016MNRAS.461..794P} found this star to have a spectral type of K3IV(e), and assigned it to the Scorpius-Centaurus OB association, with an age of 14 Myr.  For comparison, J0925 appears to be in between the two extremes of this star in terms of luminosity of its disk.

\section{Theia Groups}
\label{theiagroups}

%REMEMBER TO TAKE OUT THE GAIA DR2 NAMES AND SWAP IN THE WISE IDs FROM THIS SECTION.

The moving groups discussed above are clustered in position, velocity and age. But  \citet{2019AJ....158..122K} drew attention to the notion of diffuse moving groups, clustered in velocity and age but relatively unclustered in position. \citet{2019AJ....158..122K} used unsupervised machine learning on Gaia DR2 data to find 1901 such diffuse groups of co-moving stars, calling them ``Theia'' groups.

Subsequently, \citet[][GFMP21]{2021ApJ...915L..29G} proposed that fourteen of these Theia groups may be kinematically related to the YSAs discussed in Section \ref{NewCandidateYSAs} and \ref{Discussion_Car-Col_and_THA}. This observation led GFMP21 to suggest that some of these moving groups are possibly tidal tails from more distant open clusters. 
In the process, GFMP21 found that several of the disk candidates and co-moving stars listed in Tables \ref{table:DDinknownMGs} and \ref{table:ComoverstoDDinknownMGs} could be associated with Theia groups: Theia 92, Theia 208 and Theia 301. 
Here is a summary of these assignments suggested by GFMP21.

\smallskip
\noindent
{\bf HD 41992/WISEA J060652.79-313054.1} - GFMP21 assigned this object to Theia 208, which they argued is likely related to Columba and Carina. 
We found that this star has a position consistent with membership in Columba-Carina, but a velocity consistent with membership in Tuc-Hor, as shown in Figure  \ref{fig:vectors_TucHor_LS}. 

\smallskip
\noindent
{\bf HD 35841/WISEA J052636.59-222923.8} - GFMP21 assigned this star to Theia 208.

\smallskip
\noindent
{\bf HD 274311/Gaia DR2 4806146576925723264} - GFMP21 found this object to be part of Group 55 in \citet{2017AJ....153..257O}, which has only five members. Group 55 is near Theia 208, but not part of it.

\smallskip
\noindent
{\bf HD 44775/WISEA J062218.66-295134.7} - GFMP21 assigned this star to Theia 301; \citet{2019AJ....158..122K} and GFMP21 suggested that Theia 301 might be related to AB Doradus.

%is a member of Theia 301, which  \citet{2021arXiv210611873G} proposed is related to AB Dor.

\smallskip
\noindent
{\bf HD 40540/WISEA J055752.60-342834.1} - same as above.

%is a member of Theia 301, which  \citet{2021arXiv210611873G} proposed is related to AB Dor.

\smallskip
\noindent
{\bf HD 44510/WISEA J061903.93-535823.9} - same as above.

%is a member of Theia 301, which  \citet{2021arXiv210611873G} proposed is related to AB Dor.

\smallskip
\noindent
{\bf TYC 6518-1857-1/Gaia DR2 2898402643271131264} - same as above.

%is a member of Theia 301. \citet{2019AJ....158..122K} and \citet{2021arXiv210611873G} suggested that Theia 301 might be related to AB Doradus.

%This is a member of Theia 301 which  \citet{2021arXiv210611873G} proposes is related to AB Dor.

\smallskip
\noindent
{\bf CPD-25 1292/Gaia DR2 2911909593862390912} - same as above.

\section{CONCLUSIONS}

We examined eight nearby young moving groups using a novel VR technique, and found seven new likely disk-hosting members of these groups. Each one of these stars is likely to be a valuable target for coronagraphic imaging. 
The infrared excesses for most of our objects of interest had been previously identified in other searches, except for two stars, which Disk Detective first identified as having infrared excess (J092521.90-673224.8 and J060652.79-313054.1). 
We also identified four other likely new members of these moving groups that do not have disks, but which are comovers with other new members. We suggested that three M dwarfs, previously considered members of Tuc-Hor are better considered a separate moving group, tentatively called Smethells 165.

These findings demonstrate the value of new visual examinations of the shapes and memberships of these moving groups, and VR as a tool for performing such examinations. 
Note that none of these objects was identified by BANYAN $\Sigma$ as a high probability group member (we are saving a discussion of group members identified via BANYAN $\Sigma$ for a future paper). 
Furthermore, our VR technique identified new candidate members often found in a mass regime where isochrones are degenerate, which highlights the unique value that the VR tool demonstrates in this area.

The new proposed group members include two stars (J092521.90-673224.8 and HD 41992) with positions consistent with membership in Columba-Carina, but velocities consistent with membership in Tuc-Hor. They may have been neglected by earlier studies for exactly this reason.  One of these, {\em TYC 9196-2916-1 (J092521.90-673224.8)}, 
appears to host an extreme debris disk with a two-component infrared excess of $L_{IR,warm}/L_{\star} \approx 1.6 \times 10^{-2}$ and $L_{IR,hot}/L_{\star} \approx 2.1 \times 10^{-2}$. Its characteristics are not unusual when compared to other K-type extreme debris disks.

%We do not yet have spectroscopic evidence for accretion onto this star--but assuming that this object does belong in the category of Peter Pan Disks, it is the earliest type star to host a Peter Pan disk.

% \citet{2020MNRAS.496L.111C} modeled the survival of Peter Pan disks around both 0.1 solar mass and 1 solar mass stars, and found no reason why such a disk should not survive around a K dwarf based on their models.  However, not only are most Peter Pan disks around M dwarfs, but also a higher primordial disc fraction is seen around lower mass stars \citep{2018AJ....156...75E}.  \citet{2020MNRAS.496L.111C} suggest that these trends could reflect higher photoevaporation rates or higher disk viscosities around more massive stars. The question remains: why was this particular K dwarf exempt from these trends?

Since we completed this work, Gaia Early Data Release 3 (EDR3) has come out, and a new version of the Disk Detective citizen science project has launched. Disk Detective 2.0 utilizes the Zooniverse Panoptes platform, and targets a Gaia-selected sample of nearby systems as opposed to a photometrically-selected sample. This updated project has the express goals of finding new YSA members and new Peter Pan disks. We expect VR to help us with this second generation project as well.  

\label{CONCLUSIONS}

\acknowledgements

The authors thank Sarah Logsdon, Michaela Allen, Petr Pokorny and Veselin Kostov for helpful discussions, and Joel Kastner for valuable advice and expertise.  SH thanks Professor Hugh Hill at International Space University (ISU) for unwavering encouragement and support.

The authors acknowledge support from grant 14-ADAP14-0161 from the NASA Astrophysics Data Analysis Program and 
grant 16-XRP16\_2-0127 from the NASA Exoplanets Research Program. M.J.K. acknowledges funding from the NASA Astrobiology Program via the Goddard Center for Astrobiology. Support for this work was provided by NASA through the Space Telescope Science Institute's Directors Discretionary Research Fund (DDRF). The Space Telescope Science Institute is operated by AURA, Inc., under NASA contract NAS 5-26555.  The material is based upon work supported by NASA under award number 80GSFC21M0002.

%Based in part on observations obtained at the Gemini Observatory, acquired through the Gemini Observatory Archive and processed using the Gemini IRAF package, which is operated by the Association of Universities for Research in Astronomy, Inc., under a cooperative agreement with the NSF on behalf of the Gemini partnership: the National Science Foundation (United States), National Research Council (Canada), CONICYT (Chile), Ministerio de Ciencia, Tecnolog\'{i}a e Innovaci\'{o}n Productiva (Argentina), Minist\'{e}rio da Ci\^{e}ncia, Tecnologia e Inova\c{c}\~{a}o (Brazil), and Korea Astronomy and Space Science Institute (Republic of Korea).

%Based in part on observations at Cerro Tololo Inter-American Observatory, National Optical Astronomy Observatory (2017A-0259; PI: S. Silverberg; 2017B-0229; PI: S.Silverberg; 2018A-0292; PI: S. Silverberg), which is operated by the Association of Universities for Research in Astronomy (AURA) under a cooperative agreement with the National Science Foundation. 

This publication makes use of data products from the Wide-Field Infrared Survey Explorer, which is a joint project of the University of California, Los Angeles, and the Jet Propulsion Laboratory (JPL)/California Institute of Technology (Caltech), and NEOWISE, which is a project of JPL/Caltech. WISE and NEOWISE are funded by NASA.

This work has made use of data from the 2MASS project and the Digitized Sky Survey. 2MASS is a joint project of the University of Massachusetts and the Infrared Processing and Analysis Center (IPAC) at Caltech, funded by NASA and the NSF. The Digitized Sky Survey was produced at the Space Telescope Science Institute under U.S. Government grant NAG W-2166. The images of these surveys are based on photographic data obtained using the Oschin Schmidt Telescope on Palomar Mountain and the UK Schmidt Telescope. The plates were processed into the present compressed digital form with the permission of these institutions.

This work has made use of data from the European Space Agency (ESA) mission
{\it Gaia} (\url{https://www.cosmos.esa.int/gaia}), processed by the {\it Gaia}
Data Processing and Analysis Consortium (DPAC,
\url{https://www.cosmos.esa.int/web/gaia/dpac/consortium}). Funding for the DPAC
has been provided by national institutions, in particular the institutions
participating in the {\it Gaia} Multilateral Agreement.

%The Pan-STARRS1 Surveys (PS1) and the PS1 public science archive have been made possible through contributions by the Institute for Astronomy, the University of Hawaii, the Pan-STARRS Project Office, the Max-Planck Society and its participating institutes, the Max Planck Institute for Astronomy, Heidelberg and the Max Planck Institute for Extraterrestrial Physics, Garching, The Johns Hopkins University, Durham University, the University of Edinburgh, the Queen's University Belfast, the Harvard-Smithsonian Center for Astrophysics, the Las Cumbres Observatory Global Telescope Network Incorporated, the National Central University of Taiwan, the Space Telescope Science Institute, the National Aeronautics and Space Administration under Grant No. NNX08AR22G issued through the Planetary Science Division of the NASA Science Mission Directorate, the National Science Foundation Grant No. AST-1238877, the University of Maryland, Eotvos Lorand University (ELTE), the Los Alamos National Laboratory, and the Gordon and Betty Moore Foundation.

This research has made use of the SIMBAD database, operated at CDS, Strasbourg, France. Some of the data presented in this paper were obtained from the Mikulski Archive for Space Telescopes (MAST). STScI is operated by the Association of Universities for Research in Astronomy, Inc., under NASA contract NAS5-26555. Support for MAST for non-HST data is provided by the NASA Office of Space Science via grant NNX13AC07G and by other grants and offices. This research has made use of the VizieR catalogue access tool, CDS, Strasbourg, France.

This research has also made use of the Tool for OPerations on Catalogues And Tables (TOPCAT) graphical application  \citep{2005ASPC..347...29T}.  

%HAVE WE ALSO USED THE VIZIER SED TOOL?

%IRAF is distributed by the National Optical Astronomy Observatory, which is operated by the Association of Universities for Research in Astronomy (AURA) under a cooperative agreement with the National Science Foundation. PyRAF is a product of the Space Telescope Science Institute, which is operated by AURA for NASA. This research made use of ds9, a tool for data visualization supported by the Chandra X-ray Science Center (CXC) and the High Energy Astrophysics Science Archive Center (HEASARC) with support from the JWST Mission office at the Space Telescope Science Institute for 3D visualization.

%\software{IRAF \citep{1993ASPC...52..173T}, PyRAF, AstroPy \citep{2013A&A...558A..33A}, NumPy \citep{van2011numpy}, SciPy \citep{jones_scipy_2001}, Matplotlib \citep{Hunter:2007}, pandas \citep{mckinney}, AstroImageJ \citep{2017AJ....153...77C}, L.A.Cosmic \citep{2001PASP..113.1420V}, Spextool \citep{2003PASP..115..389V,2004PASP..116..362C}, PyVAN \citep{2019arXiv190303240L}}

\facilities{CTIO:2MASS, Sloan. WISE, Gaia}

\bibliographystyle{aasjournal}
\bibliography{apj-jour,references}

%% This command is needed to show the entire author+affilation list when
%% the collaboration and author truncation commands are used.  It has to
%% go at the end of the manuscript.
%\allauthors

%% Include this line if you are using the \added, \replaced, \deleted
%% commands to see a summary list of all changes at the end of the article.
%\listofchanges

\end{document}